\documentclass{article}
\usepackage{graphicx,epsfig}
\usepackage{graphicx,subfigure}
\usepackage{arxiv}
\usepackage{amsmath}
\usepackage{amssymb}
\usepackage{titlesec}
\usepackage[utf8]{inputenc} % allow utf-8 input
\usepackage[T1]{fontenc}    % use 8-bit T1 fonts
\usepackage{url}            % simple URL typesetting
\usepackage{booktabs}       % professional-quality tables
\usepackage{amsfonts,bm}       % blackboard math symbols
\usepackage{microtype}      % microtypography
\usepackage[mathscr]{euscript}
\usepackage[dvipsnames]{xcolor}
\pdfoutput=1
\usepackage{hyperref}
\usepackage{url}            % simple URL typesetting
\usepackage{doi}
\hypersetup{
    colorlinks=true,
    linkcolor=black,
    citecolor=black,
    filecolor=black,
    urlcolor=black,
}

\makeatother
\usepackage[numbers, compress]{natbib}
\title{Stability of surfactant-laden double-layered viscoelastic fluids flowing over an inclined plane
}
\author{
   \href{https://orcid.org/0000-0002-6439-0701}{\includegraphics[scale=0.06]{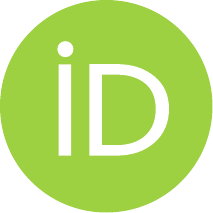}\hspace{1mm}Md. Mouzakkir Hossain}\\
   School of Energy Science and Engineering\\
   Harbin Institute of Technology\\
  Harbin 150001, China\\
  \texttt{mm\_hossain@hit.edu.cn} \\
  \And
 \href{https://orcid.org/0000-0001-9944-3490}{\includegraphics[scale=0.06]{orcid.pdf}\hspace{1mm}Mohamin B.M. Khan}\\
  Department of Mathematics\\
   Manarat Riyadh International\\
  Riyadh 12488,  Saudi Arabia\\
  \texttt{ muhaymin77@gmail.com} \\ 
\And
  \href{https://orcid.org/0000-0001-8495-786X}{\includegraphics[scale=0.06]{orcid.pdf}\hspace{1mm}Youchuang Chao}\thanks{The author to whom correspondence should be addressed}\\
  School of Energy Science and Engineering\\
   Harbin Institute of Technology\\
  Harbin 150001, China\\
  \texttt{ychao@hit.edu.cn} \\
  }

\begin{document}
\maketitle

\begin{abstract}

The linear dynamics and instability mechanisms of double-layered weakly viscoelastic fluid flowing over an inclined plane are analyzed in the presence of insoluble surfactant at both the free surface and interface. The constitutive equation of the non-Newtonian flow field follows the rheological property of Walters' $B^{''}$ model. The Orr-Sommerfeld-type boundary value problem is derived using the classical normal-mode approach and numerically solved within the framework of the Chebyshev spectral collocation method. Numerical analysis detects three distinct types of instabilities: surface, interface, and interface surfactant modes. The viscoelasticity of both the top and bottom layers strengthens the surface wave instability in the longwave region. On the other hand, the behavior of interfacial wave instability depends on both viscosity and density stratification. Stronger top-layer viscoelasticity suppresses interfacial instability, while increased bottom-layer viscoelasticity amplifies it, provided the viscosity ratio $m>1$. However, in the case of $m<1$, top-layer viscoelasticity provides strong interfacial wave stabilization in the longwave region but becomes comparatively weak in the shortwave regime. The viscoelasticity of the bottom layer has a destabilizing/stabilizing effect on the interfacial wave in the longwave/shortwave regions. Meanwhile, top-layer viscoelasticity stabilizes the interfacial surfactant mode. However, this mode can be destabilized in the vicinity of the instability threshold but is effectively stabilized far away from the onset of instability by higher bottom-layer viscoelasticity. Additionally, at high Reynolds numbers with low inclination angles, a new instability, named shear mode, emerges in both layers when the viscosity and density of the bottom layer become much stronger than those of the top layer. Both top and bottom-layer viscoelasticity exhibit the destabilizing effect, strengthening the shear wave instability of the top layer. On the contrary, the bottom-layer shear wave instability is suppressed/amplified with the increase in viscoelasticity of the top/bottom layered viscoelastic fluid. The shear wave instability exhibits significantly stronger sensitivity to the viscoelastic coefficients of both layers than all the aforementioned unstable modes, while the interface and interface surfactant modes are more sensitive than the surface mode.

\end{abstract}

% keywords can be removed

\section{Introduction}

Two-layer viscoelastic fluid flows, especially those involving free surfaces, serve as fundamental and complex systems for understanding interfacial wave dynamics relevant across a range of scientific and industrial domains. The stability characteristics of such flows are of particular interest in applications such as curtain and slide coating, where uniform layer deposition is critical (\citet{weinstein2004coating}), and in thin-film evaporators, where interfacial instabilities can enhance transport phenomena (\citet{tilley1994nonlinear}). The viscoelasticity, which is a key feature in polymer solutions, biological fluids, and engineered coatings, introduces elastic effects that can significantly alter the stability behavior of the double-layered flows. Furthermore, surfactants, which modify interfacial tension through Marangoni stresses, add complexity, either stabilizing or destabilizing the flow depending on their distribution and interaction with the underlying fluid dynamics. While previous studies have extensively examined Newtonian two-layer films with surfactants, the stability of non-Newtonian viscoelastic counterparts remains unexplored. Therefore, it is required to bridge that gap by analyzing a surfactant-laden, double-layered viscoelastic fluid, which captures instability characteristics of weak elasticity through a minimal extension of Newtonian theory. By investigating the interplay between viscoelasticity, surfactant transport, and interfacial shear, in this work, we provide deep insights into the dynamics of double-layered viscoelastic flows, which is crucial for optimizing coating processes and understanding physiological flows, such as mucus transport in pulmonary airways.

The study of two-layer fluid systems has evolved significantly since the foundational work of~\citet{benjamin1957wave} and~\citet{yih1963stability} on single-layer free-surface flows, wherein they first demonstrated the stability characteristics of the surface mode by solving the Orr-Sommerfeld-type equation using the longwave asymptotic expansion. \citet{lin1967instability} further improved the analysis by identifying an additional instability, known as shear mode instability in the single-layered film, based on the higher phase speed of shear waves compared to surface waves, when the inertia force becomes very strong with a low inclination angle. Later, a series of studies (\citet{bruin1974stability, chin1986gravity, floryan1987instabilities}, and the citations therein) were involved in deciphering the primary instability of surface and shear wave instabilities in a single-layered fluid overlying an inclined wall. It was \citet{kao1965stability, kao1965role} and \citet{kao1968role} who first extended the single-layer framework to two-layer fluid systems, capturing shear-driven instabilities at Reynolds numbers close to the instability onset, and identified two canonical unstable modes: surface and interface modes. The surface mode, originating at the top layer surface, is primarily driven by gravity, inertia, and viscosity contrast, resembling the classical instability in single-layer falling films. In contrast, the interface mode arises at the internal fluid-fluid boundary, governed by a jump at the interface in the slope of the velocity profile due to the viscosity mismatch. This mode is particularly sensitive to interfacial tension and velocity discontinuities and becomes dominant in systems with pronounced viscosity or density gradients. Later studies, including those by \citet{chen1993wave}, extended the analysis to account for inertial effects and arbitrary wavenumbers, revealing that viscosity stratification significantly influences interfacial instability, particularly when the upper layer is less viscous. Further research by \citet{hu2006inertialess} explored inertialess spatio-temporal stability, emphasizing the influence of density and viscosity stratification on two-layer instability mechanisms. In inertialess two-layer film flows, \citet{gao2007effect} considered an insoluble surfactant at both the top-layer surface and the interface and found that surface surfactants generally stabilize long-wave disturbances, while interfacial surfactants can have destabilizing effects, especially when the upper layer is less viscous. Their idea behind considering an insoluble surfactant mainly originated from earlier research on surfactant-laden single-layer Newtonian fluids, pioneered by \citet{whitaker1966stability}, who identified a stabilizing Marangoni effect. The application of insoluble surfactants has since proven highly effective in delaying the transition to turbulence. 
A huge body of evidence (\citet{pozrikidis2003effect, blyth2004effect, samanta2014effect, anjalaiah2015effect, hu2020falling, bhat2020linear,  samanta2021effect}) is available in the context of single and double-layered fluid flow that claims that insoluble surfactant stabilizes surface instability by increasing the critical Reynolds number. \citet{samanta2014effect} further revisited the work of \citet{gao2007effect} by including the inertia effect. They solved the corresponding Orr-Sommerfeld (OS) eigenvalue problem (EVP) using both analytical (longwave expansion) and numerical (Chebyshev spectral collocation) methods. His study focused on the influence of interfacial and free-surface surfactants on wave instabilities, revealing that while interfacial instability arises due to inertia, it is suppressed by interfacial surfactant when the bottom fluid layer is more viscous than the top one.
Later, \citet{bhat2020linear} conducted a detailed investigation into the instability behavior of four distinct unstable modes in a contaminated two-layered Newtonian fluid flowing over a slippery incline, identified by numerically and analytically solving the corresponding OS BVP. They observed that while the ``surface surfactant mode" remains universally stable, the ``interface surfactant mode" becomes unstable at high interfacial P\'eclet numbers.  The surface surfactant mode, localized at the free surface contaminated with insoluble surfactants, exerts a stabilizing influence by suppressing classical surface instabilities through surface tension gradients, even in strongly stratified systems. In contrast, the interface surfactant mode, which arises at the fluid-fluid interface, exhibits destabilizing behavior depending on the interfacial P\'eclet  number and viscosity ratio. While high surfactant concentrations typically enhance this instability, it can be mitigated by increased interfacial density or viscosity contrast. More recently, \citet{bhat2023linear} extended their earlier work by performing a linear stability analysis in the high Reynolds number regime. Through energy budget analysis (\citet{smith1990mechanism}), they identified the key physical mechanisms driving shear mode instabilities under varying flow parameters. Their findings revealed that shear mode instability primarily stems from base shear stress by transferring energy to the disturbance through the Reynolds stress term.

%---------------------------------------------------------------------------------------------
%---------------------------------------------------------------------------------------------
%---------------------------------------------------------------------------------------------

The aforementioned studies in the context of two-layer flows extensively investigated the stability of Newtonian fluid, encompassing both clean and surfactant-laden cases. However, the two-layered flow systems involving non-Newtonian fluids have received much less attention, despite their immense usages, such as polymer processing (\citet{petrie1976instabilities}), drag reduction (\citet{savins1967stress}), coating and lubrication applications (\citet{davalos2013stability}). There are different categories of non-Newtonian fluids, where the viscoelastic fluid is a subclass of non-Newtonian fluid that has typical features of both viscosity and elasticity. Viscoelastic fluids such as polymer solutions, micellar suspensions, and biological fluids like mucus exhibit complex rheology that cannot be captured within the Newtonian paradigm. Given the prevalence of such fluids in industrial coating processes and physiological transport systems, the absence of a comprehensive framework for surfactant-contaminated, two-layered viscoelastic liquid films constitutes a notable area to explore. Note that there are various types of models used to represent the different rheological characteristics of viscoelastic fluid, such as the second-order liquid, the Maxwell model, and Walter's $B^{''}$ model. For a detailed discussion on the rheological modeling of non-Newtonian and viscoelastic fluids, we refer to the seminal work of \citet{bird1987dynamics}. Among these models, the Walters B$^{''}$ model, a second-order fluid model, neglects second-order contributions from relaxation and retardation times. This model is most frequently used in practical problems in fluid mechanics. The reason is that the constitutive equation of this model includes only one non-Newtonian parameter, by which it is capable of capturing essential viscoelastic behavior with minimal complexity (\citet{rallison1970second,fardin2014hydrogen}). However, this model represents only weak viscoelasticity, characterized by short or rapidly decaying memory effects. Here, the concept of rapidly fading memory refers to a physical property of weakly viscoelastic fluids, where the material quickly loses the influence of its earlier deformation history. For example, colloids, suspensions, and some manmade fluids, such as polymeric fluids, fluids with additives, and liquid crystals, etc. A series of studies (\citet{gupta1967stability, shaqfeh1989stability, dandapat1978long, cheng2000walters, sadiq2005linear, uma2006dynamics, davalos2013stability, samanta2017linear, chattopadhyay2022dynamics, du2024instability}) is available for obtaining the wave properties of the viscoelastic liquid film by using different classes of 
constitutive models of viscoelasticity. Preliminary efforts of \citet{gupta1967stability} uncovered the linear stability of the viscoelastic fluid via the second-order model in the long-wave limit, and demonstrated a destabilizing effect of elasticity at low Reynolds numbers. Building on this, \citet{shaqfeh1989stability} extended the analysis to moderate Reynolds numbers using the Oldroyd-B model. They numerically solved the OS EVP and uncovered the dual role of the Weissenberg number: near the onset of instability, elasticity promotes destabilization by lowering the critical Reynolds number, while at higher inertia, it suppresses the instability by shrinking the unstable region. Later, \citet{cheng2000walters} extended the work of \citet{gupta1967stability} for Walters' liquid B$''$ and performed both linear and nonlinear analysis by deriving surface evolution equations for film thickness $h(x,t)$. It was observed that the viscoelastic coefficient destabilizes linear instability of surface waves and also strengthens the amplitude and speed of nonlinear waves in the vicinity of instability onset.  
\citet{dandapat2008bifurcation} further generalized this framework by deriving a complex Ginzburg-Landau equation near criticality and highlighted the destabilizing role of viscoelasticity. Their nonlinear bifurcation analyses revealed that the viscoelastic parameter enhances/attenuates the subcritical unstable/supercritical stable zones. \citet{pal2021} examined the linear stability of surfactant-laden flow of Walters’ B$''$ fluids over a slippery wall. Their numerical solution of the corresponding OS EVP revealed a double role of the viscoelastic parameter on surface mode instability, exhibiting a stabilizing effect near the onset of instability and a destabilizing effect far away from it. Additionally, they observed that the viscoelastic coefficient exerts a destabilizing influence on the shear mode instability at high Reynolds numbers and small inclination angles. Currently, following the Walters’ B$''$ model, various aspects of the viscoelastic flow problem have been widely investigated (\citet{chattopadhyay2022dynamics, pal2023role, du2024instability}).  
%---------------------------------------------------------------------------------------------
%---------------------------------------------------------------------------------------------
%---------------------------------------------------------------------------------------------

Thus, in this study, we adopt Walters’ B$^{''}$ as the rheological model for the double-layered viscoelastic fluid with insoluble surfactant at the free surface and interface flowing over an inclined plane. Our work extends the earlier framework of surfactant-laden two-layered Newtonian fluid down an inclined bounding wall (\citet{samanta2014effect}) by incorporating a viscoelastic property in the liquid layers (i.e., non-Newtonian fluid) governed by the Walters B$''$ constitutive equations. Here, the primary goal is to elucidate how the addition of elasticity modifies interfacial wave dynamics in double-layered systems, particularly with surfactant-induced Marangoni effects at play. 
This investigation underscores the complex interaction of fluid properties, surfactants, and boundary conditions to determine the stability of two-layer film flows.
% Notably, it mirrors the rheological complexity observed in pulmonary airway flows, where viscoelastic mucus and endogenous surfactants govern closure dynamics. 
Similar mechanisms arise in polymer-based coating technologies, where uniform film deposition depends on the interplay between elastic stresses and Marangoni effects. Additionally, this double-layered viscoelastic framework is pertinent to drug-delivery films, emulsions, and industrial multiphase flows, where coupled surfactant–elastic interactions dictate interfacial stability (\citet{han1981multiphase}). 
The classical normal mode approach is implemented to obtain the OS EVP, which is numerically solved using the Chebyshev spectral collocation method. The paper's layout is as follows: In section \ref{MF}, the governing equations of motion along with the boundary conditions related to the double-layered flow model are described, and section \ref{numerical} elaborates the numerical methods (Chebyshev spectral collocation) and their convergence. A detailed discussion of the numerical outcomes is included in the section \ref{NRD}. Finally, a conclusion is made in section \ref{CON}.

\section{Mathematical formulation} \label{MF}

\begin{figure}[ht!]
\begin{center}
\includegraphics[width=12cm]{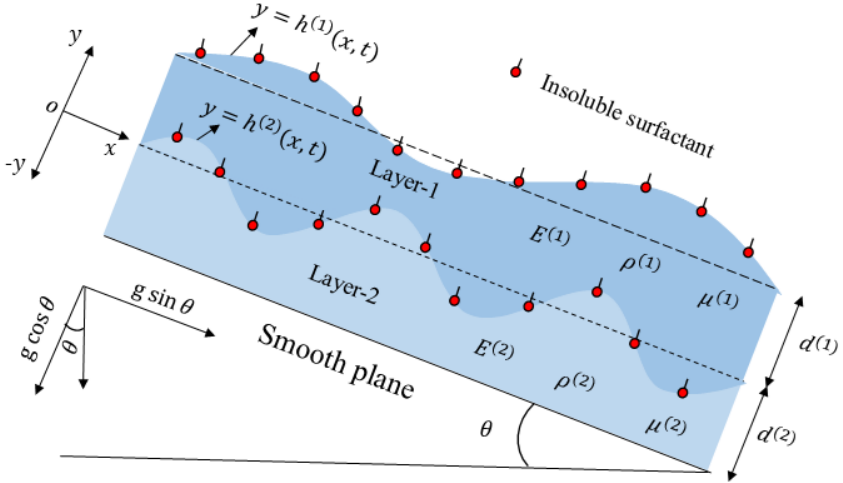}
\end{center}
\caption{Sketch of a double-layered viscoelastic fluid flow with insoluble surfactant at the top surface and interface.}\label{f1}
\end{figure}

The 2D model, as shown in Fig.~\ref{f1}, consists of incompressible, irrotational, double-layered viscoelastic liquids, with layers I and II, flowing down an incline at an angle $\theta$. The liquids contain insoluble surfactants on the free surface and the interface between the two layers. Walter's $\text{B}^{''}$ model \citet{walters1960motion} is used to describe the rheology of viscoelastic fluids I and II with viscoelastic coefficients $\mathrm{E}^{(i)}$, dynamic viscosities $\mu^{(i)}$, densities $\rho^{(i)}$, and undisturbed fluid thicknesses $d^{(i)}$, where $i = 1,\,2$.  The origin of the Cartesian coordinate frame is considered at the unperturbed liquid-liquid interface, with the $x$ and $y$ axes along streamwise and cross-stream directions of the viscoelastic liquid flow, respectively. The equations of state for the two viscoelastic fluids capturing the fluids' stress response are given as (\citet{beard1964elastico,andersson1999gravity}):  
\begin{align}
& \tau^{(i)}_{lm}=-p^{(i)}\delta_{lm}+2\mu^{(i)}e_{lm}^{(i)}-2\mathrm{E}^{(i)} \frac{\delta}{\delta t}e^{(i)}_{lm}, \quad l,\,m=x,\,y,   \end{align}
 where $\tau^{(i)}_{lm}$ is the stress tensor, $p^{(i)}$ represents the isotropic pressure, $\delta_{lm}$ denotes the Kronecker delta, $\displaystyle e^{(i)}_{lm}$ corresponds to the Newtonian stress tensor, defined as $\displaystyle e^{(i)}_{lm}=\frac{\partial u^{(i)}_l}{\partial x_m}+ \frac{\partial u^{(i)}_m}{\partial x_l}$, the term $\frac{\delta}{\delta t}e^{(i)}_{lm}$ is the polymer elastic stress, and $E^{(i)}$ is the viscoelastic coefficient.
When applying the constitutive equation, the flow must exhibit low shear rates and weak viscoelasticity. This assumption is appropriate for analyzing flows that resemble boundary layers or liquid films. For representative parameters of Walters' liquid B$''$ (\citet{walters1960motion}), a typical example is a mixture of polymethyl methacrylate in pyridine, which possesses a density of $\rho = 0.98 \times 10^3$~kg\,m$^{-3}$, a limiting viscosity of $\mu = 0.79$~N$\cdot$s\,m$^{-2}$, and a viscoelastic coefficient of $E_0 = 0.04$~N$\cdot$s$^{2}$\,m$^{-2}$. The upper-convected derivative of $e^{(i)}_{lm}$ accounting for fluid memory is:
\begin{align}
    & \frac{\delta}{\delta t}e^{(i)}_{lm}= \frac{\partial}{\partial t}e^{(i)}_{lm}+u^{(i)}_n \frac{\partial}{\partial n} e^{(i)}_{lm} -\frac{\partial}{\partial n}u^{(i)}_{m}e^{(i)}_{ln}- \frac{\partial}{\partial n}u^{(i)}_{l}e^{(i)}_{n m}, \quad n=x,~y.
\end{align}
The deformed free surface for fluid I and the interface between fluids I and II are denoted by $h^{(i)}(x,t), i=1,2$, and are covered with a monolayer of insoluble surfactants with concentrations $\Gamma^{(i)}(x,t)$, respectively. The fluid flow system in 2D is governed by the usual Navier-Stokes system of equations:
\begin{align}
     \nabla.\textbf{u}^{(i)} &= 0, \label{e4}\\
     \rho^{(i)} (\textbf{u}^{(i)}_t+\textbf{u}^{(i)}.\nabla\textbf{u}^{(i)}) &= \rho^{(i)} g+ \nabla \tau^{(i)}, 
\end{align}
 where $\textbf{u}^{(i)} = (u^{(i)},v^{(i)})$ is the fluid velocity with the gradient operator $\nabla=(\frac{\partial}{\partial x}, \frac{\partial}{\partial y})$ and $g$ is the acceleration due to the gravity. The kinematic boundary conditions on $h^{(i)}(x,t)$, $ i=1,2$, ensuring the fluid particles at an interface remain on the same interface over time via continuity of velocities, are given by
 \begin{align}
          & h^{(i)}_t = \textbf{u}^{(i)}.\nabla(h^{(i)}-y), ~~ \mbox{at}~~ y = h^{(i)}(x,t).
\end{align} 
The surfactant concentrations $\Gamma^{(i)}(x,t)$ modify the surface tension of the top layer $ \sigma^{(1)} $ when $i = 1$ and the interfacial tension $ \sigma^{(2)} $ when $i = 2$. It results in the Marangoni stresses at the top layer surface and the liquid-liquid interface that control the flow motion. Under the linear stability assumption of infinitesimal perturbations, these effects can be captured through the following linear surfactant distribution:
\begin{align}
\sigma^{(i)}(x,t) = \sigma^{(i)}_0 - \mathrm{M}^{(i)}\big(\Gamma^{(i)}(x,t) - \Gamma^{(i)}_0\big),
\end{align}
where $\sigma^{(i)}_0$ and $\Gamma^{(i)}_0$ are base surface tension and base surfactant concentration for the two layers, respectively and $\displaystyle \mathrm{M}^{(i)}= -\frac{\partial\sigma^{(i)}}{\partial\Gamma^{(i)}}\bigg|_{\Gamma^{(i)}=\Gamma^{(i)}_0}$. The evolution of the surfactant concentrations $ \Gamma^{(i)}(x,t) $ along the interfaces $ h^{(i)}(x,t) $ is governed by convection--diffusion equations (\citet{frenkel2002stokes, blyth2004evolution}), and expressed as
\begin{align}
\Gamma^{(i)}_t+u^{(i)}~\Gamma^{(i)}_x+\Gamma^{(i)}\biggl(u^{(i)}_x+u^{(i)}_y\,h^{(i)}_x\biggr)=\mathcal{D}_s\Gamma^{(i)}_{xx}~~~~~\mbox{at}~~~~~y=h^{(i)}(x,t),\label{new2}
\end{align}
where $\mathcal{D}^{(i)}_s$ represents the surfactant diffusivity on the interfaces $h^{(i)}(x,t)$.

At the free surface, the tangential component of the stress is balanced with the gradient of surface tension $\sigma^{(1)}$ (Marangoni effect), and given by
\begin{align}
& \textbf{n}^{(1)}.\tau^{(1)}.\textbf{t}^{(1)} = \nabla_s \sigma^{(1)}.\textbf{t}^{(1)}  ~~ \mbox{at}~~ y = h^{(1)}(x,t),
\end{align}
where $\displaystyle \textbf{n}^{(i)}=\frac{1}{\sqrt{1+\big(h^{(i)}_x\big)^2}}(-h^{(i)}_x,1)$ is the unit normal and $\displaystyle \textbf{t}^{(i)}=\frac{1}{\sqrt{1+\big(h^{(i)}_x\big)^2}}(1,h^{(i)}_x)$ is the unit tangential vector to the interfaces $h^{(i)}(x,t)$. Here $\nabla^{(i)}_s$ is the surface gradient operator which is defined as
$\nabla^{(i)}_s=(\textbf{I}-\textbf{n}^{(i)}\times\textbf{n}^{(i)}).\nabla$, where $\textbf{I}$ is the identity matrix and $\textbf{n}^{(i)}\times\textbf{n}^{(i)}$ is the dynamic product of the normal vector $\textbf{n}^{(i)}$ with itself.  Further, the normal stress at the free surface is balanced with the capillary pressure due to surface curvature as given in
\begin{align}
    & \textbf{n}^{(1)}.\tau^{(1)}.\textbf{n}^{(1)} = -\sigma^{(1)} \nabla.\textbf{n}^{(1)}~~ \mbox{at}~~ y = h^{(1)}(x,t),
  \end{align}
where the ambient pressure $p_\infty$ is assumed to be negligible, since the air flow at the top layer surface is passive. The corresponding balance of hydrodynamic stresses on the interface between the fluids leads to the following dynamic interface boundary conditions for tangential and normal directions (\citet{gao2007effect}), respectively, as
\begin{align}
& \textbf{n}^{(2)}.\tau^{(2)}.\textbf{t}^{(2)} -\nabla_s \sigma^{(2)}.\textbf{t}^{(2)}=\textbf{n}^{(2)}.\tau^{(1)}.\textbf{t}^{(2)}  ~~ \mbox{at}~~ y = h^{(2)}(x,t),
\end{align}
\begin{align}
    & \textbf{n}^{(2)}.\tau^{(2)}.\textbf{n}^{(2)} + \sigma^{(2)} \nabla.\textbf{n}=\textbf{n}^{(2)}.\tau^{(1)}.\textbf{n}^{(2)}~~ \mbox{at}~~ y = h^{(2)}(x,t).
  \end{align}
Continuity of both the streamwise and cross-streamwise velocity components across the interface between the fluids requires
\begin{align}
\mathbf{u}^{(1)} = \mathbf{u}^{(2)} \quad \text{at} \quad y = h^{(2)}(x,t).
\end{align}
Along the rigid substrate beneath the two-layered fluid system, the no-slip and no-penetration conditions impose
\begin{align}
\mathbf{u}^{(2)} = \mathbf{0} \quad \text{at} \quad y = -d^{(2)}.\label{elast}
\end{align}
 \subsection{Base Flow and Dimensionless Governing Equations}
 To investigate the linear instability of the present two-layer weakly viscoelastic flow system, the non-perturbed basic (or base) flow variables are assumed in the form $ (u^{(j)}, v^{(j)}) = (U^{(j)}(y), 0) $ and $ p^{(j)} = P^{(j)}(y) $, where $ j = 1,2 $ denotes the fluid layers. The base flow profiles are obtained by substituting the basic flow variables into the governing equations, together with the associated boundary conditions. Then, the average velocity $ U_c $ of the double-layer film flow on the impermeable substrate is determined by integrating the dimensional base velocities across the respective layer thicknesses and averaging over the total thickness of the film. This yields:
\begin{align}
U_c = \frac{\rho^{(1)}g\sin\theta\, (d^{(1)})^2}{\mathcal{K}\mu}\quad \text{where} \quad \frac{1}{\mathcal{K}} = \frac{1}{\delta+1} \left( \frac{1}{3} + \frac{\delta^2}{2m} + \frac{\delta}{m} + \frac{\delta^3 r}{3m} + \frac{\delta^2 r}{2m} \right),
\end{align}
where $\displaystyle\delta = \frac{d^{(2)}}{d^{(1)}}$, $\displaystyle r = \frac{\rho^{(2)}}{\rho^{(1)}}$ and $\displaystyle m = \frac{\mu^{(2)}}{\mu^{(1)}}$ are the thickness, density and viscosity ratios, respectively. The governing equations and boundary conditions (Eqs.~\eqref{e4}--\eqref{elast}) are non-dimensionalized by following scaling arguments where the characteristic velocity and length scales are taken as $U_c$ and $d^{(1)}$, respectively, for both fluid layers, while the pressures $ p^{(1)} $ and $ p^{(2)} $ are scaled by $ \rho^{(1)}U_c^2 $ and $ \rho^{(2)}U_c^2 $, respectively. Assuming a unidirectional, locally parallel base flow with fixed layer thicknesses $ d^{(i)},~i = 1,2 $, the corresponding non-dimensional base state solutions for velocity and pressure are obtained analytically, as (\citet{gao2007effect, samanta2014effect})
\begin{align}
U_b^{(1)}(y) &= \mathcal{K} \left( y - \frac{y^2}{2} + \frac{\delta(r\delta+2)}{2m} \right), \quad
U_b^{(2)}(y) = \frac{\mathcal{K}}{m} \left( y - \frac{r y^2}{2} + \frac{\delta(r\delta+2)}{2} \right), \label{base1} \\
P_b^{(1)}(y) &= \frac{\mathcal{K}}{\mathrm{Re}_1} \cot\theta\, (1-y), \quad
P_b^{(2)}(y) = \frac{\mathcal{K} \cot\theta}{r\, \mathrm{Re}_1}\, (1-ry).\label{base2}
\end{align}
 Note that the parabolic-type base velocity and linear-type pressure profiles are independent of the viscoelastic coefficient of both layers. 
Correspondingly, the dimensionless forms of equations Eqs.~\eqref{e4}--\eqref{elast} for both liquid layers are written as:
\allowdisplaybreaks
\begin{align}
&u^{(i)}_x+v^{(i)}_y=0,\label{e10}
\\
&\mathrm{Re}_i\biggl[u^{(i)}_t+u^{(i)}~u^{(i)}_x+v^{(i)}~u^{(i)}_y\biggr]=-\mathrm{Re}_i~p^{(i)}_x+\biggl[\partial_x\tau^{(i)}_{xx}+\partial_y\tau^{(i)}_{xy}\biggr]+G,
\\
&\mathrm{Re}_i\biggl[v^{(i)}_t+u^{(i)}~v^{(i)}_x+v^{(i)}~v^{(i)}_y\biggr]=-\mathrm{Re}_i~p^{(i)}_y+\biggl[\partial_x\tau^{(i)}_{yx}+\partial_y\tau^{(i)}_{yy}\biggr]-G\cot\theta,
\\
&v^{(j)}=h^{(j)}_t+u^{(j)}~h^{(j)}_x~~~~\mbox{at}~~~~y=h^{(i)}(x,t),\\
&\frac{1}{\sqrt{1+(h_x^{(1)})^2}}\biggl[\tau^{(1)}_{xy}\biggl(1-(h^{(1)}_x)^2\biggr)-\biggl(\tau^{(1)}_{xx}-\tau^{(1)}_{yy}\biggr)h_x^{(1)}\biggr]=-\frac{\mathrm{Ma}_1}{\mathrm{Ca}_1}~\Gamma^{(1)}_x~~\text{at}~~y = h^{(1)}(x,t),\\
&\frac{1}{1+(h^{(1)}_x)^2}\biggl[\tau^{(1)}_{xx}(h^{(1)}_x)^2-2\tau^{(1)}_{xy}h^{(1)}_x+\tau^{(1)}_{yy}\biggr] = \frac{[1-\mathrm{Ma}_1(\Gamma^{(1)}-1)]~h^{(1)}_{xx}}{\mathrm{Ca}_1\big[1+(h^{(1)}_x)^2\big]^{3/2}}~~\text{at}~~y=h^{(1)}(x,t),
\\
&\biggr[\tau^{(2)}_{xy}\left\{1-(h^{(2)}_x)^2\right\}-\left\{\tau^{(2)}_{xx}-\tau^{(2)}_{yy}\right\}h_x^{(2)}\biggl]+\frac{m \mathrm{Ma}_2}{\mathrm{Ca}_2}\Gamma^{(2)}_x\,\sqrt{1+ (h^{(2)}_x)^{2} }\nonumber\\
&\hspace{3cm}= \biggr[\tau^{(1)}_{xy}\left\{1-(h^{(2)}_x)^2\right\}-\left\{\tau^{(1)}_{xx}-\tau^{(1)}_{yy}\right\}h_x^{(2)}\biggl]~~~\text{at}~~~~y=h^{(2)}(x,t),
\\
&\frac{1}{1+(h^{(2)}_x)^2}\biggl[\tau^{(1)}_{xx}(h^{(2)}_x)^2-2\tau^{(1)}_{xy}h^{(2)}_x+\tau^{(1)}_{yy}\biggr]=\frac{1}{1+(h^{(2)}_x)^2}\biggl[\tau^{(2)}_{xx}(h^{(2)}_x)^2-2\tau^{(2)}_{xy}h^{(2)}_x+\tau^{(2)}_{yy}\biggr]\nonumber\\
&\hspace{5cm}-\frac{m[1-\mathrm{Ma}_2(\Gamma^{(2)}-1)]~h^{(2)}_{xx}}{\mathrm{Ca}_2\big[1+(h^{(2)}_x)^2\big]^{3/2}}~~~\text{at}~~~~y=h^{(2)}(x,t),\\
&u^{(1)}=u^{(2)}~~~~\mbox{and}~~~~v^{(1)}=v^{(2)}~~\mbox{at}~~y=h^{(2)}(x,t),
\\
&u^{(2)}=0~~~~\mbox{and}~~~~v^{(2)}=0~~\mbox{at}~~y=-\delta.\label{e19}
\end{align}
The dimensionless form of the stress tensor components $\tau^{(i)}_{lm}$ is given in the Appendix~\ref{app_1}. The ratio of inertia to the viscous force of the $i$\textsuperscript{th} layer fluid is marked by the Reynolds number $\mathrm{Re}_i=\rho^{(i)}U_cd^{(1)}/\mu^{(i)}$. The Reynolds numbers are related by $\mathrm{Re}_2 = (r/m)\mathrm{Re}_1$. $\mathrm{G}=g\sin\theta d^{(1)}/U^2_c$ defines the Galileo number (\citet{bhat2020linear, anjalaiah2013thin}). Furthermore, for the two liquid layers, the Marangoni numbers $\mathrm{Ma}_i=\mathrm{M}^{(i)}\Gamma^{(i)}/\sigma^{(i)}_0$ are associated with the surfactants and $\mathrm{Ca}_i=U_c\mu^{(i)}/\sigma^{(i)}_0$ are the Capillary numbers. Lastly, the dimensionless equation of the surfactant concentration $\Gamma^{(i)}(x,t)$  along the interfaces $ h^{(i)}(x,t) $ is given as
\begin{align}
&\Gamma^{(i)}_t+u^{(i)}~\Gamma^{(i)}_x+\Gamma^{(i)}\biggl(u^{(i)}_x+u^{(i)}_yh^{(i)}_x\biggr)=\frac{1}{\mathrm{Pe}_i}\Gamma^{(i)}_{xx}~~~~~\mbox{at}~~~~~y=h^{(i)}(x,t),\label{e20}
\end{align}
where $\mathrm{Pe}_i = U_c d^{(1)}/\mathcal{D}^{(i)}_s$ are the P\'eclet numbers. 
\subsection{Linear Stability Analysis}
A linear stability framework is used to investigate the response of the system to small-amplitude perturbations imposed on the steady base state (\citet{schmid2002stability}). In this formulation, it is assumed that the base flow varies solely in the wall-normal direction $y$, while the perturbations depend on all spatial and temporal variables, namely $x$, $y$, and $t$. The base flow profiles (Eqs.~\eqref{base1} and \eqref{base2}) are subjected to infinitesimal two-dimensional disturbances using a normal mode decomposition of the form $e^{\mathrm{i} k (x - ct)}$, where $k$ is the streamwise wavenumber and $c = c_r + \mathrm{i} c_i$ is the complex wave speed. The imaginary part of $c$ (or equivalently, the frequency $\omega_i$ in $\omega = k c = \omega_r + \mathrm{i} \omega_i$) determines the temporal behavior of the disturbance: the flow is unstable if $\omega_i > 0$, stable if $\omega_i < 0$, and neutrally stable when $\omega_i = 0$. Using this approach, the perturbed flow fields for $i=1,2$ are expressed as:
\begin{equation}
u^{(i)} = U_b^{(i)} + \hat{u}^{(i)},~~v^{(i)} = 0 + \hat{v}^{(i)},~~p^{(i)} = P_b^{(i)} + \hat{p}^{(i)},~~h^{(i)} = 1 + \hat{h}^{(i)}~~\text{and}~~\Gamma^{(i)} = 1 + \hat{\Gamma}^{(i)}\label{eq28}
\end{equation}
where the perturbations take the standard normal form
\begin{align}
\left(\hat{u}^{(i)},~ \hat{v}^{(i)},~ \hat{h}^{(i)},~ \hat{\Gamma}^{(i)}\right) = \left( \mathcal{D}\varphi^{(i)}(y),~ -\mathrm{i}k\varphi^{(i)}(y),~ \eta^{(i)}(y),~ \zeta^{(i)}(y) \right)\cdot e^{\mathrm{i}k(x - ct)}.\label{eq29}    
\end{align}
Here, $\varphi^{(i)}(y)$ denotes the stream function amplitude, the differential operator $\mathcal{D} \equiv d/dy$, $\eta^{(i)}$ is the amplitude of the deformation $\hat{h}^{(i)}$, and $\zeta^{(i)}$ is the amplitude of the perturbation surface surfactant (for $i=1$) and interface surfactant (for $i=2$) concentrations.\\
Upon substituting the linear perturbation form (Eq.~\eqref{eq28}) into the dimensionless governing Eqs.~\eqref{e10}-\eqref{e20} and then implementing the normal mode form (Eq.~\eqref{eq29}) in the linearized perturbation form, yields the following modified Orr-Sommerfeld equations (\citet{samanta2014effect, anjalaiah2015effect, bhat2020linear}), which describe the evolution of the stream function amplitudes
\begin{equation}
\biggl[1 - \mathrm{i}k\gamma_i\mathrm{Re}_i\left(U^{(i)}_b - c\right)\biggr]\left(\mathcal{D}^2-k^2\right)^2\varphi^{(i)}= \mathrm{i}k\mathrm{Re}_i\biggl[\left(U^{(i)}_b - c\right)\left(\mathcal{D}^2-k^2\right) - \mathcal{D}^2U^{(i)}_b\biggr]\varphi^{(i)},\label{orr}
\end{equation}
where $0\leq y\leq 1$ for the top fluid layer ($i=1$) and $-\delta\leq y\leq 0$ for the bottom fluid layer ($i=2$), and $\gamma_i$ is the non-dimensional viscoelastic parameter (with $\gamma_1$ for the top layer and $\gamma_2$ for the bottom layer) given by $\displaystyle \gamma_i = \frac{\mathrm{E}^{(i)}}{\rho^{(i)}(d^{(i)})^2}$. The corresponding linearized versions of the boundary conditions are evaluated  as follows:
\begingroup
\allowdisplaybreaks
\begin{align}	
\varphi^{(i)}+\left(U_b^{(i)}-c\right)\eta^{(i)}&=0,\label{freeorr}\\
\mathcal{D}\varphi^{(i)}+\mathcal{D}U_b^{(i)}\eta^{(i)}+\left(U_b^{(i)}-c-\frac{\mathrm{i}k}{\mathrm{Pe}_i}\right)\zeta_i&=0,\label{ee16c}
\end{align}
where $y=1$ for $i=1$ and $y=0$ for $i=2$. Specifically, at $y=1$, the following equations hold:
\begin{align}
&\biggl[\biggl\{ 1 - \mathrm{i} k \gamma_1 \mathrm{Re}_1 \left( U^{(1)}_b - c \right) \biggr\}\left(\mathcal{D}^2 + k^2 \right) +\mathrm{i} k~\gamma_1~\mathrm{Re}_1~\mathcal{D}^2U^{(1)}_b\biggr] \varphi^{(1)}+\mathcal{D}^2U^{(1)}_b\eta^{(1)} \nonumber\\
& \hspace{9cm}+ik\frac{\mathrm{Ma}_1}{\mathrm{Ca}_1}\zeta^{(1)} = 0,\\
&\biggl[\biggl\{1 - \mathrm{i} k \mathrm{Re}_1 \gamma_1 \left( U^{(1)}_b - c \right) \biggr\} \biggl(\mathcal{D}^3  - 3 k^2\mathcal{D}\biggr)  - \mathrm{i} k \mathrm{Re}_1 \left( U^{(1)}_b - c \right)\mathcal{D} -\mathrm{i}k\gamma_1 \mathrm{Re}_1\mathcal{D}^2U^{(1)}_b\mathcal{D}\biggr]\varphi^{(1)}&\nonumber\\
&\hspace{7cm}-ik\biggl[\mathcal{K}\cot{\theta}+\frac{k^2}{\mathrm{Ca}_1}\biggr]\eta^{(1)}=0.\label{ee16e}
\end{align}
At $y=0$, corresponding to the fluid-fluid interface, the following equations hold:
\begin{align}
&\mathcal{D}\varphi^{(1)}-\mathcal{D}\varphi^{(2)}+(m-1)\mathcal{D}U_b^{(2)}\eta^{(2)}=0,\label{eq35}
\\
&\varphi^{(1)}-\varphi^{(2)}=0,\label{eq36}\\
&\biggl[\biggl\{1-\mathrm{i}k\gamma_1\mathrm{Re}_1\biggl(U_b^{(1)}-c\biggr)\biggr\}\left(\mathcal{D}^2+k^2\right)-\mathrm{i}k\mathrm{Re}_1\gamma_1\biggl(2\mathcal{D}U_b^{(1)}\mathcal{D}-\mathcal{D}^2U_b^{(1)}\biggr)\biggr]\varphi^{(1)}\nonumber\\
&= \biggl[m\biggl\{1 - \mathrm{i}k\gamma_2 {\mathrm{Re}_2}\biggl(U_b^{(2)} - c\biggr)\biggr\}\left(\mathcal{D}^2  + k^2 \right) -\mathrm{i}km\mathrm{Re}_2\gamma_2\biggl(2\mathcal{D}U_b^{(2)}\mathcal{D}-\mathcal{D}^2U_b^{(2)}\biggr)\biggr]\varphi^{(2)}\nonumber\\
&+\biggl[\left(m\mathcal{D}^2U_b^{(2)} - \mathcal{D}^2U_b^{(1)}\right) -2\mathrm{i}k\biggl\{m\,\gamma_2\,\mathrm{Re}_2(\mathcal{D}U_b^{(2)})^2-\gamma_1\mathrm{Re}_1(\mathcal{D}U_b^{(1)})^2\biggr\}\biggr]\eta^{(2)}\nonumber\\
&\hspace{9cm}+ \mathrm{i}km\frac{{\mathrm{Ma}_2}}{{\mathrm{Ca}_2}}\zeta^{(2)},\\	
&\biggl[\biggl\{1-\mathrm{i}k\gamma_1\mathrm{Re}_1\left(U_b^{(1)}-c\right)\biggr\}\left(\mathcal{D}^3-3k^2\mathcal{D}\right)-\mathrm{i}k\gamma_1\mathrm{Re}_1\biggl\{\mathcal{D}^2U_b^{(1)}\mathcal{D}-\mathcal{D}U_b^{(1)}\left(\mathcal{D}^2-k^2\right)\biggr\}\nonumber\\
&-\mathrm{i}k\mathrm{Re}_1\biggl\{\left(U_b^{(1)}-c\right)\mathcal{D}-\mathcal{D}U_b^{(1)}\biggr\}\biggr]\varphi^{(1)}=\biggl[m\biggl\{1-\mathrm{i}k\gamma_2\mathrm{Re}_2\left(U_b^{(2)}-c\right)\biggr\}\left(\mathcal{D}^3-3k^2\mathcal{D}\right)\nonumber\\
&-\mathrm{i}km\gamma_2\mathrm{Re}_2\biggl\{\mathcal{D}^2U_b^{(2)}\mathcal{D}-\mathcal{D}U_b^{(2)}\left(\mathcal{D}^2-k^2\right)\biggr\}-\mathrm{i}km\mathrm{Re}_2\biggl\{\left(U_b^{(2)}-c\right)\mathcal{D}-\mathcal{D}U_b^{(2)}\biggr\}\biggr]\varphi^{(2)}\nonumber\\
&\hspace{7cm}-\mathrm{i}k\left[\mathcal{K}(r-1)\cot{\theta}+\frac{mk^2}{\mathrm{Ca}_2}\right]\eta^{(2)}
\end{align}
Finally, the equations corresponding to the bottom boundary $y=-\delta$ are 
\begin{align}
&\mathcal{D}\varphi^{(2)}=0\quad \mbox{and} \quad \varphi^{(2)}=0.\label{bottomorr}
\end{align}
\endgroup
The governing equations (Eqs.~\eqref{orr}-\eqref{bottomorr}) reduce to the Orr-Sommerfeld eigenvalue problem (OS EVP) for surfactant-laden double-layered Newtonian flow over a slippery inclined plane (\citet{bhat2020linear}) in the limit of vanishing slip parameter when the viscoelastic coefficients $\gamma_1$ and $\gamma_2$ become negligible in the current flow problem. Furthermore, when both insoluble surfactants (at the top layer surface and liquid-liquid interface) and viscoelastic effects are neglected, the above boundary value problem (Eqs.~\eqref{orr}-\eqref{bottomorr}) exactly recovers the classical OS EVP for two-layer Newtonian flow over an inclined plane as derived by \citet{kao1968role} and  \citet{chen1993wave}.

\section{Numerical Method and Convergence Verification}\label{numerical}

The modified Orr-Sommerfeld equations, along with the boundary conditions, are numerically solved using the Chebyshev spectral collocation method (\citet{schmid2002stability, canuto2012spectral}). Based on this method, the functions $\varphi^{(i)},\eta^{(i)},$ and $\zeta^{(i)}, i=1,2$ in the Orr-Sommerfeld Eqs.~\eqref{orr}-\eqref{bottomorr} are approximated by truncated Chebyshev series,
\begin{equation}
    \varphi^{(i)}(y) = \sum_{j=0}^{N} \varphi^{(i)}_j T_j(y), \quad \eta^{(i)}(y) = \sum_{j=0}^{N} \eta^{(i)}_j T_j(y), \quad \zeta^{(i)}(y) = \sum_{j=0}^{N} \zeta^{(i)}_j T_j(y),\label{cheb}
\end{equation}
where $T_j(y)$ are Chebyshev polynomials defined on $[-1,1]$, and $\varphi^{(i)}_j$, $\eta^{(i)}_j$ and $\zeta^{(i)}_j$   are unknown coefficients. The domain is discretized using Gauss–Lobatto collocation points $y_j = \cos(\pi j/N)$, $j = 0, 1, \ldots, N$, and the governing system is evaluated at these points.
Thus, the linear system, comprising Eqs.~\eqref{orr} and boundary conditions \eqref{freeorr}- \eqref{bottomorr}, is cast into a generalized eigenvalue problem of the form
\begin{equation}
    \mathcal{A}X = c\mathcal{B}X, \label{evprob}
\end{equation}
where $\displaystyle X = \left[\zeta^{(1)},\eta^{(1)},\varphi_0^{(1)},\varphi^{(1)}_1,\dots,\varphi^{(1)}_N,\zeta^{(2)},\eta^{(2)},\varphi^{(2)}_0,\varphi^{(2)}_1,\dots,\varphi^{(2)}_N\right]^T$ is the eigenvector of order $(2N+6)\times 1$ and the corresponding eigenvalue $c$ is the complex wave speed. 
% The functions $\varphi^{(i)},\eta^{(i)},$ and $\zeta^{(i)}, i=1,2$ are approximated by truncated Chebyshev series,
% \begin{equation}
%     \varphi^{(i)}(y) = \sum_{j=0}^{N} \varphi^{(i)}_j T_j(y), \quad \eta^{(i)}(y) = \sum_{j=0}^{N} \eta^{(i)}_j T_j(y), \quad \zeta^{(i)}(y) = \sum_{j=0}^{N} \zeta^{(i)}_j T_j(y),\label{cheb}
% \end{equation}
% where $T_j(y)$ are Chebyshev polynomials defined on $[-1,1]$, and $\varphi^{(i)}_j$, $\eta^{(i)}_j$ and $\zeta^{(i)}_j$   are unknown coefficients. The domain is discretized using Gauss–Lobatto collocation points $y_j = \cos(\pi j/N)$, $j = 0, 1, \ldots, N$, and the governing system is evaluated at these points. The eigenvalue problem is subsequently solved using the QZ algorithm (\citet{canuto2012spectral}).
The generalized matrices $\mathcal{A}$ and $\mathcal{B}$, each of order $(2N+6)\times(2N+6)$, contain differential operators and are structured based on the linear system of Eqs.~\eqref{orr} - \eqref{bottomorr}.
% \begin{equation}\label{matrixEV}
%     \mathcal{A} = \begin{bmatrix}
%         A_{11} & \mathrm{O} \\
%         \mathrm{O} & A_{22}
%     \end{bmatrix}, \quad 
%     \mathcal{B} = \begin{bmatrix}
%         B_{11} & \mathrm{O} \\
%         \mathrm{O} & B_{22}
%     \end{bmatrix},
% \end{equation}
% with
% \begin{align*}
%     A_{ii} &= \left(1 - \mathrm{i}k\gamma_i\mathrm{Re}_iU^{(i)}_b\right)\left(\mathcal{D}^2-k^2\right)^2 - \mathrm{i}k\mathrm{Re}_i\left(U^{(i)}_b\left(\mathcal{D}^2-k^2\right) - \mathcal{D}^2U^{(i)}_b\right), \\
%     B_{ii} & = -\mathrm{i}k\mathrm{Re}_i\left(\mathcal{D}^2-k^2\right)- \mathrm{i}k\gamma_i\mathrm{Re}_i\left(\mathcal{D}^2-k^2\right)^2,
% \end{align*}
% where $D^k = d^k/dy^k$, and $\mathrm{O}$ is the null operator. 
% Since the eigenvalue system consists of fourth- and second-order ordinary differential equations, it is closed by incorporating six boundary conditions (Eqs.~\eqref{obc1}–\eqref{obc3}) for different flow configurations.
The boundary conditions are imposed by replacing the appropriate rows in $\mathcal{A}$ and $\mathcal{B}$ with boundary operators for the bottom boundaries and adding appropriate rows for kinematic and dynamic conditions at the free surface and fluid-fluid interface.   
In our numerical simulations, we primarily focus on detecting the eigenvalues $c$ of the generalized EVP (Eq.~\ref{evprob}) with $c_i>0$, as these correspond to unstable modes ($\omega_i>0$), which is crucial for stability analysis. The eigenvalues with $c_i<0$ provide the stable modes ($\omega_i<0$), which are not relevant to our analysis. Also, we have numerically computed the marginal stability curve ($\omega_i=0$) by vanishing $c_i=0$ for a given set of flow parameters. 

To verify the numerical accuracy and spectral convergence, we follow the approach outlined by \citet{tilton2008linear} and \citet{samanta2017role}. For a given $N$, the relative error $E_N$ is computed using the discrete $L^2$ norm $\left(\|\cdot\|\right)$as
\begin{equation}
    E_N = \frac{\|c_{N+1} - c_N\|_2}{\|c_N\|_2},
\end{equation}
where $c_N$ and $c_{N+1}$ are vectors of eigenvalues corresponding to the twenty least stable modes computed with $N$ and $N+1$ Chebyshev polynomials, respectively.

\begin{figure}[ht!]
\begin{center}
\subfigure[]{\includegraphics[width=8.2cm]{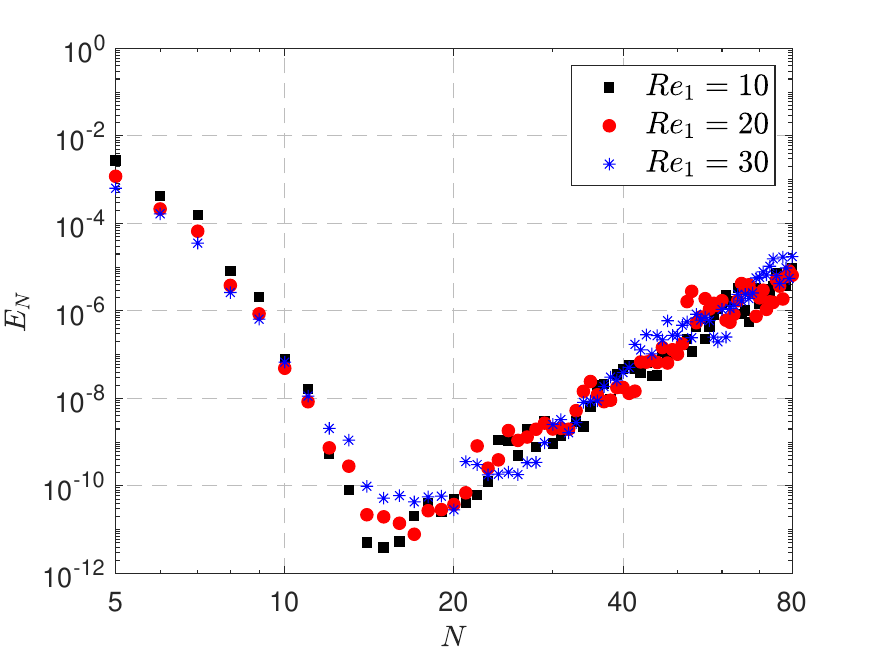}} 
\subfigure[]{\includegraphics[width=8.2cm]{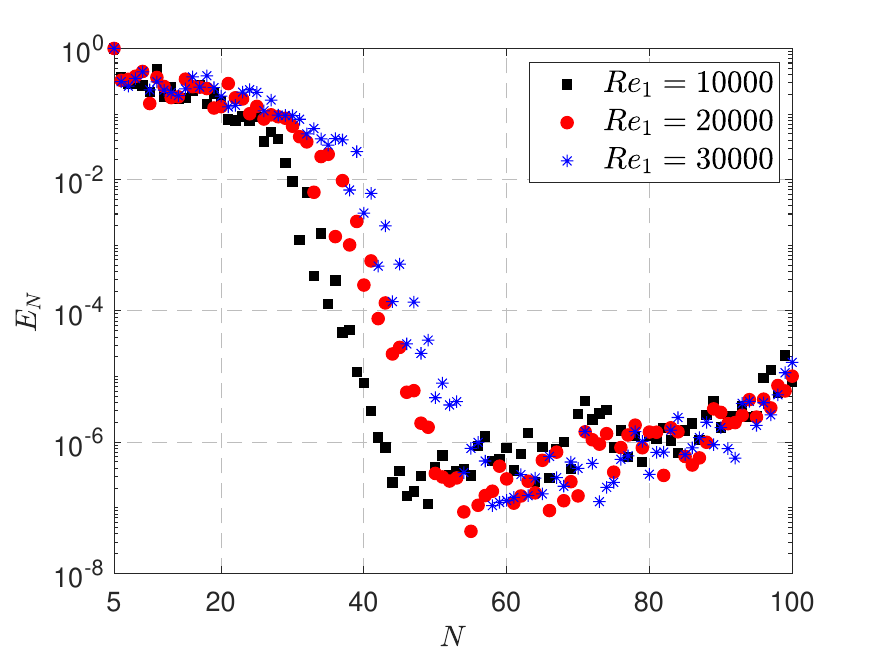}}
\end{center}
\caption{The variation of space convergence with the order of Chebyshev polynomials for moderate values of $Re_1$ when the flow parameters are $\gamma_1=0.1$, $\gamma_2=0.1$, $k=0.3$, $r=1$, $m=1.5$, $\delta=1$, $\mathrm{Ca}_1=1$, $\mathrm{Ca}_2=1$, $\mathrm{Ma}_1=0.1$, $\mathrm{Ma}_2=3.0$, $\theta=0.2~\textrm{rad}$. (b) The variation of space convergence with the order of Chebyshev polynomials for vary high values of $Re_1$ when the flow parameters are $\gamma_1=\gamma_2=10^{-5}$, $k=0.6$, $r=6$, $m=6$, $\delta=1$, $\mathrm{Ca}_1=1$, $\mathrm{Ca}_2=1$, $\mathrm{Ma}_1=1$, $\mathrm{Ma}_2=1$, and $\theta=1^{\circ}$.  }\label{err1}
\end{figure}

\begin{figure}[ht!]
\begin{center}
\subfigure[]{\includegraphics[width=8.2cm]{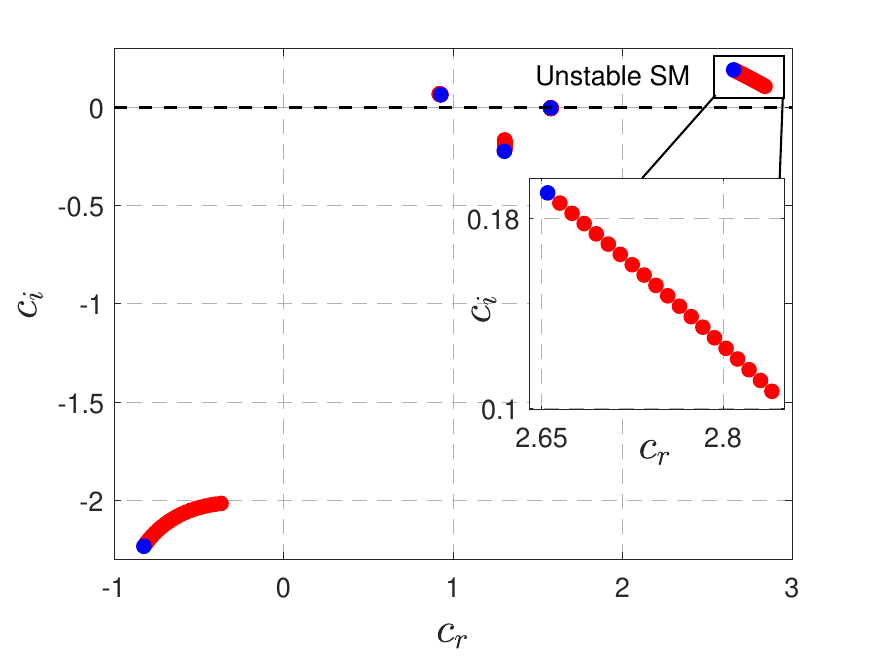}} 
\subfigure[]{\includegraphics[width=8.2cm]{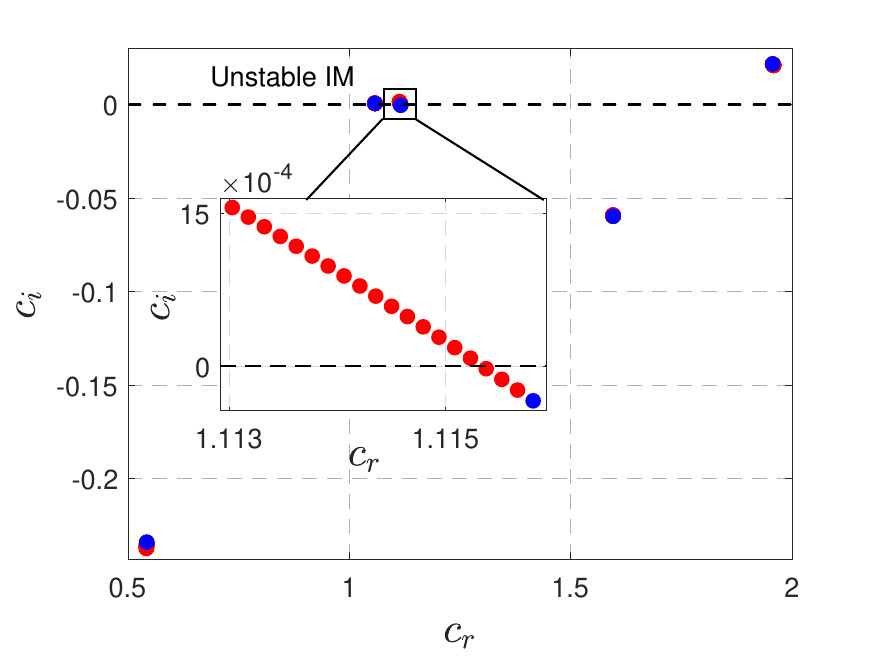}}\hspace{0cm}
\subfigure[]{\includegraphics[width=8.2cm]{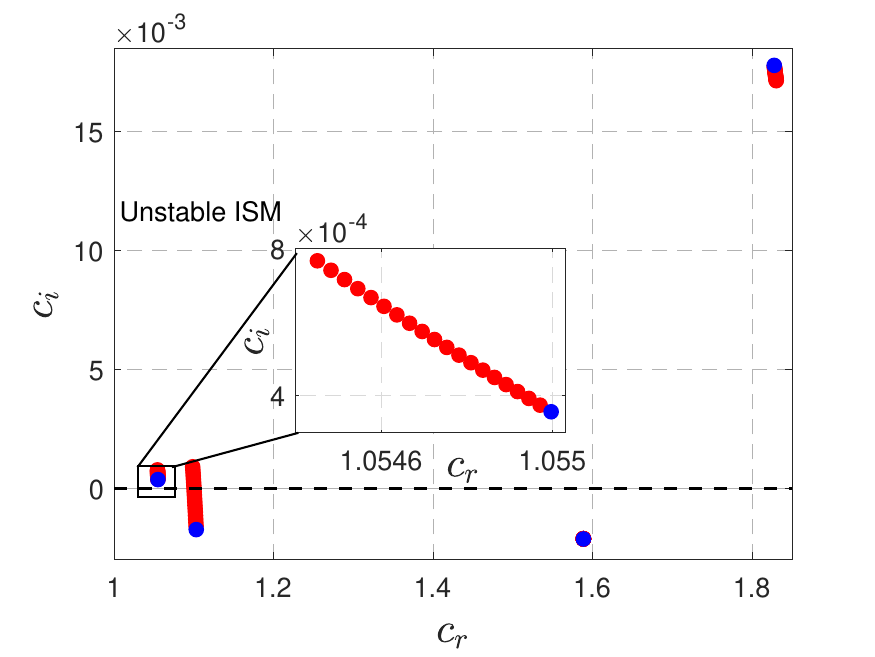}}
\subfigure[]{\includegraphics[width=8.2cm]{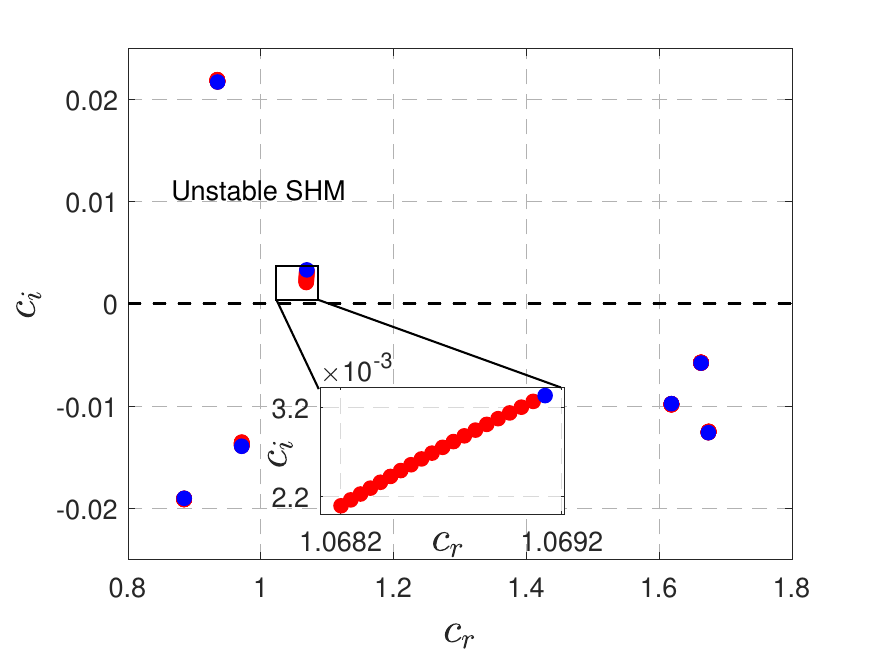}}
\end{center}
\caption{The effect of top-layer viscoelasticity $\gamma_1$ on the eigenvalues in $(c_r,c_i)$ plane when (a) $k=0.1$, $Re_1=5$, $Ma_1=0.5$, $Ma_2=3$, $\gamma_2=0.1$, $r=1$, $m=1.5$, and $\theta=0.2~\textrm{rad}$, (b) $k=0.5$, $Re_1=30$, $Ma_1=0.5$, $Ma_2=0.005$, $\gamma_2=0.001$, $r=1$, $m=1.5$, and $\theta=0.2~\textrm{rad}$, (c) $k=0.7$, $Re_1=50$, $Ma_1=0.01$, $Ma_2=0.01$, $\gamma_2=0.001$, $r=1$, $m=1.5$, $Pe_1=1000$, $Pe_2=500$, and $\theta=0.2~\textrm{rad}$, and (d) $k=1$, $Re_1=30000$, $Ma_1=1$, $Ma_2=1$, $\gamma_2=1\times 10^{-5}$, $r=6$, $m=6$, and $\theta=0.017~\textrm{rad}$. The remaining parameters are $\delta=1$, $\mathrm{Ca}_1=1$ and $\mathrm{Ca}_2=1$. Here $\gamma_1\in[0,~0.4]$ in (a), $\gamma_1\in[0,~0.004]$ in (b) and (c), and $\gamma_1\in[0,~4\times 10^{-5}]$ in (d). The solid blue circular shapes mark the optimal complex wave speed $c$.  }\label{f3}
\end{figure}

The convergence behavior of the spectral scheme is assessed by monitoring the relative error $E_N$ of the eigenvalue spectrum with increasing collocation points $N$. Figures~\ref{err1}(a) and (b) show the variation of $E_N$ with $N$ for low and high Reynolds numbers, respectively. For low Reynolds numbers, the spectral method achieves high accuracy with as few as $N = 15$ to $20$ Chebyshev polynomials (see, Fig.~\ref{err1}(a)), yielding relative errors in the range $\mathcal{O}(10^{-12})$ to $\mathcal{O}(10^{-10})$. In contrast, for flows with high Reynolds numbers, a higher resolution of $N = 50$–$60$ is required to maintain a comparable accuracy, with relative errors (see, Fig.~\ref{err1}(b)) remaining within $\mathcal{O}(10^{-8})$ to $\mathcal{O}(10^{-5})$. Thus, these results highlight two major points: the accurate numerical results at high Reynolds numbers demand more Chebyshev polynomials, and the Chebyshev spectral collocation method offers reliable convergence across a wide range of flow regimes. It is worth noting that the numerical solution of the eigenvalue problem may yield spurious eigenvalues because of homogeneous boundary conditions (Eqs.~\eqref{eq35}-\eqref{eq36} and Eq.~\eqref{bottomorr}) used in the rows of the matrix $\mathcal{A}$. However, these spurious eigenvalues are mapped to the arbitrary irrelevant stable modes by carefully choosing the complex multiple for the corresponding rows of the matrix $\mathcal{B}$. In this way, one can avoid spurious eigenvalues from the matrix eigenvalue problem Eq.~\eqref{evprob}.

Now, we have determined the reasonable parameter ranges for a pair of immiscible viscoelastic fluid flows before discussing the numerical results. To estimate the parameter ranges used in this study, we consider the properties of polymethyl methacrylate in pyridine (\citet{andersson1999gravity, walters1960motion}) as a top layer, which has a density $\rho^{(1)} = 0.98 \times 10^3~$kg m$^{-3}$, dynamic viscosity of $\mu^{(1)} = 0.79~$N s m$^{-2}$, surface tension $\sigma^{(1)}=40\times 10^{-3}~$N m$^{-1}$ and the dimensional viscoelastic coefficient $E_0 = 0.04~$N s$^{2}$ m$^{-2}$ at $25^{\circ}$. As a bottom layer, a typical example is Silicone Oil-Polyisobutylene (PIB) Mixtures, whose density is $\rho^{(2)} = 0.9 \times 10^3~$kg m$^{-3}$, dynamic viscosity of $\mu^{(2)} = 0.005~$N s m$^{-2}$, surface tension $\sigma^{(1)}=21\times 10^{-3}~$N m$^{-1}$ and the dimensional viscoelastic coefficient of $E_0 = 0.05~$N s$^{2}$ m$^{-2}$. Also, a typical mixture of  Polybutene–kerosene (\citet{kubinski2024extensional}), where $10-30\%$ polybutene $+$ $70-90\%$ kerosene (by weight), which has a density $\rho^{(1)} = 0.8-0.9 \times 10^3~$kg m$^{-3}$, dynamic viscosity of $\mu^{(1)} = 0.02-0.06~$N s m$^{-2}$, surface tension $\sigma^{(1)}= 27-32\times 10^{-3}~$N m$^{-1}$, and the dimensional viscoelastic coefficient $E_0 = 0.0001 -0.01~$N s$^{2}$ m$^{-2}$.  Now, if we consider the mean thickness $d_1=10^{-2}~$m of the top layer, then the value of the dimensionless top-layered viscoelastic coefficient $\gamma_1=0.4$ and the value of the dimensionless bottom-layered viscoelastic coefficient $\gamma_2=0.5$. Thus, the numerical analysis is performed in the current work for the viscoelasticity $\gamma_i$ of both layers from the suitable range $0$ to $0.4$ (\citet{mukhopadhyay2020waves, pal2021, chattopadhyay2022dynamics, du2024instability}).

The numerical solution of the OS-BVP (Eqs.~\eqref{orr} - \eqref{bottomorr}) detects at most four distinct unstable modes:  SM (surface mode), IM (interface mode), ISM (interface surfactant mode), and SHM (shear mode) in Fig.~\ref{f3}, based on the imaginary part of the complex wave speed relation $c_i|_{\textrm{SM}}>c_i|_{\textrm{IM}}>c_i|_{\textrm{ISM}}>c_i|_{\textrm{SHM}}$ in different flow parameter regimes (\citet{samanta2014effect, bhat2020linear, bhat2023linear}). The SM is associated with instability at the top surface, whereas the IM arises from instability at the interface between the two fluids. When insoluble surfactants are present, an additional interfacial surfactant/Marangoni mode (ISM) emerges alongside the IM. Physically, this mode arises from local variations in interfacial tension resulting from surfactant concentration gradients. These variations drive Marangoni flows, which can either amplify instability by directing fluid toward the troughs and crests of perturbed waves or suppress disturbances by redistributing fluid away from them. Beyond these modes, a distinct shear mode (SHM) is linked to the top and bottom fluid layers that can become unstable under conditions of strong inertial forces and low inclination angle. In Fig.~\ref{f3}(a), we have detected the most unstable SM in the eigenspectrum with varying top-layer viscoelasticity $\gamma_1\sim \mathcal{O}(10^{-1})$ when $k=0.1$, $Re_1=5$, $Ma_1=0.5$, $Ma_2=3$, $\gamma_2=0.1$, $r=1$, $m=1.5$, and $\theta=0.2~\textrm{rad}$. It is found that the imaginary part $c_i$ of the complex wave speed enhances as the top layer viscoelasticity $\gamma_1$ increases. That means the increment in temporal growth rate (i.e., equivalent to $c_i$) with increasing $\gamma_1$ assures the destabilizing behavior of the most unstable SM at the considered flow parameters. Next, we have identified the most unstable IM in the eigenspectrum, as in Fig.~\ref{f3}(b), which is significantly varies with $\gamma_1$ (here $\gamma_1\sim \mathcal{O}(10^{-3})$) when the other flow parameters are considered to be fixed at $k=0.5$, $Re_1=30$, $Ma_1=0.5$, $Ma_2=0.005$, $\gamma_2=0.001$, $r=1$, $m=1.5$, and $\theta=0.2~\textrm{rad}$. Here, the value $c_i$ associated with IM decreases as $\gamma_1$ increases, thereby exhibiting a stabilizing effect of $\gamma_1$ on IM with fixed parameter values. Next, we have repeated the numerical simulation for the eigenspectrum (see Fig.~\ref{f3}(c)) with varying $\gamma_1\sim \mathcal{O}(10^{-3})$ when $k=1$, $Re_1=50$, $Ma_1=0.01$, $Ma_2=0.01$, $\gamma_2=0.001$, $r=1$, $m=1.5$, $Pe_1=1000$, $Pe_2=500$, and $\theta=0.2~\textrm{rad}$. Here, the most unstable ISM, whose complex part $c_i$ attenuates as soon as $\gamma_1$ increases. This implies a stabilizing behavior of ISM under the influence of the top layer viscoelasticity $\gamma_1$ at the considered flow parameter regions. Point to note that both IM and ISM are highly sensitive to the top-layer's viscoelastic parameter $\gamma_1$ compared to the SM. Finally, in Fig.~\ref{f3}(d), we have found the variation of the most unstable SHM with $\gamma_1$ in the eigenspectrum result when the constant parameter values are $k=1$, $Re_1=30000$, $Ma_1=1$, $Ma_2=1$, $\gamma_2=1\times 10^{-5}$, $r=6$, $m=6$, and $\theta=0.017~\textrm{rad}$. In this case, the value $c_i$ corresponding to the temporal growth rate rapidly increases as $\gamma_1\sim \mathcal{O}(10^{-5})$ increases. Therefore, it is expected that the top layer viscoelasticity $\gamma_1$ has a destabilizing effect on the unstable SHM. Another important finding from Fig.~\ref{f3} is that compared to both IM and ISM, the SHM is highly sensitive with respect to the top-layer viscoelasticity $\gamma_1$. A systematic comparison of the identified unstable modes is conducted in subsection~\ref{COMP}, examining their stability dominance in the wide range of flow parameters and also the sensitivity to the viscoelasticity. Besides the numerical outcomes related to the behavior of different unstable modes from the eigenspectrum results in Fig.~\ref{f3} are limited to the fixed flow parameters. In the subsequent section, we have discussed in detail the behavior of all the identified unstable modes emerging in the contaminated double-layered viscoelastic fluid overlying an inclined plane on the large scale of different flow parameter regions.

\section{Numerical results and discussion}\label{NRD}

\subsection{Results for the surface mode (SM)}

% \begin{figure}[ht!][h!]
% \begin{center}
% \subfigure[$Ma_1=0$]{\includegraphics[width=5.6cm]{FIG_4/SM_r_k_Ma1_0_diff_gamma2.pdf}}
% \subfigure[$Ma_1=0.5$]{\includegraphics[width=5.6cm]{FIG_4/SM_r_k_Ma1_0.5_diff_gamma2.pdf}}
% \subfigure[$Ma_1=1$]{\includegraphics[width=5.6cm]{FIG_4/SM_r_k_Ma1_1_diff_gamma2.pdf}}
% \end{center}
% \caption{The variation of unstable boundary lines ($\omega_i=0$) of the SM in ($\displaystyle k-1/r$) plane for various values of bottom-layer viscoelasticity $\gamma_2$ when (a) $Ma_1=0$, (b) $Ma_1=0.5$, and (c) $Ma_1=1$. Here, the values of $\gamma_1=0$ and the viscosity ratio $m=0.4$. The remaining parameters are the same as in Fig.~\ref{f6}. }\label{f7}
% \end{figure}

\begin{figure}[ht!]
\begin{center}
\subfigure[]{\includegraphics[width=8.2cm]{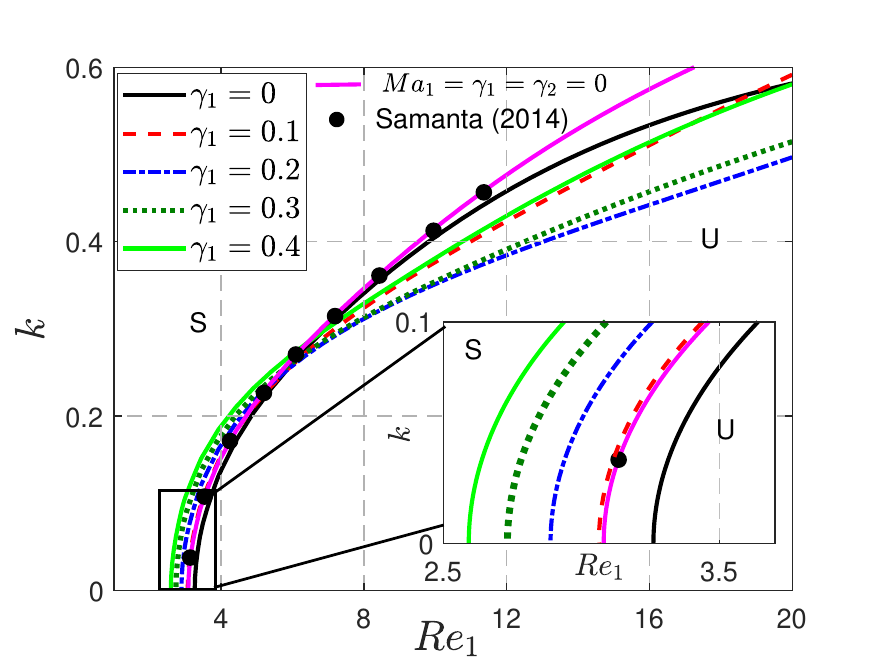}}
\subfigure[]{\includegraphics[width=8.2cm]{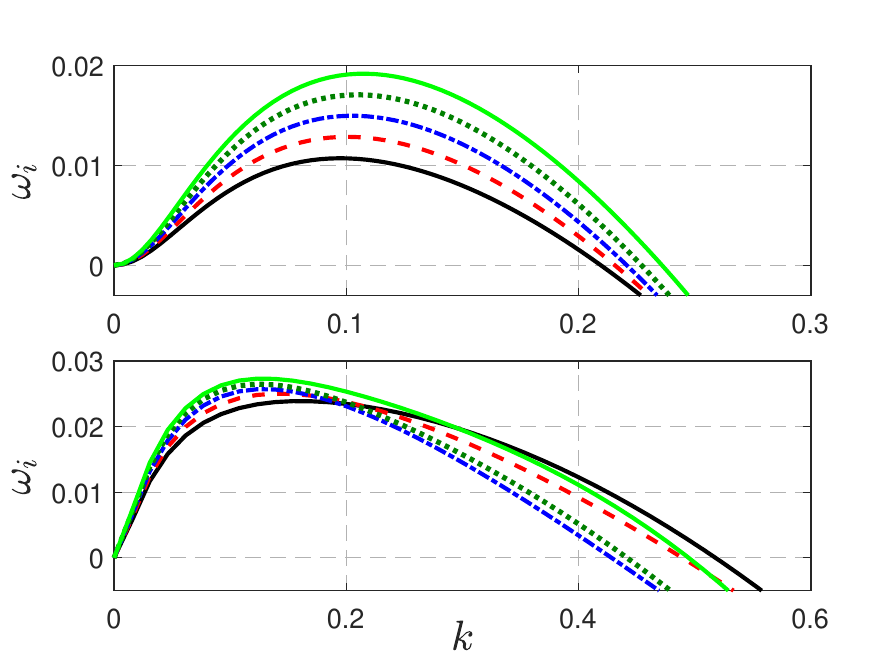}}
\end{center}
\caption{(a) The unstable boundary lines ($\omega_i=0$) of the SM in ($Re_1-k$) plane for varying top-layer viscoelasticity $\gamma_1$, (b) the corresponding temporal growth rate curves when $Re_1=5$ (top), and $Re_1=15$ (bottom). Here, the fixed value $\gamma_2=0.1$ with the remaining parameters $\gamma_2=0.1$, $r=1$, $m=1.5$, $\delta=1$, $Ca_1=1$, $Ca_2=1$, $Ma_1=0.5$, $Ma_2=3.0$, $\theta=0.2~\textrm{rad}$, $Pe_1=\infty$ and $Pe_2=\infty$. The magenta solid line is the stability boundary line for $Ma_1=\gamma_1=\gamma_2=0$, and the blue solid circular symbols represent the results of \citet{samanta2014effect}. }\label{f4}
\end{figure}

In this subsection, we have discussed the behavior of SM in the surfactant-laden two-layered non-Newtonian viscoelastic fluid. The stability boundaries related to the SM in the $(Re_1-k)$ are displayed in Fig.~\ref{f4}(a) with varying top-layer viscoelasticity $\gamma_1$. Here, the symbols `U' and `S' mark the unstable and stable regions, respectively. It is found that the current neutral stability results as in Fig.~\ref{f4}(a), match well with the available result of the double-layered Newtonian liquid over an inclined plane (\citet{samanta2014effect}) when the limiting values  $Ma_1$, $\gamma_1$, and $\gamma_2$ $\rightarrow 0$. 
The neural curve of the SM exhibits a dual nature for the top-layer viscoelastic coefficient $\gamma_1$ when a moderate Reynolds number is considered. 
As in  Fig.~\ref{f4}(a), when the Reynolds number becomes low, the unstable longwave range of the SM expands significantly as long as $\gamma_1$ enhances, followed by the successive reduction of the corresponding critical Reynolds number $Re_1^c$. As a consequence, the top-layered viscoelastic coefficient $\gamma_1$ has a destabilizing effect on the SM in the longwave region with low inertia force. However, for moderate values of the Reynolds number, while the unstable longwave region expands, the unstable region in the short-wave range shows non-monotonic behavior for higher values of $\gamma_1$. This non-monotonic nature of SM is possible due to the interplay between elastic stresses and viscous dissipation.  Hence, for moderate $Re_1$ values, the top-layer viscoelastic coefficient $\gamma_1$ has a destabilizing effect on the SM in the vicinity of the instability onset. However, the unstable SM behaves non-monotonically under the influence of $\gamma_1$ far away from the instability threshold. This destabilization nature of SM in the longwave region is similar to the behavior of SM in the single-layered viscoelastic fluid over an incline~\citep{pal2021linear}. To strengthen these facts, we have shown the corresponding temporal growth rate result as a function of $k$ for the Reynolds number $Re_1=5$ (Fig.~\ref{f4}(b)(top)) and  $Re_1=15$ (Fig.~\ref{f4}(b)(bottom)) considered from the unstable region. For low Reynolds number $Re_1=5$, the longwave instability of the SM grows as $\gamma_1$ increases, thereby confirming the destabilizing effect of $\gamma_1$. However, for a comparatively higher Reynolds number $Re_1=15$, the top-layer viscoelasticity $\gamma_1$ strengthens the maximum growth rate in the longwave region, whereas the non-monotonic trend is observed in the short-wave range. This assures the dual characteristics of SM at moderate values of $Re_1$. Hence, the above growth rate results of SM are fully consistent with the fact illustrated in Fig.~\ref{f4}(a). 

\begin{figure}[ht!]
\begin{center}
\subfigure[]{\includegraphics[width=8.2cm]{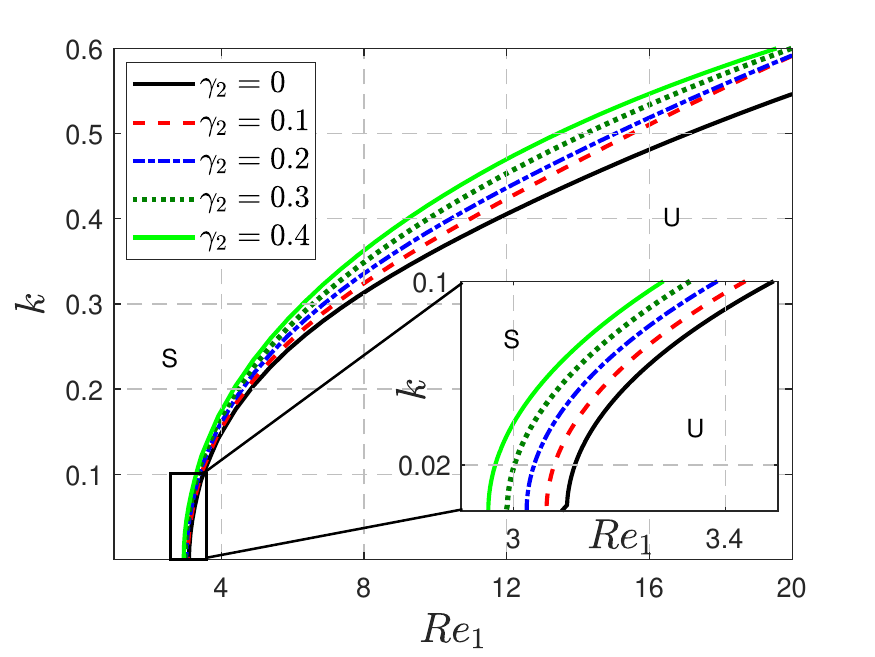}} 
\subfigure[]{\includegraphics[width=8.2cm]{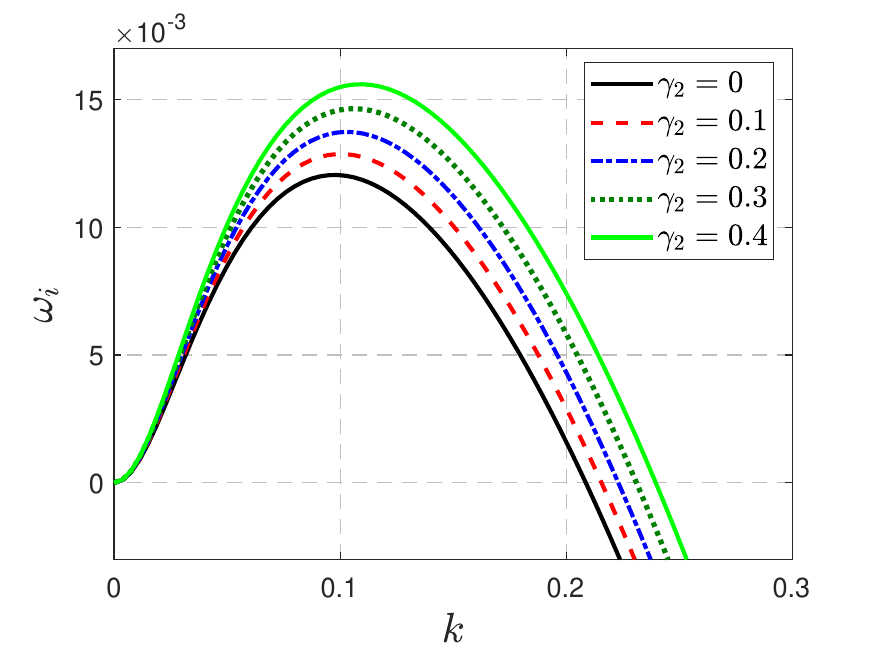}}
\end{center}
\caption{(a) The unstable boundary lines ($\omega_i=0$) of the SM in ($Re_1-k$) plane for varying bottom-layer viscoelasticity $\gamma_2$, (b) the corresponding temporal growth rate curves when $Re_1=5$. Here, the fixed value $\gamma_1=0.1$ with the remaining parameters as in Fig.~\ref{f4}.  }\label{f5}
\end{figure}

On the other hand, the bottom-layer viscoelasticity $\gamma_2$ shows the destabilizing effect on the SM (see Fig.~\ref{f5}(a)), followed by the successive expansion of the corresponding unstable region. This fact is further ensured by growth rate results in Fig.~\ref{f5}(b). The maximum growth rate of the SM amplifies with the increase of $\gamma_1$ owing to the rise in unstable wave number domain. Thus, the viscoelasticity of both layers individually boosts the SM instability due to the advection of base flow by the perturbation velocity components via the elastic stresses.

\begin{figure}[ht!]
\begin{center}
\subfigure[]{\includegraphics[width=5.4cm]{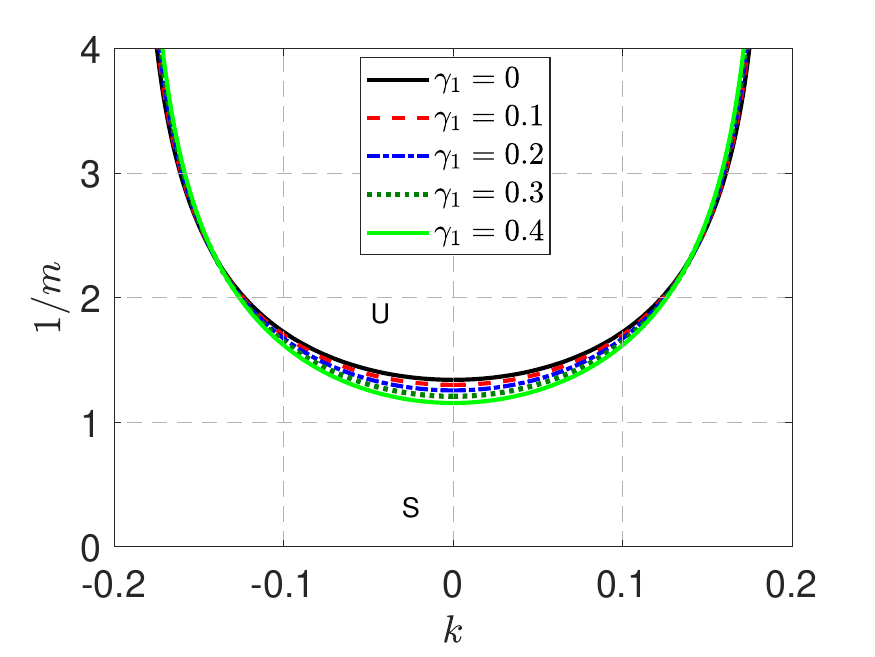}}
\subfigure[]{\includegraphics[width=5.4cm]{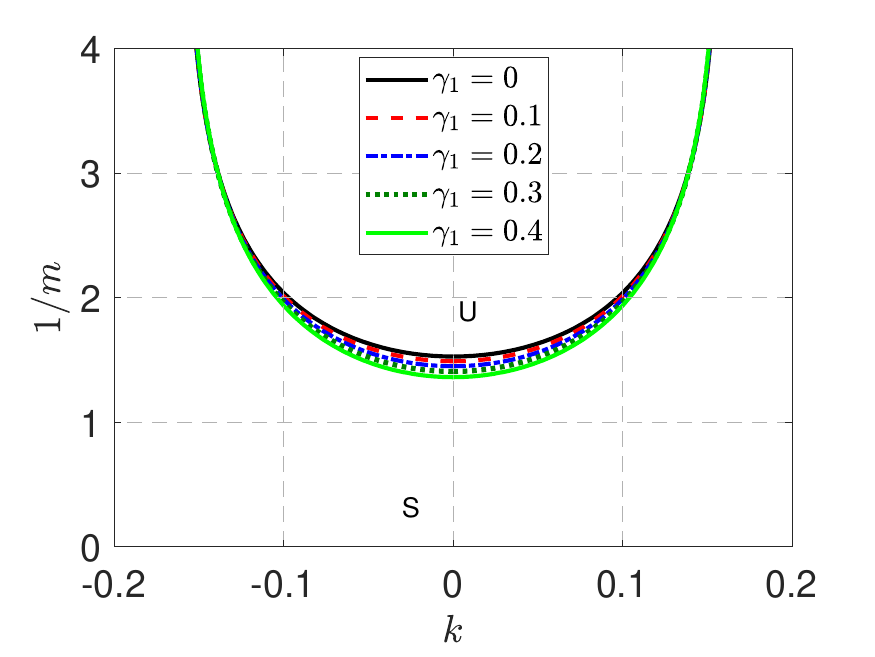}}
\subfigure[]{\includegraphics[width=5.4cm]{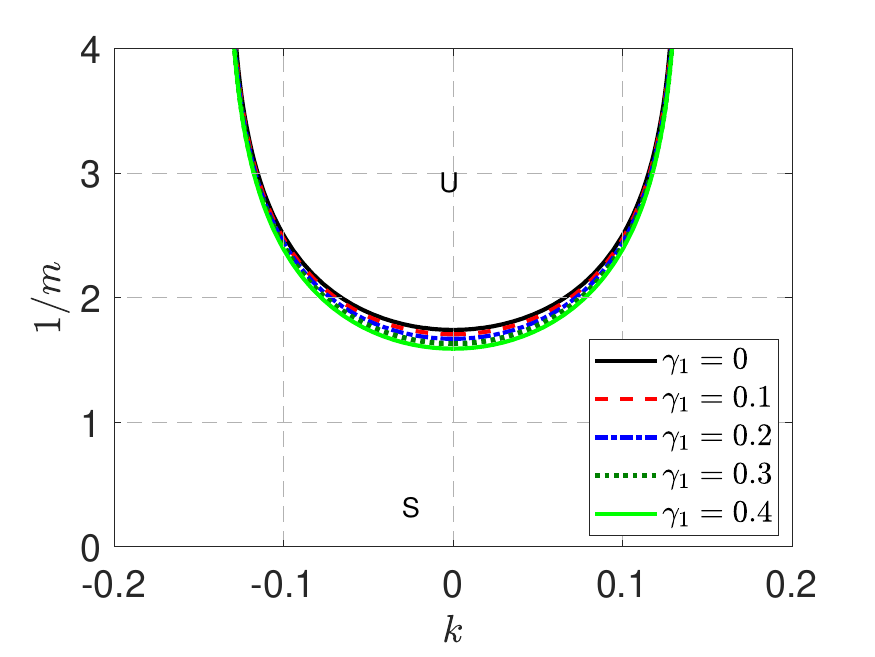}} 
\subfigure[]{\includegraphics[width=5.4cm]{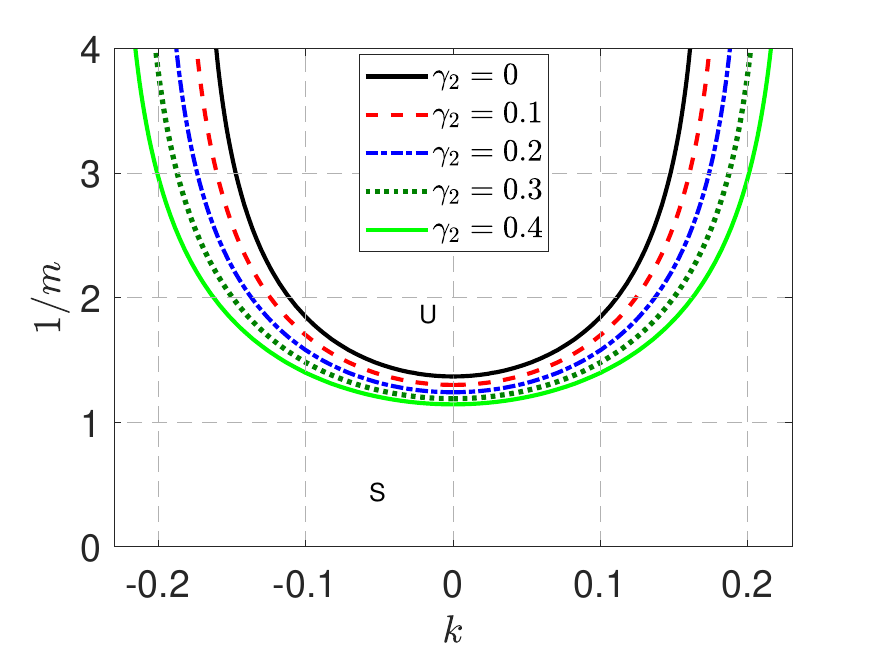}}
\subfigure[]{\includegraphics[width=5.4cm]{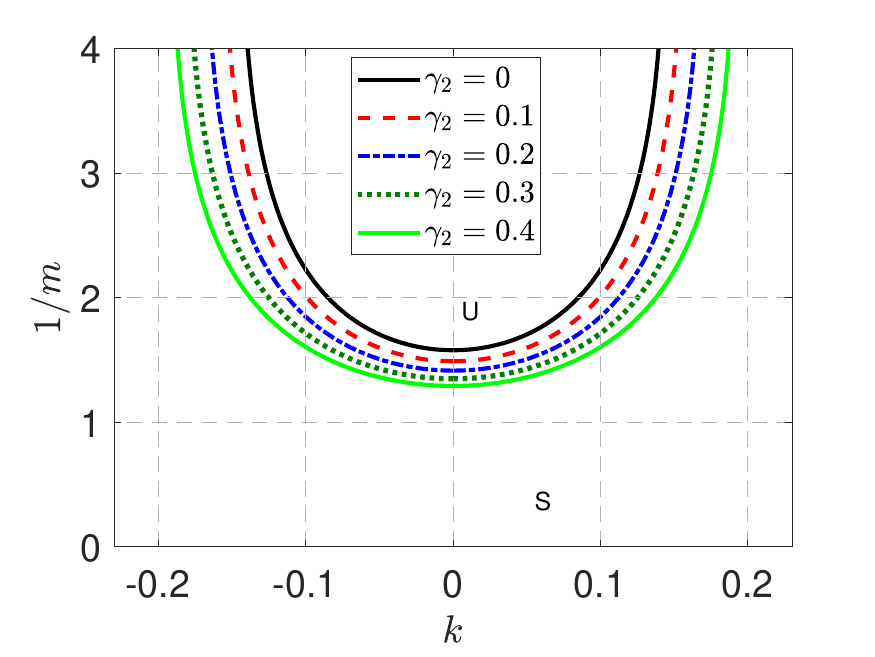}}
\subfigure[]{\includegraphics[width=5.4cm]{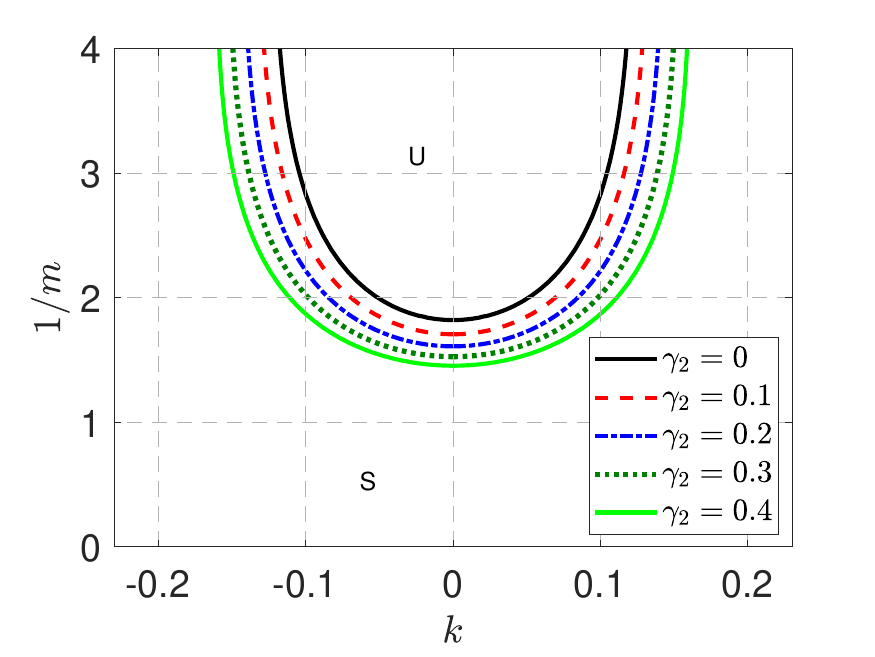}}
\end{center}
\caption{The variation of unstable boundary lines ($\omega_i=0$) of the SM in ($\displaystyle k-1/m$) plane for various values of (a)-(c) top-layer viscoelasticity $\gamma_1$ when $\gamma_2=0.1$ and (d)-(f) bottom-layer viscoelasticity $\gamma_2$ when $\gamma_1=0.1$. Here, (a)-(d) for $Ma_1=0$, (b)-(e) for $Ma_1=0.5$, and (c)-(f) for $Ma_1=1$. Here $Re_1=2$ with the remaining parameters as in Fig.~\ref{f4}. }\label{f6}
\end{figure}

\begin{figure}[ht!]
\begin{center}
\subfigure[]{\includegraphics[width=5.4cm]{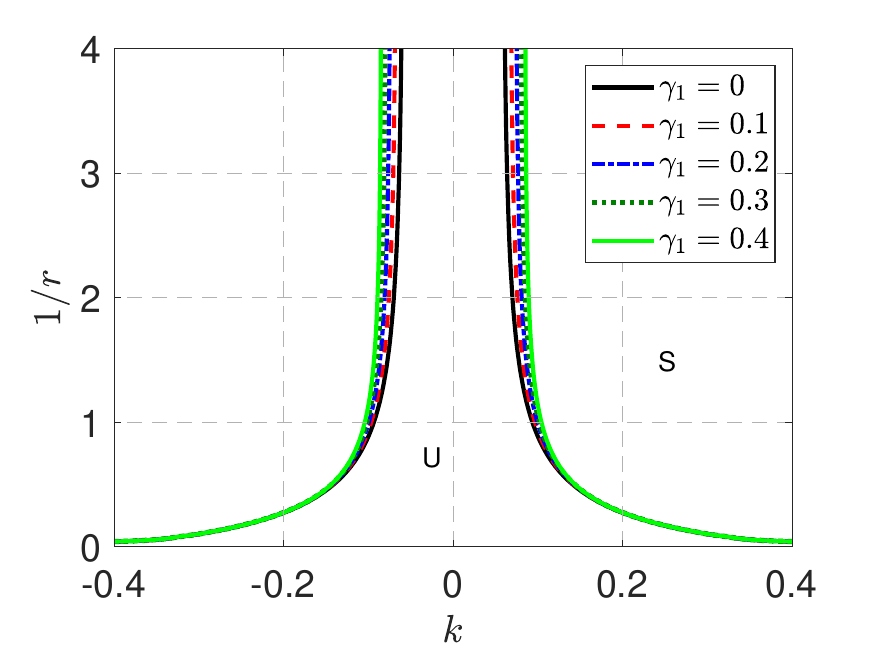}}
\subfigure[]{\includegraphics[width=5.4cm]{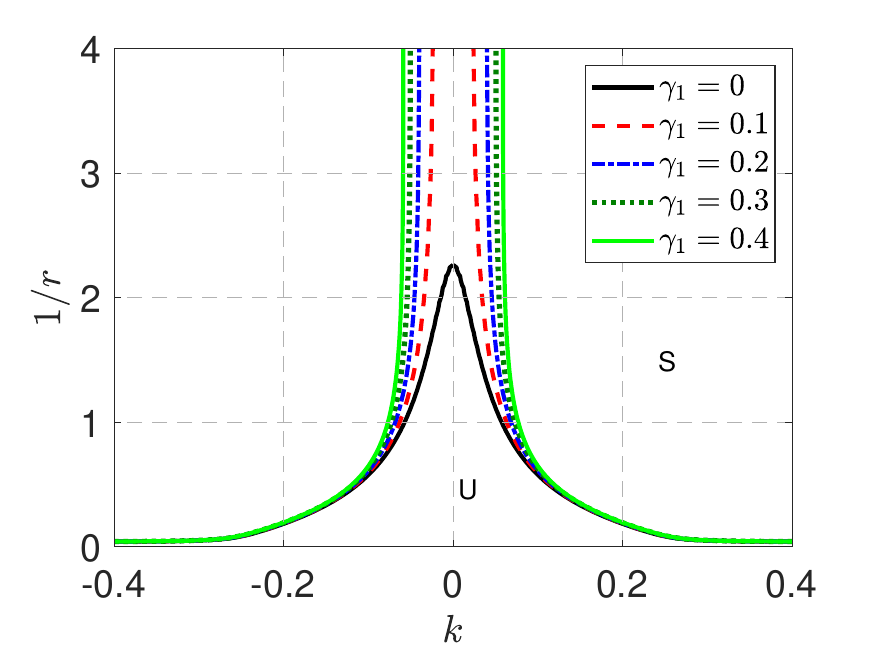}}
\subfigure[]{\includegraphics[width=5.4cm]{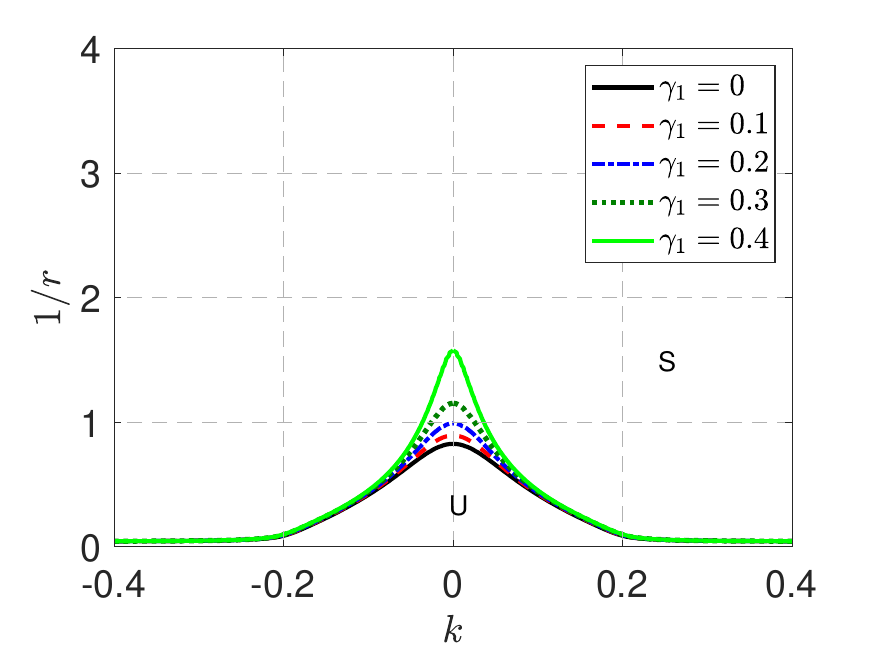}} \hspace{0cm}
\subfigure[]{\includegraphics[width=5.4cm]{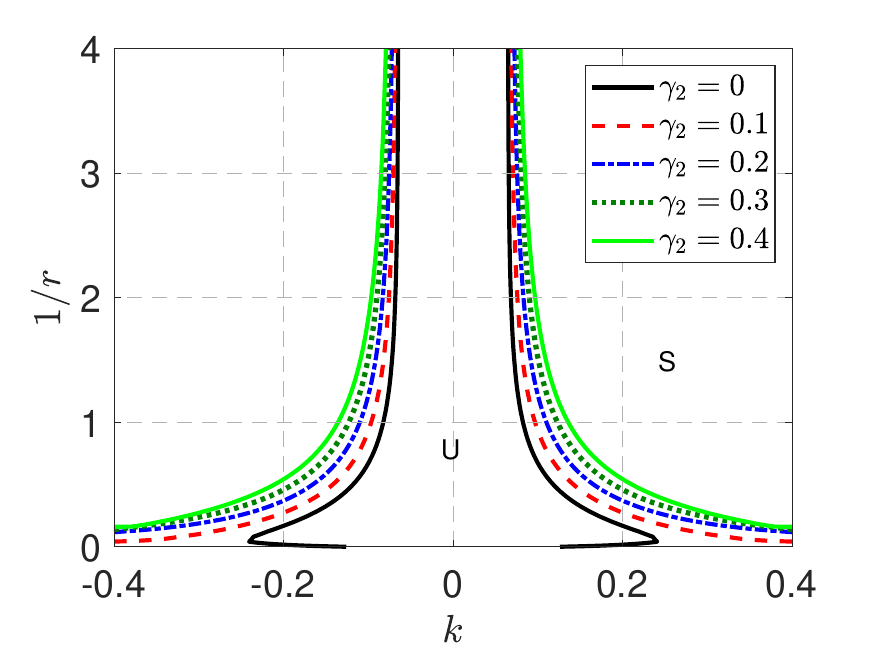}}
\subfigure[]{\includegraphics[width=5.4cm]{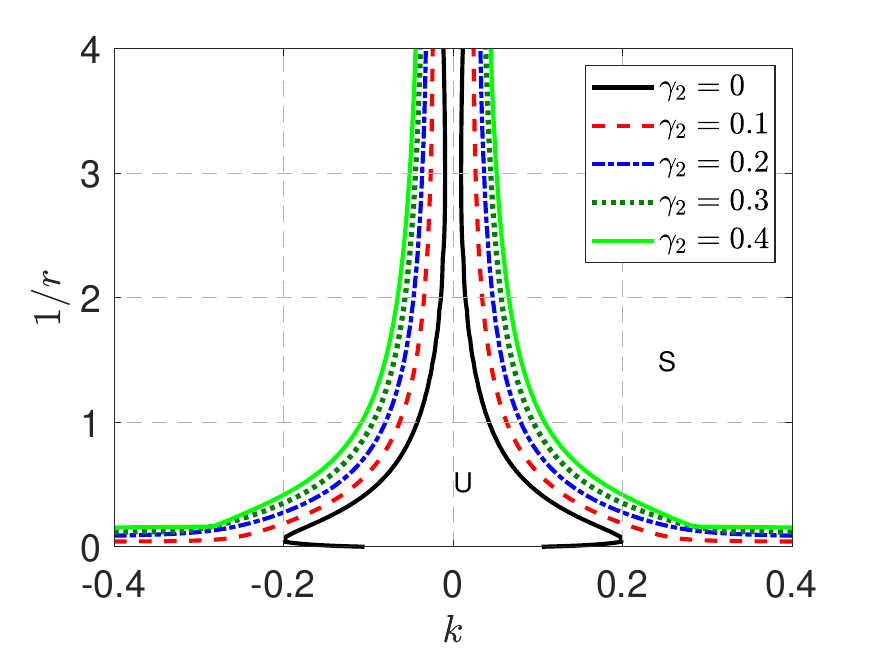}}
\subfigure[]{\includegraphics[width=5.4cm]{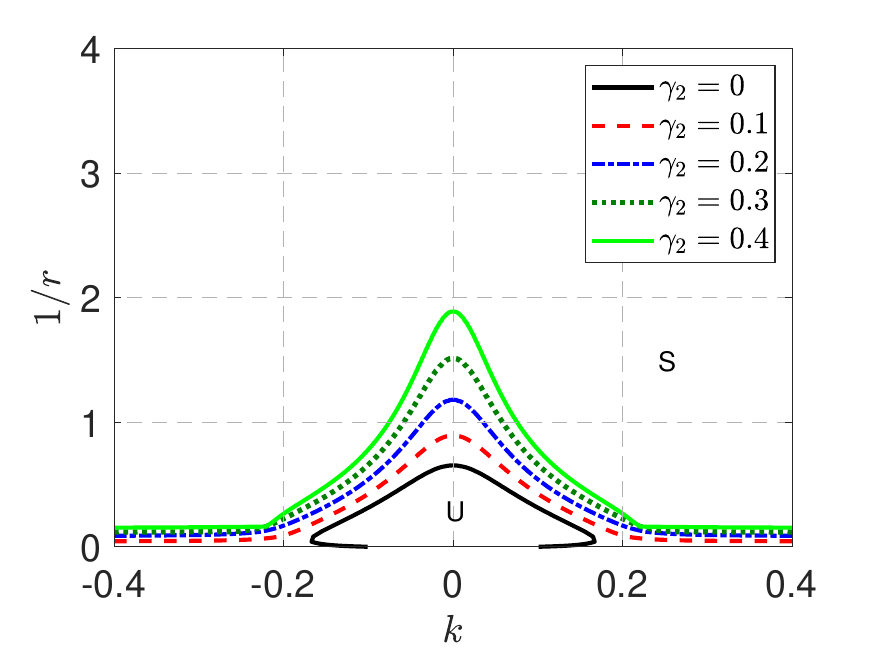}}
\end{center}
\caption{The variation of unstable boundary lines ($\omega_i=0$) of the SM in ($\displaystyle k-1/r$) plane for various values of (a)-(c) top-layer viscoelasticity $\gamma_1$ when $\gamma_2=0.1$ and (d)-(f) bottom-layer viscoelasticity $\gamma_2$ when $\gamma_1=0.1$. Here, (a)-(d) for $Ma_1=0$, (b)-(e) for $Ma_1=0.5$, and (c)-(f) for $Ma_1=1$. The value of the viscosity ratio $m=0.6$ with the remaining parameters as in Fig.~\ref{f6}.  }\label{f7}
\end{figure}

Now, a key question then arises: How does the surface mode (SM) behave when the viscoelastic coefficient varies, particularly in the case of the variation of the viscosity ratio $m$ and the density ratio $r$? To address this, the evolution of the unstable regions associated with SM in the $k-1/m$ and $k-1/r$ planes is illustrated in Fig.~\ref{f6} and Fig.~\ref{f7}, respectively, as the magnitude of the viscoelastic coefficients changes for different Marangoni numbers $Ma_1$. In Fig.~\ref{f6}, the range of the viscosity ratio $1/m\in[0, 4]$ reflects all scenarios where the bottom layer viscosity is lower than, equal to, or higher than the upper layer. The SM becomes fully damped when the viscosity of the lower layer is significantly higher than that of the upper layer and increases rapidly compared to the upper layer. This finding is consistent with the results in Fig.~\ref{f4}(a), where the SM is stable at $Re_1=2$. The surface waves instability relies on the interfacial shear between the two layers to amplify disturbances. When the bottom layer is much more viscous than the top layer, its stronger resistance to flow (due to larger viscous stresses) suppresses deformation, effectively damping perturbations and preventing them from developing into instabilities. Furthermore, as soon as $1/m$ increases, the unstable region of the SM expands rapidly, indicating that the SM instability intensifies as the bottom layer viscosity decreases relative to the top layer. The reason is that when the viscosity of the lower layer is reduced, it offers less resistance to the motion of the liquid-liquid interface. This fact allows more energy to be transmitted from the bottom layer to the top layer rather than being dissipated. As a result, the surface wave instability in the top layer will be stronger because less energy is lost to viscous damping in the bottom layer.  Furthermore, for both free surface ($Ma_1=0$), as shown in Fig.~\ref{f6}(a) and contaminated surface ($Ma_1\neq0$), as shown in Figs.~\ref{f6}(b) and (c), the higher viscoelastic coefficient $\gamma_1$ of the top-layer strengthens the SM instability in the longwave region, while exhibiting the weak impact in the shortwave zone for higher values of $1/m$. 
Thus, it is evident that regardless of whether $m<1$, $m=1$, or $m>1$, the viscoelasticity of the top layer can intensify the primary instability of the SM near the instability threshold. Moreover, on comparing the marginal stability curves in Figs.~\ref{f6}(a)-(c), it is revealed that the unstable bandwidth of the SM instability for the clean surface (i.e., $Ma_1=0$) is smaller than the contaminated surface ($Ma_1=0.5$), and it further shrinks as the Marangoni number increases to  $Ma_1=1$. This trend arises because insoluble surfactants induce Marangoni stresses that oppose surface perturbations. When the top-layer surface is deformed, spatial variations in surfactant concentration create tangential stress gradients ($Ma_1$ effects), which resist wave motion and suppress instability growth. Consequently, a higher $Ma_1$ promotes the stabilizing mechanism, leading to a narrower unstable region and weaker instability of the top-layer contaminated surface compared to a clean/uncontaminated top-layer surface. On the other hand, for $1/m\in [0,~4]$, the unstable region of the SM expands as the bottom layer's viscoelasticity $\gamma_2$ increases, confirming the destabilizing nature of SM. This behavior persists for both clean ($Ma_1=0$ in Fig.~\ref{f6}(d)) and surfactant-contaminated surfaces ($Ma_1\neq0$ in Fig.~\ref{f6}(e) and (f)) in the two-layer viscoelastic flow down an inclined plane. Additionally, as anticipated, the Marangoni force $Ma_1$ induced by the insoluble surfactant at the top surface decreases the unstable bandwidth of the SM. Consequently, the Marangoni force $Ma_1$ acts to suppress the surface wave instability enhanced by the bottom layer's viscoelastic coefficient $\gamma_2$. These findings suggest that the insoluble surfactants on the top-layer surface can effectively control the primary instability of surface waves in two-layer immiscible viscoelastic fluids.   

Next, we plot the stability boundary lines ($\omega_i=0$) of the SM in the $k-1/r$ plane to examine how the viscoelasticity of both layers individually affects the surface wave instability across the density ratio range $1/r\in[0,~4]$. Here, the results corresponding to the SM are presented for both clean (see Figs.~\ref{f7}(a) and (d)) and surfactant-contaminated surfaces (see Figs.~\ref{f7}(b) and (e) for $Ma_1=0.5$, and Figs.~\ref{f7}(c) and (f) for $Ma_1=1$). The unstable bandwidth of SM becomes maximum when $1/r\ll 1$ (i.e., when the top-layer density is much lower than the bottom layer), and it rapidly reduces with the increase in $1/r$. Therefore, with the increase in top-layer density, the SM instability loses its strength, which results in a weak flow rate of the top-layered viscoelastic fluid. 
Besides,  when the bottom layer is highly dense compared to the top layer, the viscoelastic parameter $\gamma_1$ of the top layer has a very weak impact on the SM instability. However, a significant destabilizing effect can be observed once the top layer becomes denser than the bottom layer. Therefore, the influence of the top layer's viscoelasticity ($\gamma_1$) is strongly dependent on the density ratio: its effect is minimal when the bottom layer is much denser (i.e, for a lower value of $1/r$) but becomes markedly destabilizing when the top layer is the denser one (i.e, for a high value of $1/r$). The unstable zone generated by the SM diminishes substantially (on comparing Figs.~\ref{f7}(a), (b), and (c)) with higher Marangoni force $Ma_1$, assuring the stabilizing effect of the top-layer's insoluble surfactant on the primary instability of the surface wave. The SM stabilizes completely for $Ma_1 = 1$, transitioning the top-layer flow to a laminar state when $1/r > 1.5$ (i.e., the top layer is highly denser than the bottom layer). On the other hand, increasing the bottom-layer viscoelasticity $\gamma_2$ intensifies the SM instability (see Figs.~\ref{f7}(d), (e), and (f)), which is followed by the successive enhancement of the unstable SM bandwidth due to the advection of the base flow by the perturbed velocity components via the elastic stresses of the bottom-layered viscoelastic fluid. Notably, the Marangoni force $Ma_1$ induced by the top-layer surfactant impedes the increment of the SM instability raised by the bottom-layered viscoelasticity $\gamma_2$. This is because of the reduction of the unstable bandwidth of SM instability for a stronger Marangoni force $Ma_1$. Thus, we notice that while bottom-layer viscoelasticity consistently enhances surface wave instability regardless of density and viscosity stratification, top-layer surfactants can counteract this effect by both narrowing the instability range in $1/m$ and $1/r$ space and reducing the top-layer-driven instability.

 \subsection{Results for the interface mode (IM)}
\begin{figure}[ht!]
\begin{center}
\subfigure[]{\includegraphics[width=8.2cm]{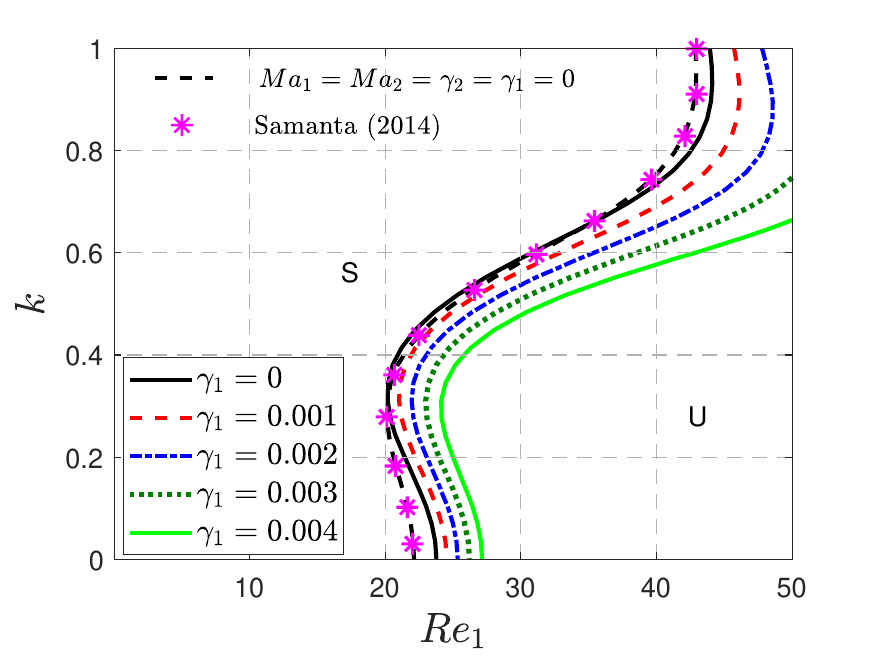}}
\subfigure[]{\includegraphics[width=8.2cm]{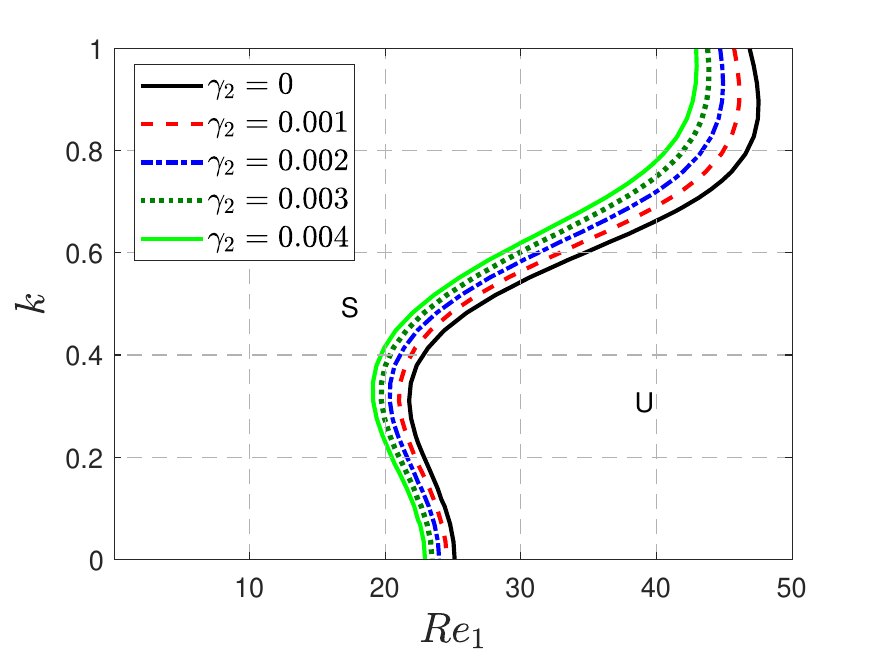}}
\end{center}
\caption{The variation of unstable boundary lines ($\omega_i=0$) of the IM in ($\displaystyle Re_1-k$) plane for various values of (a) top-layer viscoelasticity $\gamma_1$ when $\gamma_2=0.001$ and (b) bottom-layer viscoelasticity $\gamma_2$ when $\gamma_1=0.001$  The remaining parameters are  $m=1.5$, $r=1$, $\delta=1$, $\theta=0.2~\textrm{rad}$, $Ma_1=0.5$, $Ma_2=0.005$, $Ca_1=1$, $Ca_2=1$, $Pe_1=\infty$ and $Pe_2=\infty$. The black dashed line is the current results with the limiting value $Ma_1=Ma_2=\gamma_1=\gamma_2=0$ and the red asterisk symbols are the result of \citet{samanta2014effect}. }\label{f8}
\end{figure}

In this subsection, the numerical analysis is performed to examine the instability behavior of the unstable interface mode (IM), identified in Fig.~\ref{f3}(b), in the double-layered viscoelastic fluid flow model. In Fig.~\ref{f8}, the marginal stability curves of the IM are demonstrated in the $(Re_1-k)$ for different values of (a) $\gamma_1$ (as shown in Fig.~\ref{f8}(a)) and $\gamma_2$ (as shown in Fig.~\ref{f8}(b)). The parameter values $m=1.5$, $\delta=1$, $Ma_1=0.5$, $Ma_2=0.005$, $Ca_1=1$, and $Ca_2=1$ are set in this numerical analysis. It is found that the stability boundary lines of the IM, as in Fig.~\ref{f8}(a), perfectly match with the previous numerical outcomes of double-layered Newtonian fluid flowing over an inclined plane (\citet{samanta2014effect}) when the limiting parameter values are $Ma_1=0$, $Ma_2=0$, and $\gamma_1=\gamma_2=0$ (i.e., Newtonian case). The viscoelastic parameter $\gamma_1$ of the top-layer fluid shrinks the unstable zone induced by the IM, which is followed by the successive increment of the critical Reynolds number $Re^c_1$ for the IM instability. 
Thus, the viscoelastic parameter $\gamma_1$ exhibits a stabilizing nature on the IM. Contrapositively, an opposite trend (i.e., destabilizing behavior) of the interfacial instability is observed under the influence of the bottom-layer viscoelasticity $\gamma_2$, as evidenced by the expansion of the neutral curves in Fig.~\ref{f8}(b). This destabilizing impact is assured by the reduction in the critical Reynolds number ($Re_1^c$) as $\gamma_2$ increases, indicating that stronger bottom-layer viscoelasticity promotes interfacial instability. The jump in normal stresses across the interface is a primary driver of interfacial instability in double-layer systems. Increasing the viscoelastic coefficients of the bottom layer enhances interfacial instability because it increases the discontinuity in elastic stresses at the interface, amplifies disturbance growth, and enables more efficient transfer of stored elastic energy to interfacial waves, thereby promoting instability.

\begin{figure}[ht!]
\begin{center}
\subfigure[]{\includegraphics[width=8.2cm]{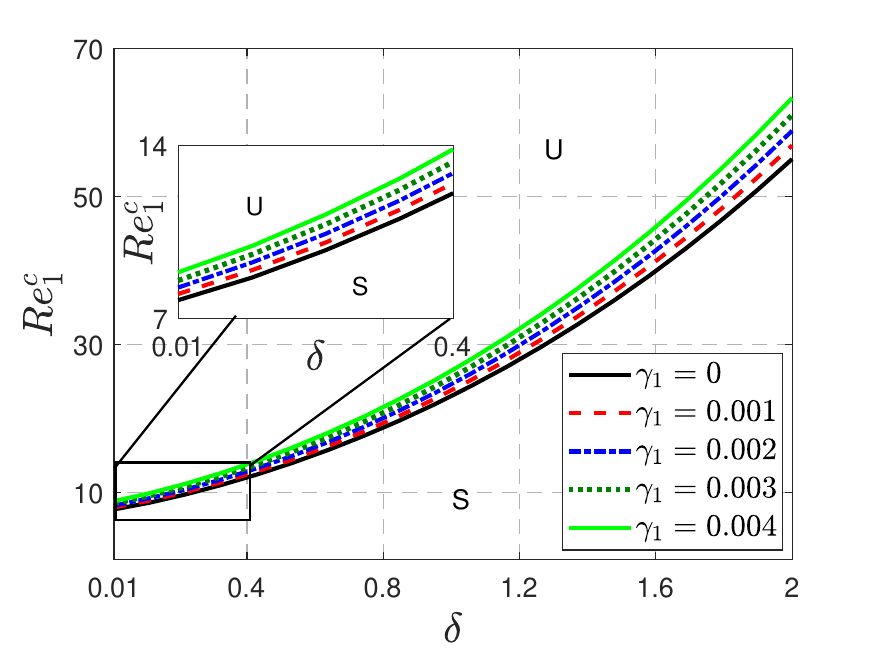}} 
\subfigure[]{\includegraphics[width=8.2cm]{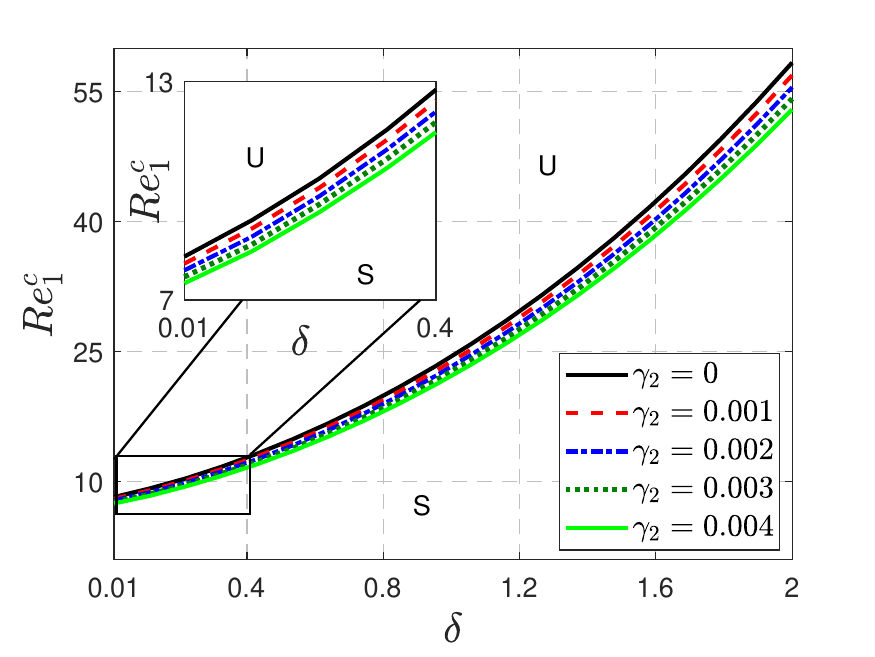}}
\end{center}
\caption{(a) The effect of top-layer viscoelasticity $\gamma_1$ on the critical Reynolds number ($Re_1^c$) curve of IM as a function of depth ratio $\delta$ when bottom-layer viscoelasticity $\gamma_2=0.001$ and (b) the effect of bottom-layer viscoelasticity $\gamma_2$ on the critical Reynolds number ($Re_1^c$) curve as a function of depth ratio $\delta$ when top-layer viscoelasticity $\gamma_1=0.001$. Here, the value of $Ma_2=0.005$ and the remaining parameters are the same as in Fig.~\ref{f8}. }\label{f10}
\end{figure}

It should be noted that the interfacial instability is highly sensitive to the viscoelastic parameter as compared to the SM instability of the double-layered fluid flow model. The liquid-liquid interface is governed by the jump in stresses (i.e,  viscous and elastic). Viscoelasticity introduces normal stress differences and shear-dependent relaxation, which directly alter the interfacial stress balance. Thus, a small change in the elasticity of both layers can significantly modify the interfacial traction condition, amplifying sensitivity. On the other hand, the top-layer surface is governed by weaker air-fluid interactions and bulk rheology. The top layer surface is primarily influenced by air-fluid interactions (e.g., surface tension, air viscosity), which are often weaker than fluid-fluid coupling. Therefore, the top layer surface lacks the shear-elastic feedback present at the interface, making it less responsive to viscoelasticity.

Now, to examine the behavior of interfacial instability of the double-layered viscoelastic fluid flow across the wide range of depth ratio $\delta\in[0.01,\,2]$, we have demonstrated the critical Reynolds number $Re_1^c$ ($k\rightarrow 0$) as a function of depth ratio $\delta$ in Fig.~\ref{f10}. Here Fig.~\ref{f10}(a) shows the influence of top-layer viscoelasticity $\gamma_1$, while Fig.~\ref{f10}(b) displays the effect of bottom-layer viscoelasticity $\gamma_2$ on the function $Re_1^c(\delta)$. For $\delta<1$  (indicating a thinner bottom layer relative to the top layer), increasing $\delta$  corresponds to a progressive thickening of the bottom layer relative to the top layer. Conversely, for $\delta>1$  (where the bottom layer is thicker than the top layer), increasing $\delta$  leads to further relative thickening of the bottom layer. Irrespective of $\delta<1$ or $\delta=1$ or $\delta>1$, the value of $Re^c_1$ corresponding to the IM monotonically increases with the increase of depth ratio $\delta$. This confirms that the intensity of the interfacial instability in the double-layered viscoelastic flow field weakens as long as the lower layer depth increases. This scenario is observed regardless of whether the viscoelasticity of the top layer $\gamma_1$ (see Fig.~\ref{f10}(a)) or the bottom layer $\gamma_2$ (see Fig.~\ref{f10}(b)) is varied. An increasing depth in the bottom layer spreads shear stresses over a larger volume, dampening interfacial disturbances more effectively, even though the fluid's viscoelasticity (viscous and elastic stresses) remains constant. Moreover, the function $Re_1^c(\delta)$ increases as the viscoelastic parameter $\gamma_1$ increases (see Fig.~\ref{f10}(a)), showing a stabilizing effect on IM. However, as soon as $\gamma_2$ enhances, the function $Re_1^c$ reduces, confirming its destabilizing effect on the IM. Thus, one can conclude that interface instability in the longwave zone, which intensifies with the increase of bottom layer viscoelasticity $\gamma_2$, can be mitigated by increasing the relative thickness of the lower layer in the two-layered viscoelastic flow system.

\begin{figure}[ht!]
\begin{center}
\subfigure[]{\includegraphics[width=5.4cm]{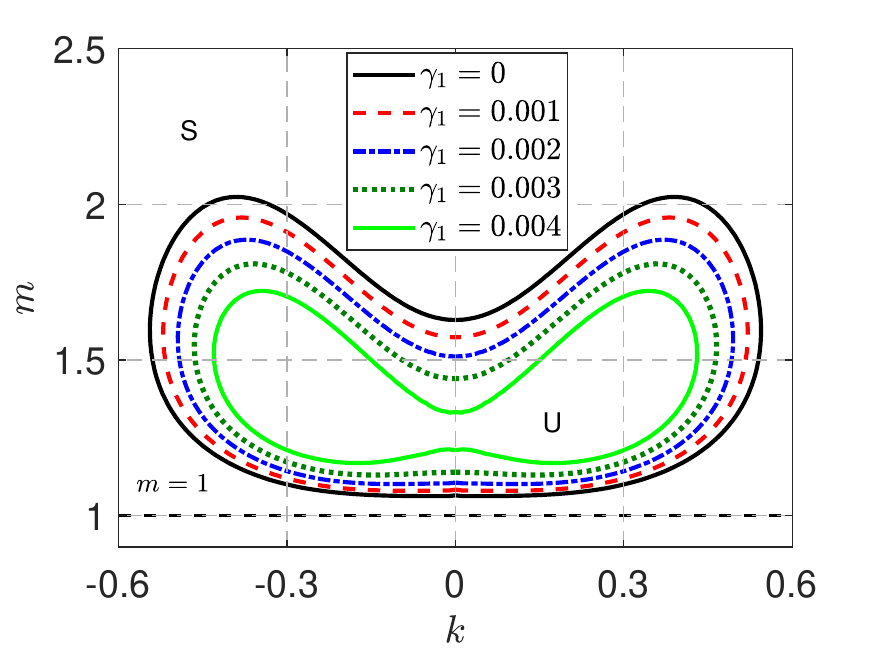}}
\subfigure[]{\includegraphics[width=5.4cm]{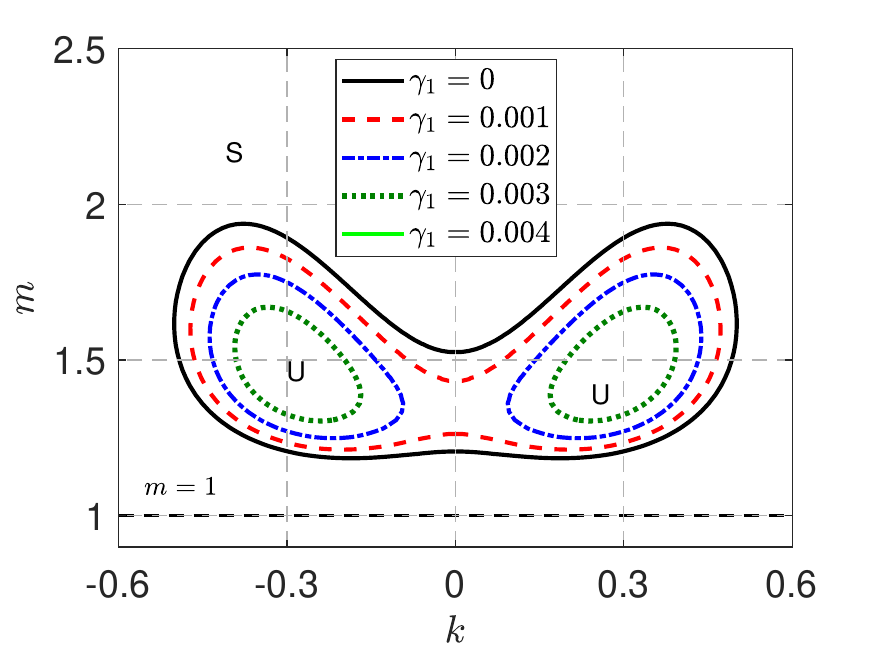}}
\subfigure[]{\includegraphics[width=5.4cm]{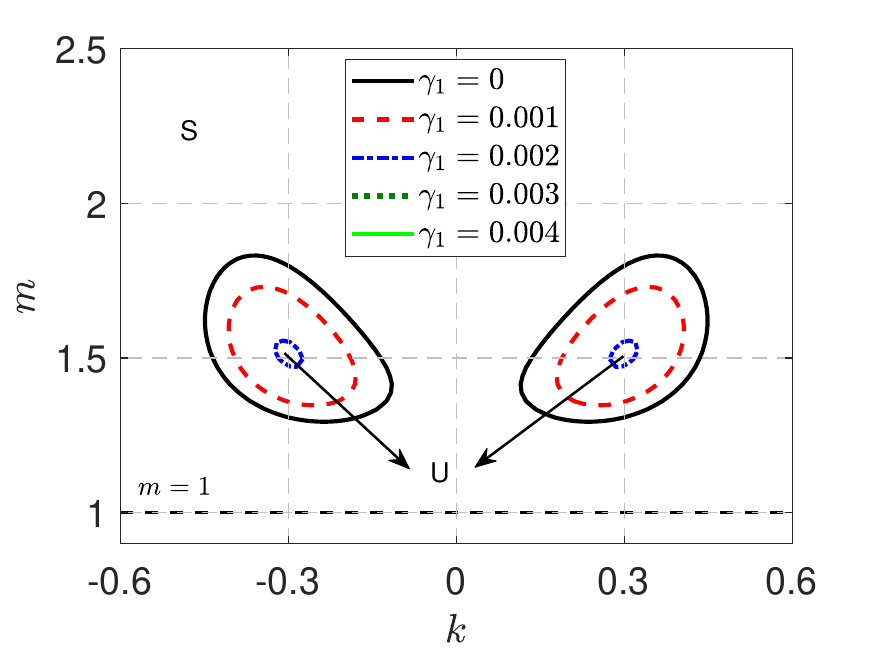}} \hspace{0cm}
\subfigure[]{\includegraphics[width=5.4cm]{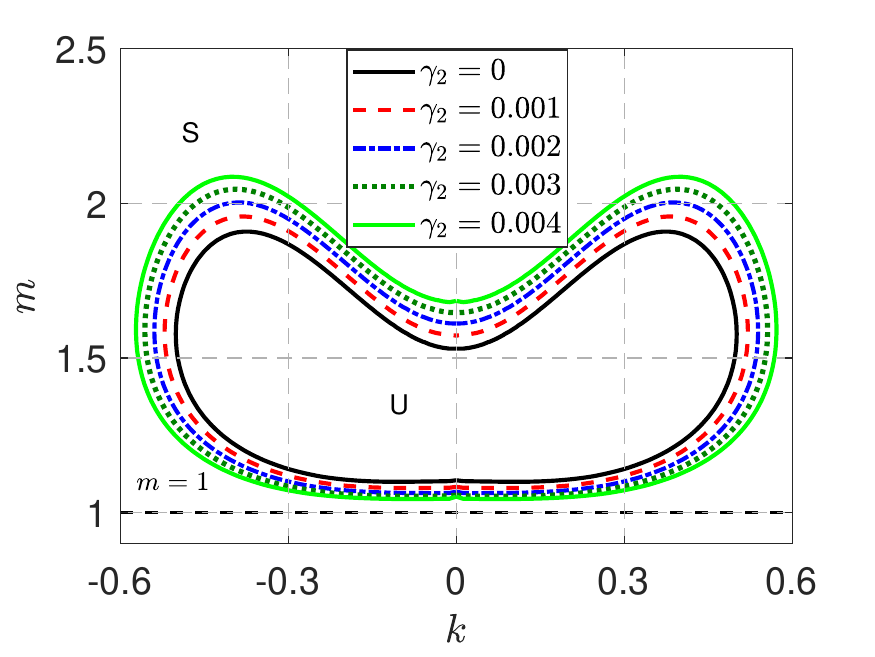}}
\subfigure[]{\includegraphics[width=5.4cm]{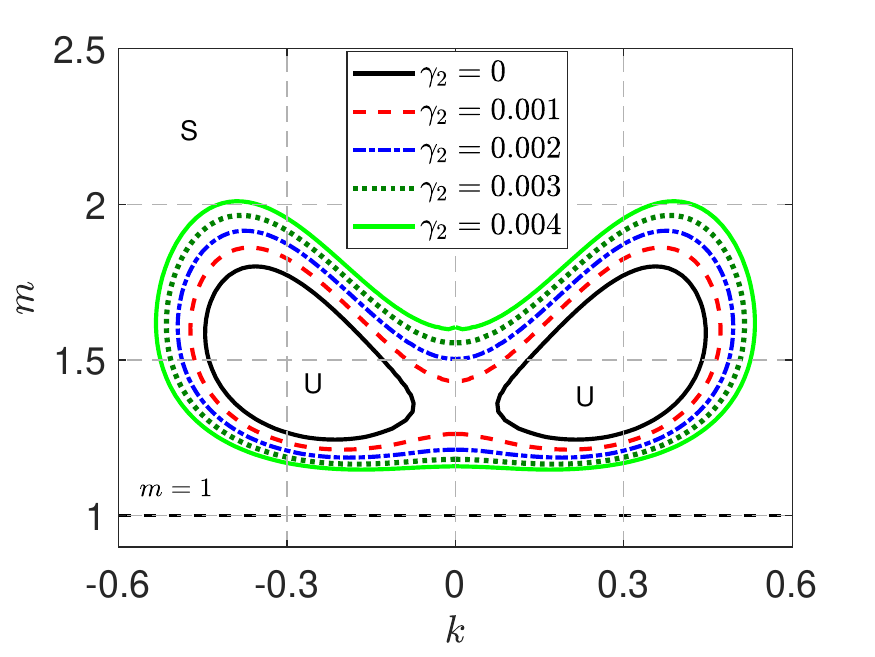}}
\subfigure[]{\includegraphics[width=5.4cm]{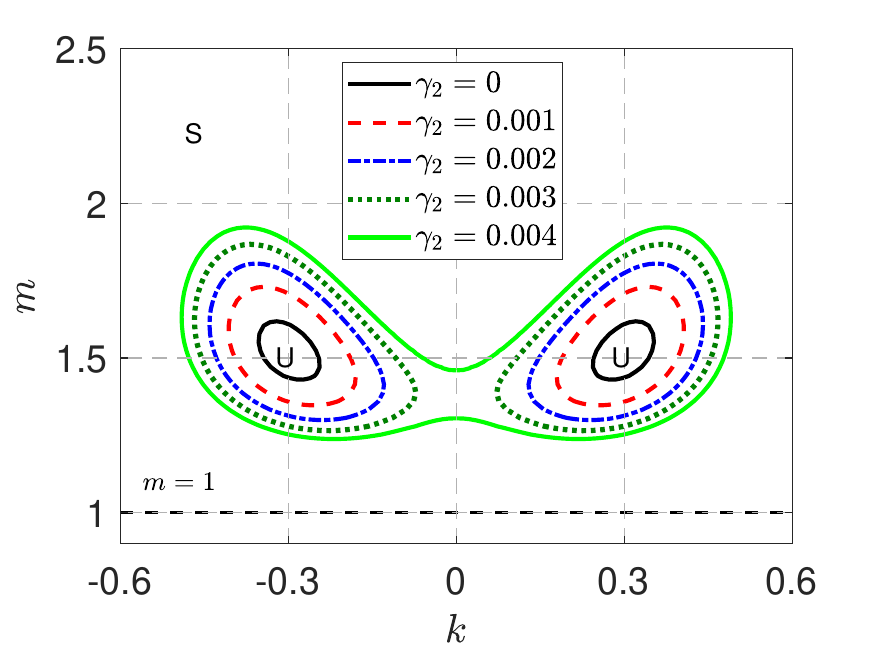}}
\end{center}
\caption{The effect of top-layer viscoelasticity $\gamma_1$ on the unstable boundary lines ($\omega_i=0$) of the IM in ($\displaystyle k-m$) plane with change in interfacial Marangoni force $Ma_2$ in the case of  $m>1$. The value of the bottom-layer viscoelasticity  $\gamma_2=0.001$ in (a)-(c) and top-layer viscoelasticity  $\gamma_1=0.001$ in (d)-(f). Here $Ma_2=0.001$ in (a) and (d), $Ma_2=0.005$ in (b) and (e), and $Ma_2=0.009$ in (c) and (f). The remaining parameters are $\gamma_2=0$, $Re_1=24$, $r=1$, $Ma_1=0.5$,  $Ca_1=1$, $Ca_2=1$, $\theta=0.2~\textrm{rad}$, $Pe_1=\infty$ and $Pe_2=\infty$. }\label{f11}
\end{figure}

\begin{figure}[ht!]
\begin{center}
\subfigure[]{\includegraphics[width=5.4cm]{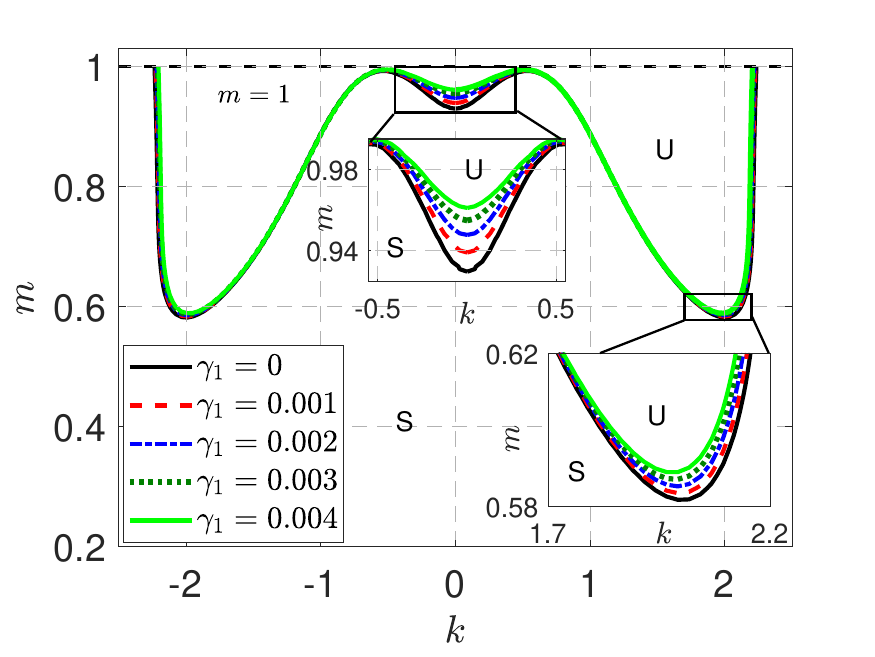}}
\subfigure[]{\includegraphics[width=5.4cm]{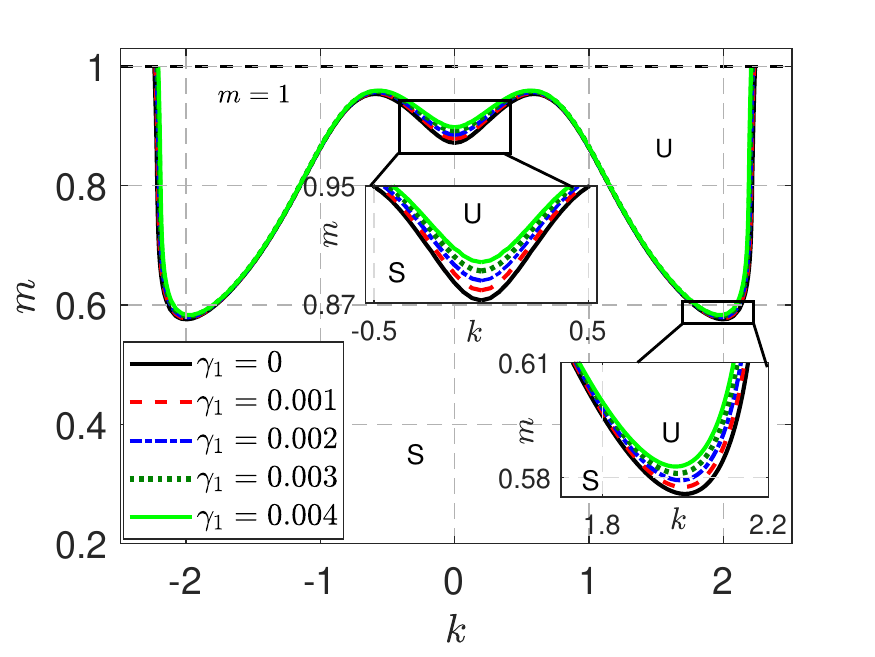}}
\subfigure[]{\includegraphics[width=5.4cm]{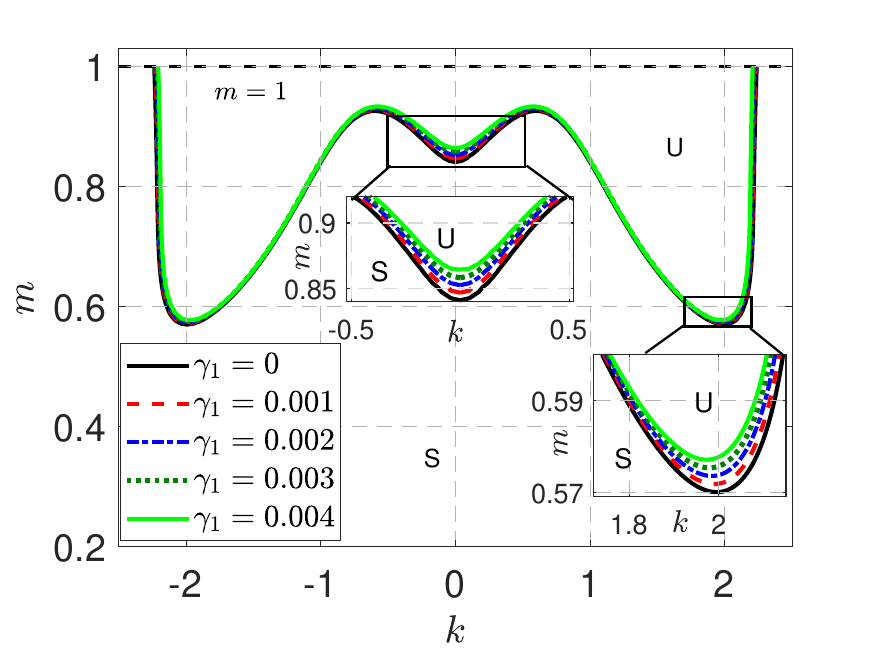}} \hspace{0cm}
\subfigure[]{\includegraphics[width=5.4cm]{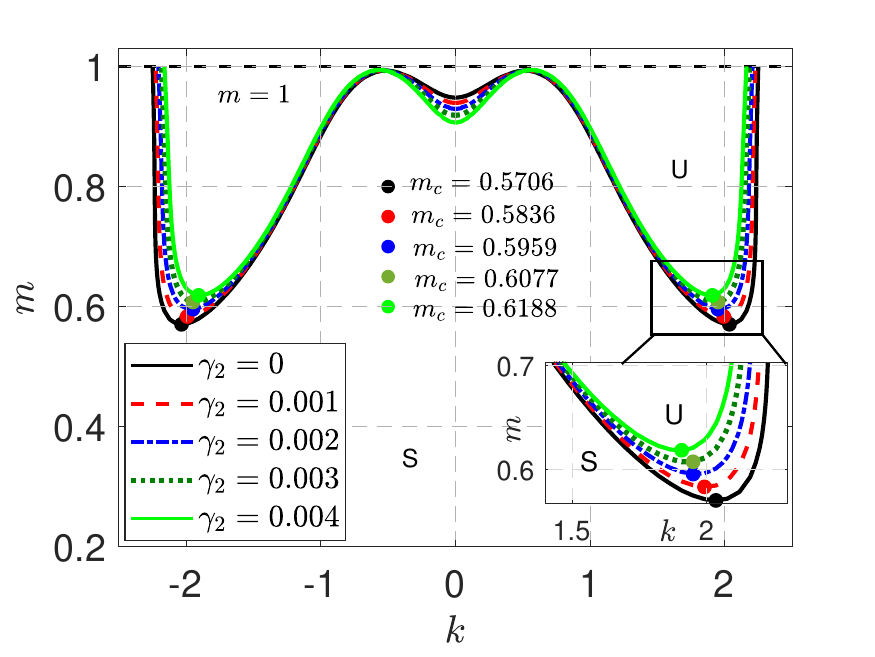}}
\subfigure[]{\includegraphics[width=5.4cm]{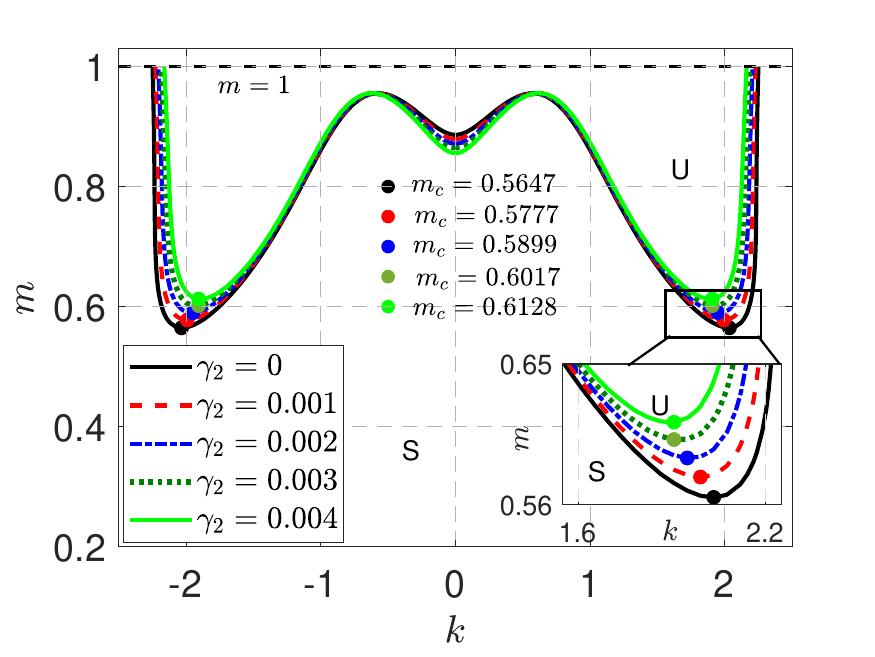}}
\subfigure[]{\includegraphics[width=5.4cm]{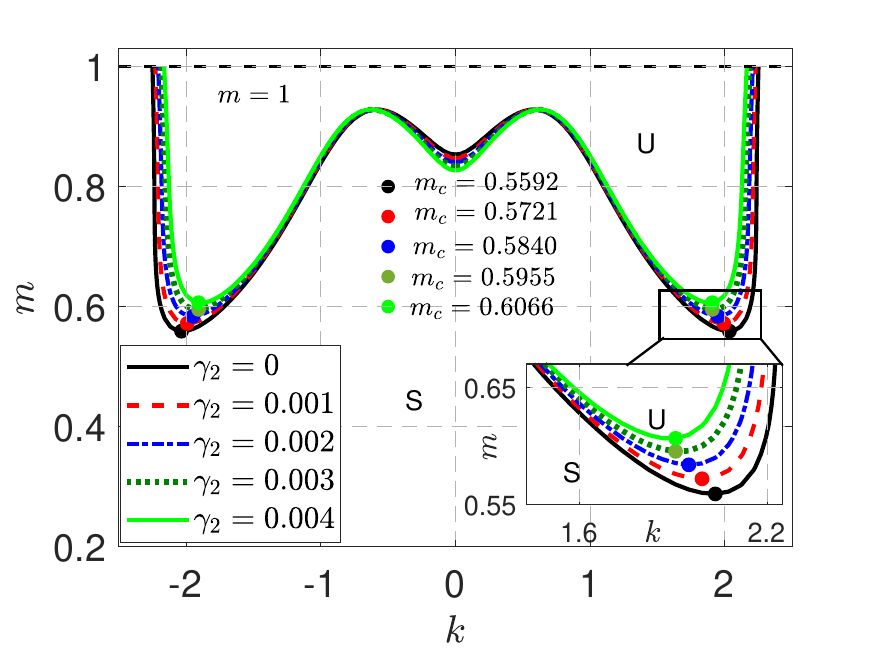}}
\end{center}
\caption{The effect of top-layer viscoelasticity $\gamma_1$ on the unstable boundary lines ($\omega_i=0$) of the IM in ($\displaystyle k-m$) plane with change in $Re_1$ in the case of  $m<1$. The value of the top-layered viscoelasticity  $\gamma_1=0.001$ in (a)-(c) and bottom-layered viscoelasticity  $\gamma_2=0.001$ in (d)-(f). Here $Ma_2=0.001$ in (a) and (d), $Ma_2=0.005$ in (b) and (e), and $Ma_2=0.009$ in (c) and (f). The remaining parameters are the same as in Fig.~\ref{f11}. }\label{f12}
\end{figure}

The numerical simulation is again performed to examine the characteristics of unstable IM in the disjoint regions of viscosity ratio $m>1$ and $m<1$. To do so, the neutral stability curves in the $(k,m)$ plane with varying viscoelasticity $\gamma_1$ are portrayed in Fig.~\ref{f11} for different Marangoni forces $Ma_2$ when $m>1$. Here, the dashed line $m=1$ divides the unstable region into $m>1$ and $m<1$, indicating that the region $m>1$ (i.e., the less viscous fluid is flowing over a higher viscous fluid) occupies the unstable region of the IM. Note that the subcritical instability arises for each viscoelastic parameter $\gamma_1$ (see Fig.~\ref{f11}(a)) in the $k-m$ domain with $m>1$, when the liquid-liquid interface is contaminated by the insoluble surfactant with Marangoni force $Ma_2=0.001$. This is due to the two-time emergence of the marginal criteria $k=0$ ($Re_1^c$). Now, if we increase the Marangoni force to $Ma_2=0.005$, the interfacial instability (see Fig.~\ref{f11}(b)) emerging in the longwave region fully diminishes when $\gamma_2\geq 0.002$ and the subcritical instability exists for $\gamma_2\leq 0.001$. The longwave instability fully disappears, and only the short-wave instability remains once the Marangoni force further increases to $Ma_2=0.009$ for each top-layer viscoelasticity $\gamma_1$. Thus, it is clear that by contaminating the interface of the double-layered viscoelastic flow field, it is possible to control the longwave IM instability and make the liquid-liquid interface fully stable in the longwave region. Now, if we fix the upper layer viscoelasticity $\gamma_1$ and increase the lower layer viscoelasticity $\gamma_2$, an opposite characteristic (i.e., destabilization nature) of IM for all $Ma_2$ values is observed in Figs.~\ref{f11}(d)-(f), followed by an amplification of the corresponding unstable region. Furthermore, the subcritical instability is found for each $\gamma_2$ value when the Marangoni force $Ma_2=0.001$, induced by the interfacial surfactant. Once the Marangoni force increases to $Ma_2=0.005$, the subcritical instability remains for $\gamma_2\geq 0.001$. The longwave instability fully disappears, and only the short-wave instability remains for $\gamma_2\leq 0.003$ as the Marangoni force increases to $Ma_2=0.009$. Thus, placing an insoluble surfactant at the liquid-liquid interface plays a crucial role in stabilizing the IM instability in the double-layer viscoelastic fluid.

On the other hand, for the unstable region $m<1$ (i.e., the bottom layer viscosity is comparatively lower than the top layer), subcritical instability ceases to exist  (see Fig.~\ref{f12}). In this regime, for all applied Marangoni forces $Ma_2$ across the liquid-liquid interface, the shortwave instability of the IM is stronger than the longwave instability. Additionally, for each $Ma_2$ value, a higher viscoelasticity $\gamma_1$ in the top layer weakly stabilizes the interfacial instability in the shortwave zone due to the gradual decrease in the critical viscosity ratio ($m_c$). However, a significant stabilizing effect in the shortwave zone can be achieved if one can enhance the increasing rate of $\gamma_1$. A more pronounced stabilizing impact is observed in the longwave region (see Fig.~\ref{f12}(a)-(c)). Meanwhile, for each $Ma_2$ value, the bottom layer's viscoelasticity $\gamma_2$ exhibits a dual influence (see Fig.~\ref{f12}(d)-(f)): as $\gamma_2$ increases, longwave IM instability intensifies due to a rise in critical viscosoity ratio $m_c$, whereas shortwave instability weakens as $m_c$ increases. Moreover, when $m<1$, a potent Marangoni force $Ma_2$ imposed at the liquid-liquid interface has a comparatively weak but destabilizing effect on the interfacial instability. This is because of the continuous reduction of the critical values of the viscosity ratio  ($m_c$) as $Ma_2$ increases for each tested value of the viscoelasticity in both layers. Note that this destabilizing Marangoni effect directly contrasts with the stability behavior observed in the case of $m>1$, highlighting the fundamental role of viscosity stratification in determining double-layered stability dynamics. 
The dual role of the interfacial surfactant on the IM is primarily due to the interplay between viscosity stratification, surfactant-induced Marangoni stresses, and the underlying flow dynamics. When $m>1$, the interface becomes stiffer, and perturbations will be damped more effectively due to the high viscosity in the lower layer. Any local interface stretching (due to a perturbation) reduces the surfactant concentration, thereby increasing the local surface tension. This creates a Marangoni stress that opposes the perturbation, aiding stabilization. On the other hand, when $m<1$, the interface is more easily deformed by perturbations. The top layer (now more viscous) tends to dominate the flow response. The Marangoni force still acts to oppose local stretching, but the less viscous lower layer cannot effectively damp the induced flow perturbations. The low viscosity of the lower layer allows for stronger velocity gradients near the interface, and the Marangoni-induced flows may feed energy into the perturbation, leading to destabilization.

\subsection{Results for the interface surfactant mode (ISM)}
\begin{figure}[ht!]
\begin{center}
\subfigure[]{\includegraphics[width=8.2cm]{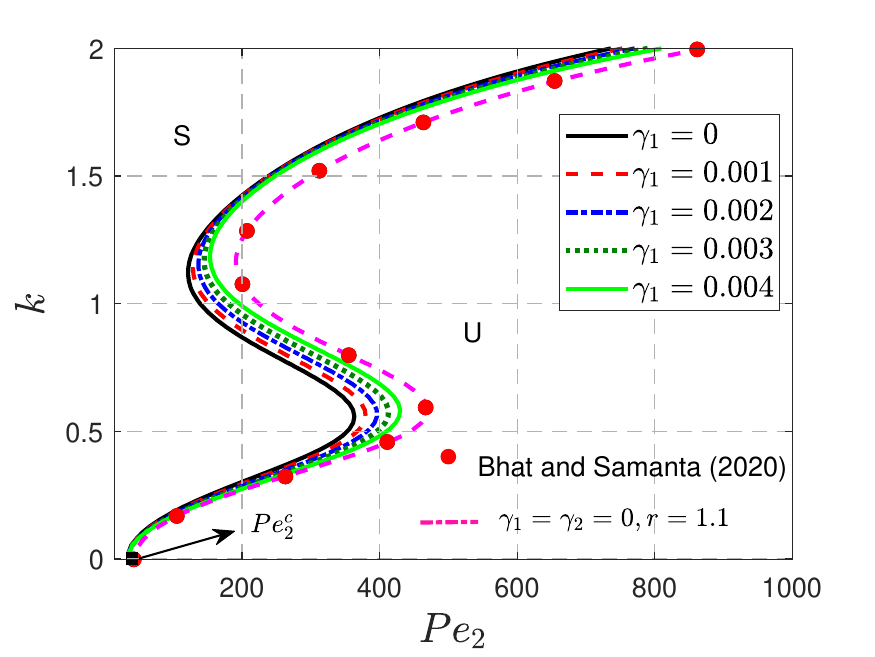}}
\subfigure[]{\includegraphics[width=8.2cm]{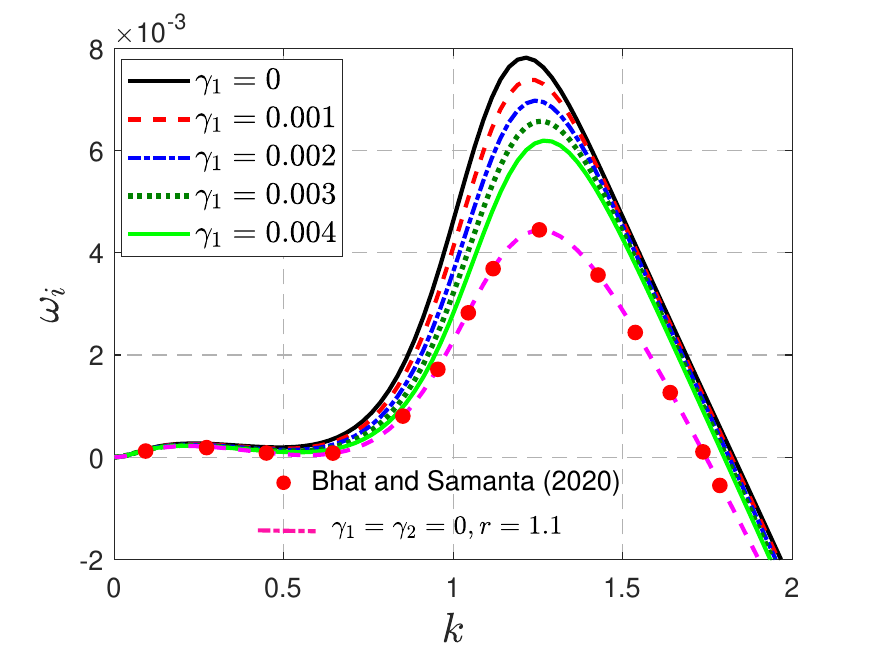}}
\end{center}
\caption{ (a) The unstable boundary lines ($\omega_i=0$) of the ISM in ($\displaystyle Pe_2-k$) plane for varying top-layer viscoelasticity $\gamma_1$ and (b) the corresponding growth rate when $Pe_1=500$. Here the fixed value $\gamma_2=0.001$ with remaining parameters $Re_1=50$, $m=1.5$, $r=1$, $\delta=1$, $Ma_1=0.01$, $Ma_2=0.01$, $Ca_1=1$, $Ca_2=1$, $\theta=0.2~\textrm{rad}$, and $Pe_1=1000$. The dash-dotted magenta line is the result for $r=1.1$, $\gamma_1=0$, the red circular symbols result from \citet{bhat2020linear}, and the black rectangular shape marks the critical p\'eclet number $Pe^c_2$ of the interfacial surfactant.  }\label{f15}
\end{figure}

In this subsection, we have discussed the behavior of ISM under the influence of the viscoelasticity in both layers. The stability boundaries corresponding to the ISM are plotted in the ($Pe_2,\,k$) plane, as shown in Fig.~\ref{f15}(a), for varying top-layered viscoelasticity $\gamma_1$ when the flow parameters are $\gamma_2=0.001$, $Re_1=50$, $m=1.5$, $r=1$, $\delta=1$, $Ma_1=0.01$, $Ma_2=0.01$, $Ca_1=1$, $Ca_2=1$, $\theta=0.2~\textrm{rad}$. The neutral curve result agrees very well with the result of \citet{bhat2020linear} in the limit $\gamma_1,\ \gamma_2\rightarrow 0$ and $r=1.1$.
\begin{figure}[ht!]
\begin{center}
\subfigure[]{\includegraphics[width=5.4cm]{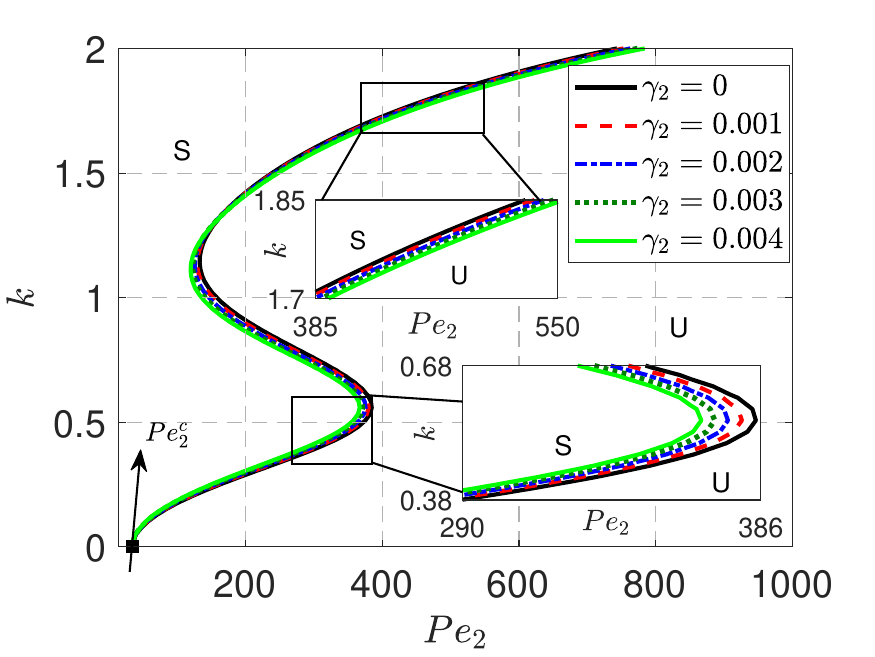}}
\subfigure[]{\includegraphics[width=5.4cm]{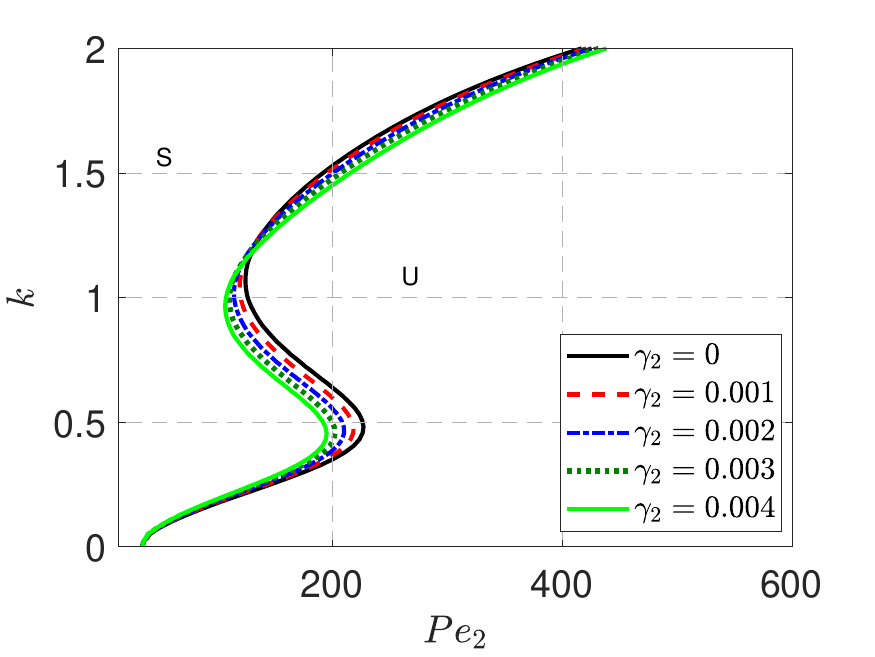}}
\subfigure[]{\includegraphics[width=5.4cm]{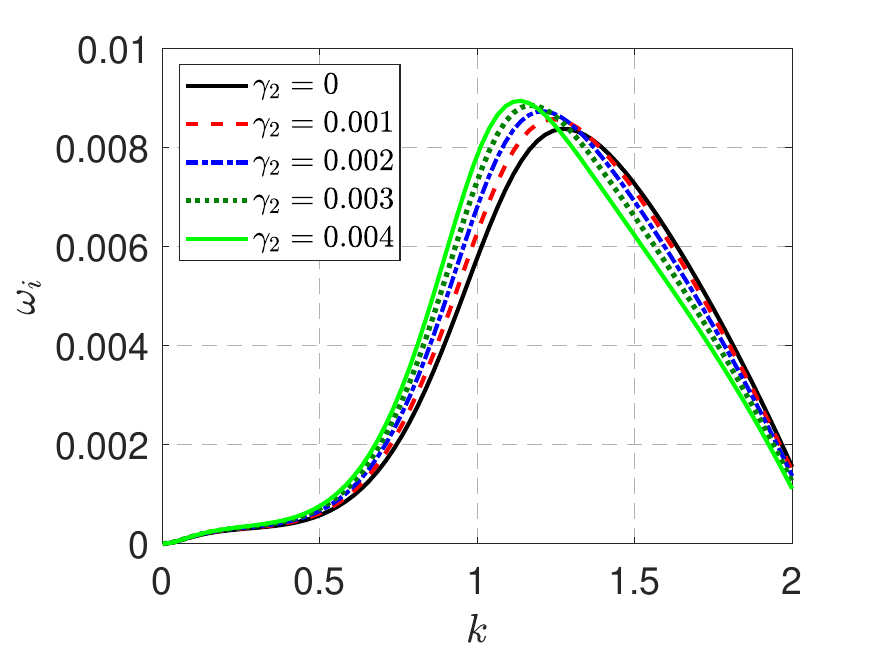}}
\end{center}
\caption{The unstable boundary lines ($\omega_i=0$) of the ISM in ($\displaystyle Pe_1-k$) plane for varying bottom-layer viscoelasticity $\gamma_2$ when (a) $\delta=1$ and (b) $\delta=0.5$. (c) The growth rate curves are corresponding to (b) when $Pe_2=100$. Here, the fixed value $\gamma_2=0.001$ with remaining parameters as in Fig.~\ref{f15}. }\label{f16}
\end{figure}
Here, even the non-zero bottom-layer viscoelasticity is fixed at $\gamma_2=0.001$, the result is validated with the work of \citet{bhat2020linear} as its effect on ISM is very weak, which is further confirmed in Fig.~\ref{f16}(c). It is noticed that the top-layered viscoelasticity $\gamma_1$ has a negligible influence on the ISM in the longwave region due to the critical P\'eclet number $Pe^c_2$ being independent of it, but has a significant effect in the finite wavenumber zone. The parameter $\gamma_1$ has a stabilizing impact on the ISM in the finite wave number region, followed by the shrinking of the corresponding unstable domain. This indicates that stronger viscoelasticity $\gamma_1$ weakens the interfacial surfactant instability in the finite wavenumber range. The corresponding growth rate results with $Pe_2=500$, as in Fig.~\ref{f15}(b), further supports these observations. Here, the temporal growth rate result also matches well with the work of \citet{bhat2020linear} in the limiting values $\gamma_1,\ \gamma_2\rightarrow 0$ and $r=1.1$. Moreover, the parameter $\gamma_1$ has a negligible effect on the ISM's growth rate in the longwave zone. However, the maximum growth rate decreases within a finite wavenumber range, which further supports the stabilizing behavior of ISM. Note that $\gamma_1$ does not always exert a stabilizing effect. The viscosity ratio $m$ can alter the instability behavior of the ISM under the influence of $\gamma_1$ (see Fig.~\ref{f222}).

Furthermore, Fig.~\ref{f16}(a) illustrates the stability boundaries in the ($Pe_2,k$ ) plane for varying bottom-layer viscoelasticity ($\gamma_2$).  As of $\gamma_1$, the parameter $\gamma_2$ has a negligible effect on the longwave ISM instability. Besides, the parameter $\gamma_2$ has a weak, as previously claimed, but dual influence on ISM: it exhibits a destabilizing effect at the shortwave zone (expanding the unstable region) while showing the opposite trend (shrinking the unstable region) beyond this zone. Note that, in the finite wavenumber region, compared to the strong impact of top-layer viscoelasticity ($\gamma_1$) on ISM, the role of $\gamma_2$ remains relatively minor in the considered flow parameter domain. However, one can achieve the significant dual effect of $\gamma_2$ on the ISM, as portrayed in Fig.~\ref{f16}(b), by changing the depth ratio to $\delta=0.6$ of the double-layered flow field. The thickness ratio $\delta$ has a negligible impact on the ISM instability in the longwave zone. However, the ISM instability in the finite wavenumber region expands as soon as the thickness ratio $\delta$ changes from $1$ to $0.6$. That means the thinner the lower layer, the stronger the ISM instability in the finite wavenumber zone.  The significant dual behavior of ISM is further validated by showcasing the corresponding temporal growth rate result in Fig.~\ref{f16}(c) when $Pe_2=100$. Here, increasing $\gamma_2$ enhances instability growth within a shortwave range but suppresses it beyond this range, reinforcing the observed dual effect.

\subsection{Results for the shear mode (SHM)}
\begin{figure}[ht!]
\begin{center}
\subfigure[]{\includegraphics[width=5.4cm]{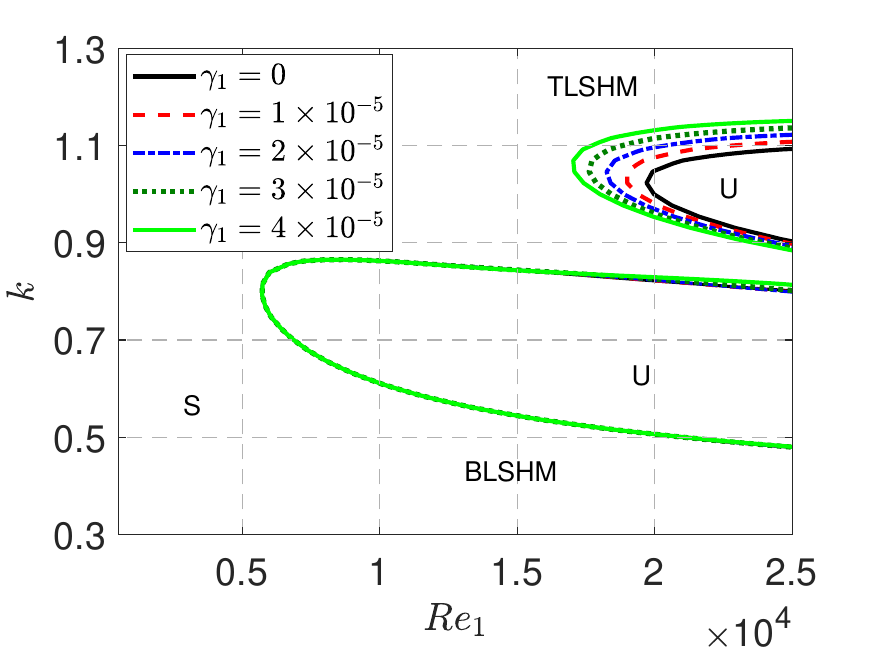}}
\subfigure[]{\includegraphics[width=5.4cm]{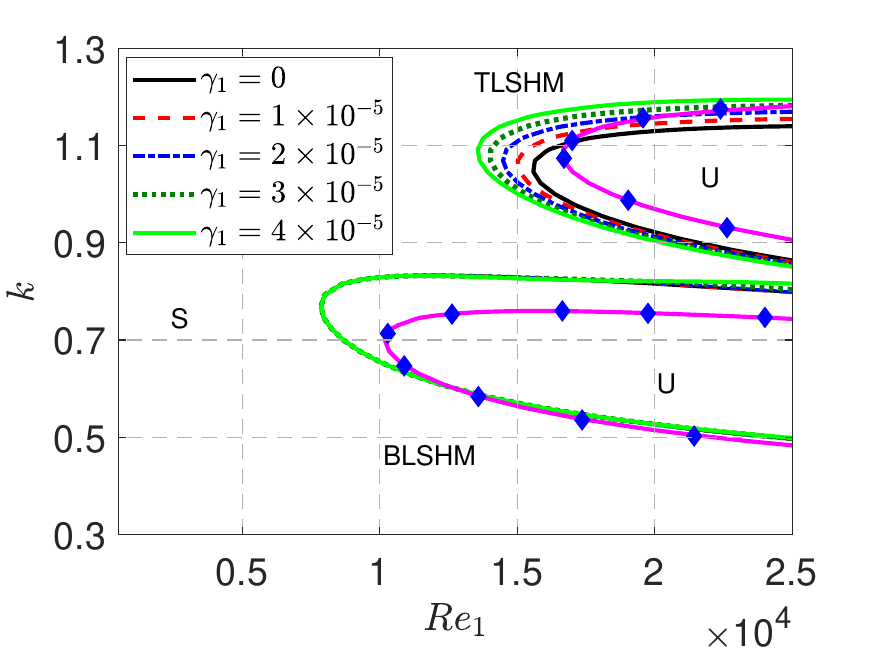}}
\subfigure[]{\includegraphics[width=5.4cm]{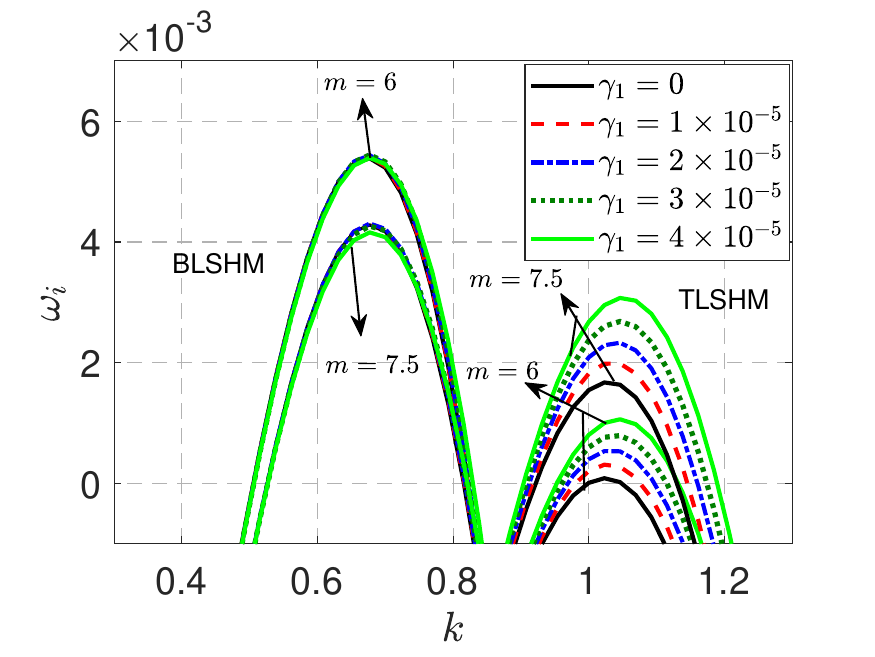}}
\end{center}
\caption{The effect of top-layer viscoelasticity $\gamma_1$ on the unstable boundary lines ($\omega_i=0$) of the SHM in ($\displaystyle Re_1-k$) plane when the viscosity ratio (a) $m=6$ and (b) $m=7.5$ and (c) the corresponding growth rate ($\omega_i$) when $Re_1=20000$. The remaining parameters are $\gamma_2=1\times 10^{-5}$,  $r=6$, $Ma_1=1$,  $Ma_2=1$, $Ca_1=1$,  $Ca_2=1$, $\theta=0.02~\textrm{rad}=1^{\circ}$, $Pe_1=\infty$ and $Pe_2=\infty$. Here, the solid magenta line represents the result with $\gamma_2=Ma_1=Ma_2=0$ and $r=5$, and the solid diamond blue shapes are the results of \citet{bhat2020linear}. }\label{f17}
\end{figure}

This subsection delineates the stability behavior of the SHM for both top and bottom layers of the double-layered viscoelastic fluid flow model over an inclined plane. When the density and viscosity of the lower layer are much higher than those of the upper layer, the SHM instability in both layers emerges with a very strong inertia force $Re_1$, and a very small inclination angle $\theta$ of the bounding wall. The neutral stability boundaries in the domain $Re_1-k$ plane of the SHM are presented in Fig.~\ref{f17} for different top-layered viscoelasticity $\gamma_1$ when  $m=6$ (see, Fig.~\ref{f17}(a)) and $m=7.5$ (see, Fig.~\ref{f17}(b)) with the parameter values $\gamma_2=1\times 10^{-5}$, $r=6$, $Ma_1=1$,  $Ma_2=1$, $Ca_1=1$,  $Ca_2=1$, and $\theta=0.02~\textrm{rad}$. The unstable boundary line fully matches with result of \citet{bhat2020linear} (see Fig.~\ref{f17}(b)) when $Ma_1=Ma_2=\gamma_2=0$ and $r=5$. The unstable region for the BLSHM (bottom-layered shear mode) exists in the shortwave range, whereas the unstable region for the TLSHM (top-layered shear mode) emerges in a comparatively higher wavenumber region. The wavenumber regions for the existence of SHM for both layers in the double-layered viscous fluid were also predicted by \citet{bhat2020linear, bhat2023linear}. Here the values of $\gamma_1$ is considered of $\mathcal{O}(10^{-5})$.
It is worth noting that the SHM instability for the bottom layer is much stronger than the TLSHM. Irrespective of the viscosity ratio $m=6$ (Fig.~\ref{f17}(a)) and $m=7.5$ (Fig.~\ref{f17}(b)), the parameter $\gamma_1$ significantly enhances the bandwidth of the unstable region of SHM related to the top-layer, which confirms the destabilizing effect of $\gamma_1$ on the TLSHM. In contrast, the viscoelastic parameter $\gamma_1$ shows the comparatively very weak impact of the BLSHM instability. However, considering high $\gamma_1$ values, one can significantly reduce the BLSHM instability (see Fig.~\ref{f21}). Now, as we increase the viscosity ratio from $m=6$ (Fig.~\ref{f17}(a)) to $m=7.5$ (Fig.~\ref{f17}(b)), the unstable region of the TLSHM amplifies, but the BLSHM shrinks for each viscoelasticity $\gamma_1$. So that increasing the bottom layer's viscosity weakens the shear wave instability of the bottom layer and strengthens the top-layered shear wave instability. A similar effect of the viscosity ratio $m$ on the SHM's primary instability in the double-layered Newtonian flow field was detected by \citet{bhat2020linear}. As the viscosity ratio $m$ increases, the bottom-layer's higher viscosity than the top-layer induces stronger velocity gradients (shear) near the liquid-liquid interface. So, the top layer (less viscous) experiences enhanced shear-induced disturbances, amplifying its shear wave instability. Meanwhile, the high-viscosity bottom-layered fluid naturally dissipates shear wave instability more effectively due to stronger viscous damping. The lower layer’s inertia is suppressed, reducing its shear wave instability. 
Thus, changing the viscosity of the double-layered viscoelastic flow field, one can regulate the shear wave instability and slow down the transition to turbulence of both shear layers. The effect of the viscoelastic parameter $\gamma_1$ on the SHM instability related to both layers with the change in viscosity ratio $m$ is further confirmed by the corresponding temporal growth rate curves $\omega_i$ as a function of $k$ in Fig.~\ref{f17}(c) when the unstable Reynolds number $Re_1=20000$. As expected, the growth rate of the BLSHM is much higher than that of the TLSHM. The top-layer viscoelastic parameter $\gamma_1$ has a weak effect on the BLSHM instability because the corresponding growth rate does not have a significant change for higher values of $\gamma_1$. On the other hand, the primary instability of the SHM associated with the top layer intensifies as the corresponding growth rate reduces for higher values of the viscoelastic parameter $\gamma_1$. Moreover, an increasing viscosity ratio $m$ increases the top-layer shear growth, while a reversed trend is observed for bottom-layer shear growth. Therefore, the nature of the growth rate with the change of viscosity ratio $m$ for different top-layered viscoelasticity $\gamma_1$ is fully consistent with the results from Figs.~\ref{f17}(a) and (b).       

\begin{figure}[ht!]
\begin{center}
\subfigure[]{\includegraphics[width=5.4cm]{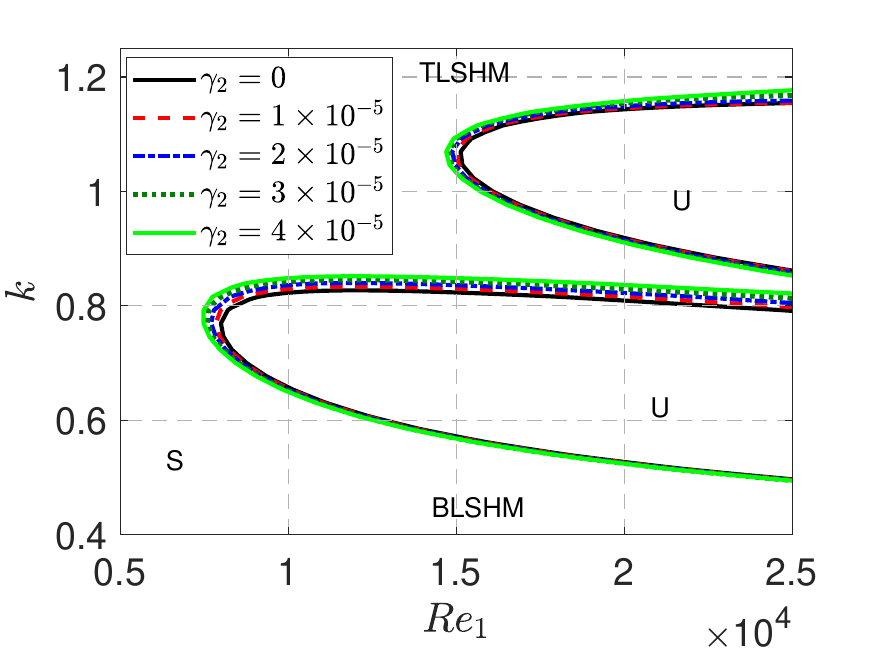}}
\subfigure[]{\includegraphics[width=5.4cm]{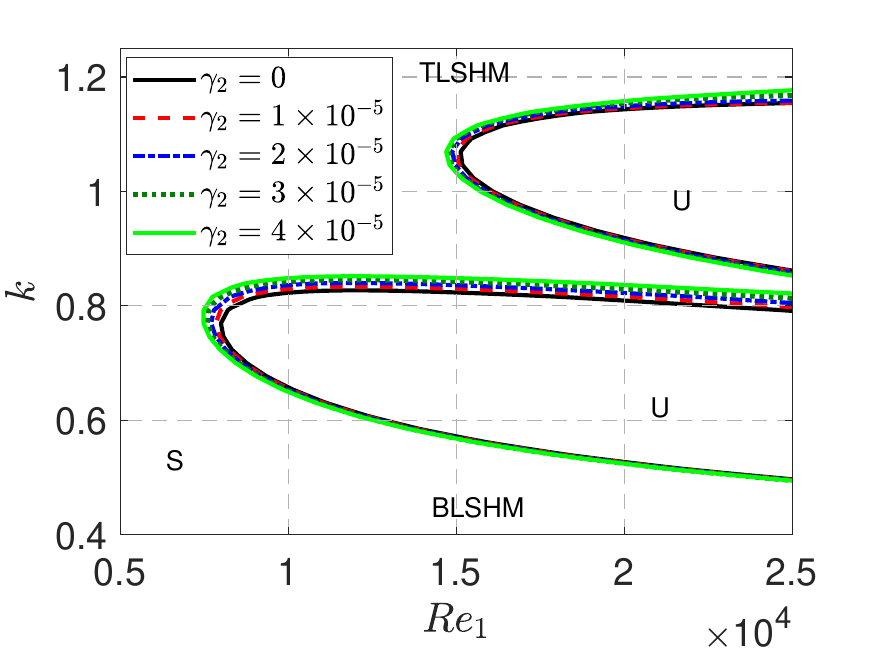}}
\subfigure[]{\includegraphics[width=5.4cm]{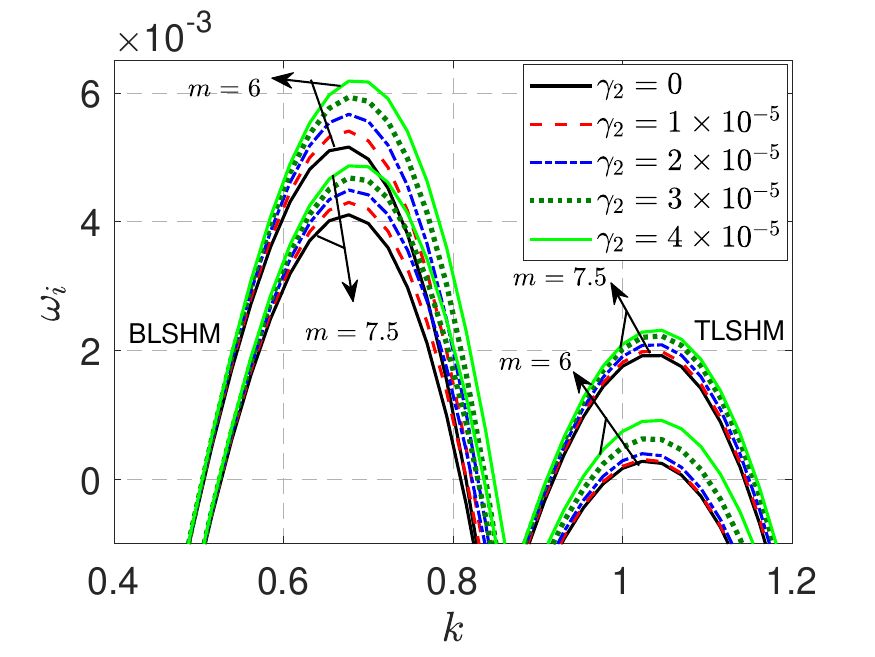}}
\end{center}
\caption{The effect of bottom-layer viscoelasticity $\gamma_2$ on the unstable boundary lines ($\omega_i=0$) of the SHM in ($\displaystyle Re_1-k$) plane when the viscosity ratio (a) $m=6$ and (b) $m=7.5$ and (c) the corresponding growth rate ($\omega_i$) when $Re_1=20000$. Here the value of $\gamma_1=1\times 10^{-5}$ with the remaining parameters are same as in Fig.~\ref{f17}.}\label{f18}
\end{figure}

On the other side, the bottom-layered viscoelastic parameter $\gamma_2$ has a destabilizing effect on the shear wave instability of both layers, i.e., the unstable regions correspond to both TLSHM and BLSHM expand as $\gamma_2$ enhances (see Fig.~\ref{f18})(a). Also, on increasing the viscosity ratio $m$ (i.e., from $m=6$ in Fig.~\ref{f17} to $m=7.5$ in Fig.~\ref{f17}(b)), the TLSHM instability intensifies followed by the increment of corresponding unstable bandwidth, while the bottom-layer's shear wave instability weakens followed by the decrement of corresponding unstable bandwidth for each value of $\gamma_2$. The temporal growth rate curves $\omega_i$, as in Fig.~\ref{f18}(c), corresponding to Figs.~\ref{f18}(a) and (b), further strengthen the destabilizing characteristic of the viscoelastic parameter $\gamma_2$ and the dual behavior of the viscosity ratio $m$. That means the bottom-layer viscoelasticity $\gamma_2$ successively enhances the temporal growth rate along with the unstable wavenumber domain for both BLSHM and TLSHM. Another novel finding from Fig.~\ref{f17} and \ref{f18} is that, compared to the SM and IM, the SHM instabilities for both top and bottom layers are highly sensitive with respect to the viscoelastic parameters.
\begin{figure}[ht!]
\begin{center}
\subfigure[]{\includegraphics[width=8.2cm]{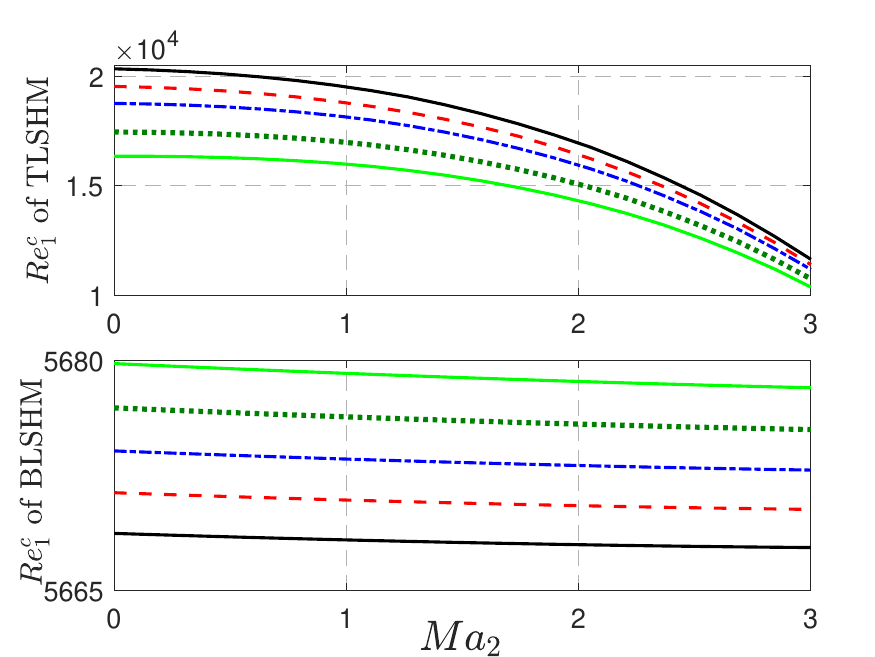}}
\subfigure[]{\includegraphics[width=8.2cm]{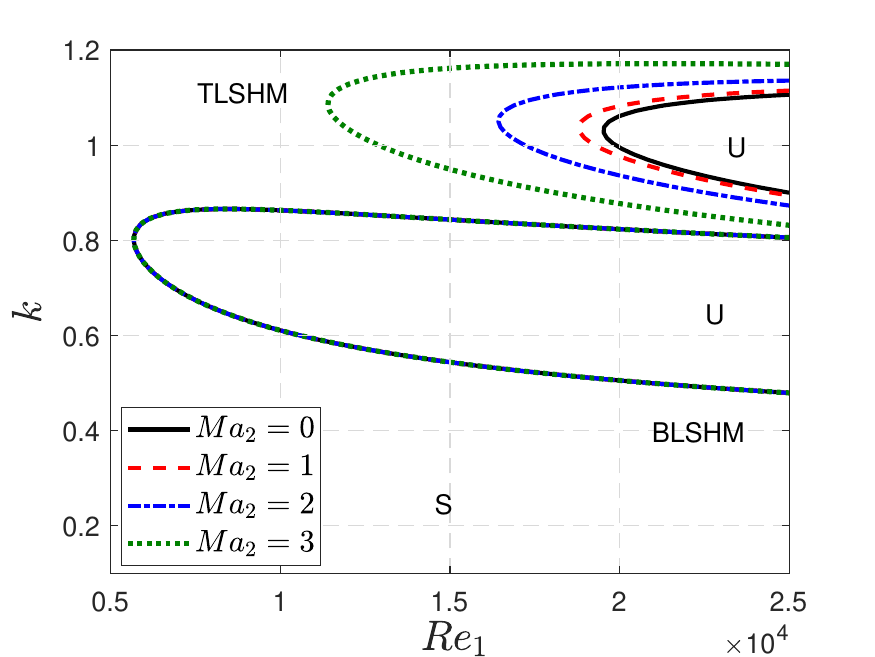}}
\end{center}
\caption{ (a) The stability boundary lines for SHM for different $Ma_2$ values when $\gamma_1=1\times 10^{-5}$. (b) The effect of top-layer viscoelasticity $\gamma_1$ on the critical Reynolds number $Re^c_1$ (for TLSHM at the top and BLSHM at the bottom) as a function of $Ma_2$. The black, red, blue, green, and magenta lines for $\gamma_1=0,~1\times 10^{-5},~2\times 10^{-5},~3\times 10^{-5},~\textrm{and}~4\times 10^{-5}$, respectively.  Here, the value of $\gamma_2=1\times 10^{-5}$ with the remaining parameters as in Fig.~\ref{f17}.}\label{f199}
\end{figure}

\begin{figure}[ht!]
\begin{center}
\subfigure[]{\includegraphics[width=8.2cm]{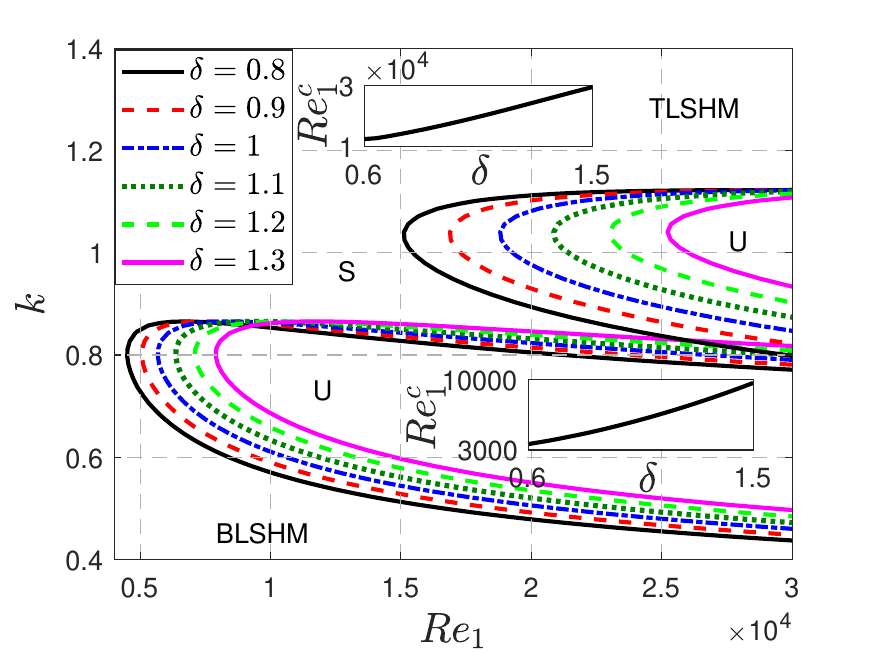}}
\subfigure[]{\includegraphics[width=8.2cm]{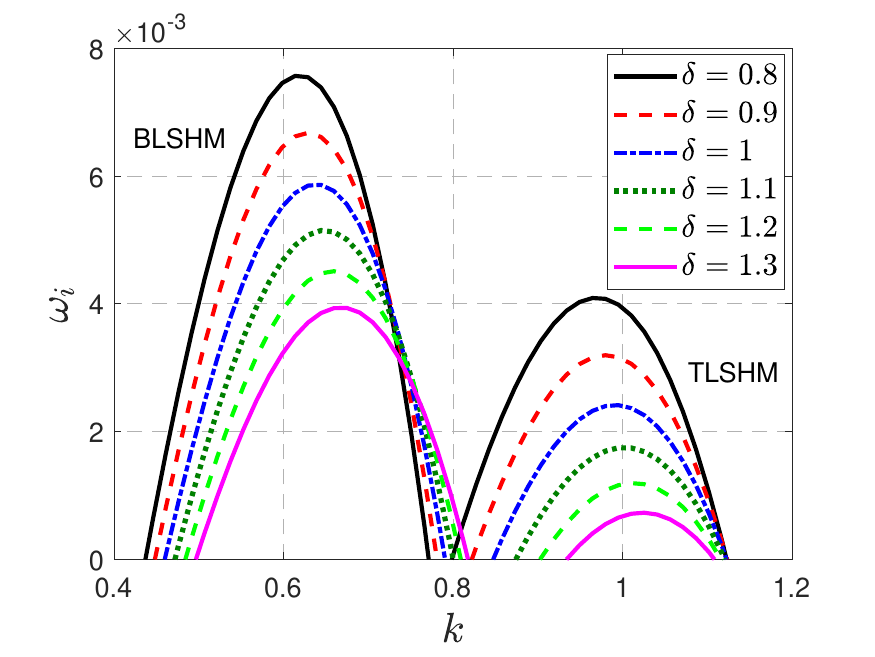}}
\end{center}\vspace{-0.5cm}
\caption{ The effect of thickness ratio $\delta$ on the unstable boundary lines ($\omega_i=0$) of the SHM in ($\displaystyle Re_1-k$) plane when the viscosity ratio and (c) the corresponding growth rate ($\omega_i$) when $Re_1=30000$. Here the value of $\gamma_1=\gamma_2=1\times 10^{-6}$, $m=6$, $r=6$ with the remaining parameters are same as in Fig.~\ref{f17}.}\label{f201}
\end{figure}

In Newtonian double-layered fluids, the destabilizing effect of interfacial surfactant-induced Marangoni force ($Ma_2$) on the TLSHM and its negligible influence on the BLSHM were conclusively demonstrated by \citet{bhat2020linear}. However, analogous investigations for non-Newtonian systems, particularly viscoelastic fluids, remain unexplored. To address this gap, Fig.~\ref{f199} presents our analysis of the interfacial Marangoni force ($Ma_2$) on shear-layer instabilities in a double-layered viscoelastic falling film. The critical Reynolds number ($Re^c_1$) for the TLSHM as a function of $Ma_2$ is depicted in Figure~\ref{f199}(a) (see the top subfigure) for different values of top layer viscoelasticity $\gamma_1$. It reveals that $Re^c_1$ is maximum for a clean interface ($Ma_2 = 0$) and decreases monotonically with increasing $Ma_2$ for all values of $\gamma_1$. In contrast, as shown in the bottom subfigure of Fig.~\ref{f199}(a), the decay rate of the function $Re^c_1$ for the BLSHM is very weak with $Ma_2$ across all $\gamma_1$ values. This confirms that the Marangoni force exerts minimal influence on the bottom-layer instability, a behavior consistent with Newtonian systems. To further validate these findings, we have portrayed the marginal stability curves of the shear mode (SHM) for varying $Ma_2$ by choosing a random viscoelastic parameter $\gamma_1=1\times 10^{-5}$ from Fig.~\ref{f199}(a). The results clearly demonstrate the destabilization effect of the Marangoni force on the TLSHM, as the unstable region expands with the higher Marangoni force $Ma_2$. In contrast, the BLSHM shows remarkable insensitivity to $Ma_2$, with no appreciable change in its unstable region across all tested values. This behavior aligns precisely with the $Re^c_1$ trends observed in Fig.~\ref{f199}(a). These findings mirror the behavior reported for Newtonian double-layered fluids by \citet{bhat2020linear}, reinforcing the conclusion that interfacial Marangoni forces predominantly affect the top-layer instability while leaving the bottom-layer dynamics essentially unchanged.

Fig.~\ref{f201} illustrates the behavior of unstable SHM in both viscoelastic layers as the thickness ratio ($\delta$) varies. The stability boundary curves in Fig.~\ref{f201}(a) demonstrate that the TLSHM instability weakens as $\delta$ increases. This is due to the gradual shrinkage of the unstable region, followed by the increase in the critical Reynolds number $Re^c_1$ (see the inset plot). Specifically, a thicker lower layer relative to the upper layer weakens TLSHM instability. In contrast, the parameter $\delta$ exhibits a dual role for BLSHM: the unstable region initially contracts within a certain wavenumber range (stabilizing effect) but expands beyond this range (destabilizing effect).
This stabilizing influence on TLSHM and the dual role of $\delta$ in BLSHM are further corroborated by the growth rate results in Fig.~\ref{f201}(b). For BLSHM, the maximum growth rate decreases significantly up to a critical wavenumber but increases thereafter with higher $\delta$, underscoring its dual effect. Meanwhile, larger $\delta$ values attenuate the growth rate of TLSHM, reinforcing its stabilizing behavior. These growth rate trends align perfectly with the stability boundaries depicted in Fig.~\ref{f201}(a).

\subsection{Competition among the unstable modes}\label{COMP}

\begin{figure}[ht!]
\begin{center}
\subfigure[]{\includegraphics[width=5.4cm]{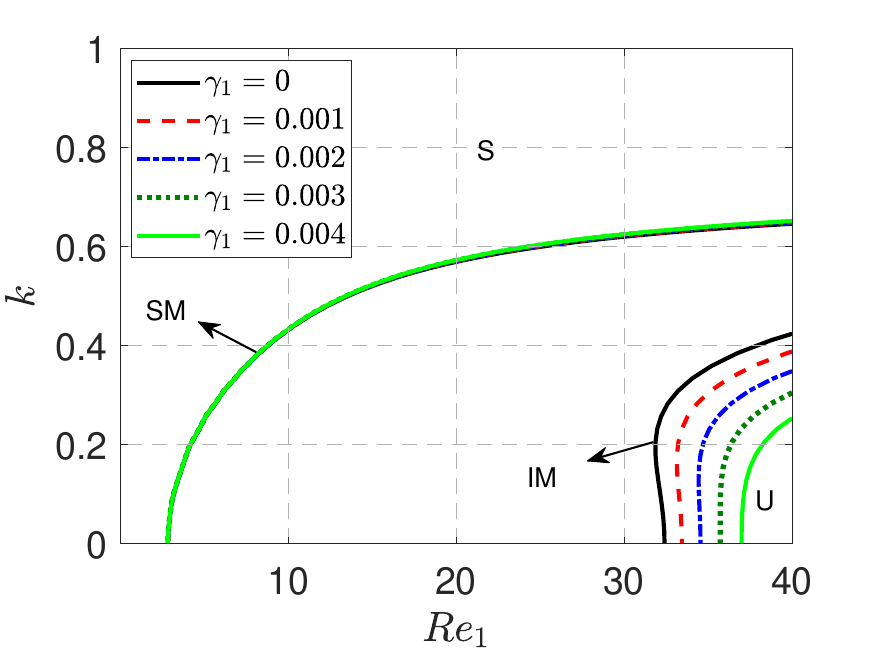}}
\subfigure[]{\includegraphics[width=5.4cm]{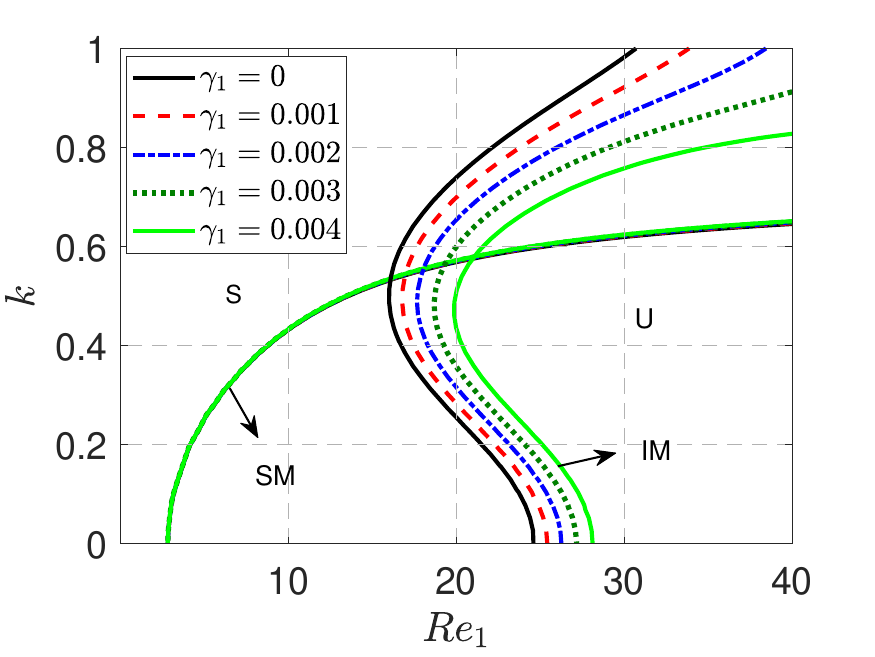}}
\subfigure[]{\includegraphics[width=5.4cm]{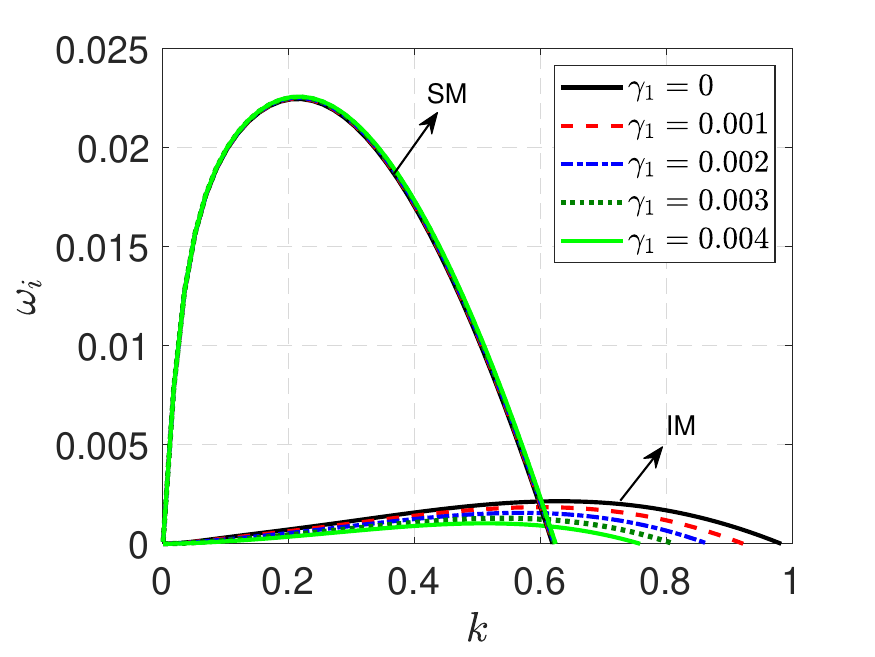}}
\end{center}
\caption{ The stability boundary lines related to SM and IM in $(Re_1,k)$ plane with varying $\gamma_1$ when (a) $Ca_2=1$ and (b) $Ca_2=4$. (c) The growth rate corresponding to (b), when  $Re_1=30$.  Here $\gamma_2=0.001$ with the remaining parameters are $\delta=1$, $r=1$, $m=1.5$, $\mathrm{Ca}_1=1$, $Ma_1=0.5$, $Ma_2=0.05$, $\theta=0.2~\textrm{rad}$, $Pe_1=\infty$ and $Pe_2=\infty$.  }\label{f20}
\end{figure}

This subsection focuses on comparing all the unstable modes identified in double-layered viscoelastic fluids flowing over an inclined wall. Thus, the numerical simulation is repeated to plot the unstable modes in the same flow parameter windows. In Fig.~\ref{f20}, the stability boundary lines with varying $\gamma_1$ for the unstable SM and IM are demonstrated in the same $Re_1-k$ window when $Ca_2=1$ (Fig.~\ref{f20}(a)) and when $Ca_2=4$ (Fig.~\ref{f20}(b)). Here top layer viscoelasticity $\gamma_1\sim \mathcal{O}(10^{-3})$. The novel finding is that the top layer viscoelasticity $\gamma_1\sim \mathcal{O}(10^{-3})$ has a negligible impact on the SM, but has a significant stabilizing effect on the IM, followed by decreasing the associated unstable region. This confirms that the IM is more sensitive compared to the SM. Moreover, as in Fig.~\ref{f20}(a), the onset of IM instability remains far away from that of SM when $Ca_2=1$. Meanwhile, the unstable SM boundary curves fully occupy the unstable boundary curves of IM, confirming that SM is fully dominant over IM. However, when $Ca_2=4$, the unstable $k-$ bandwidth of the IM expands to higher wavenumber regions (see Fig.~\ref{f20}(b)), but still the unstable $k-$ bandwidth of the SM remains almost the same. 
Thus, for a higher Capillary number, the SM loses its dominance over IM in the higher wavenumber region. This fact is further validated by displaying the corresponding temporal growth rate curves in Fig.~\ref{f20}(c). It is observed that SM's growth rate is significantly higher than that of IM in the longwave region; however, beyond this region, SM becomes fully stable, whereas IM remains unstable. Therefore, changing the interfacial Capillary force, one can alter the instability dominance between the SM and IM in the double-layered viscoelastic fluid flow model.

\begin{figure}[ht!]
\begin{center}
\subfigure[]{\includegraphics[width=5.4cm]{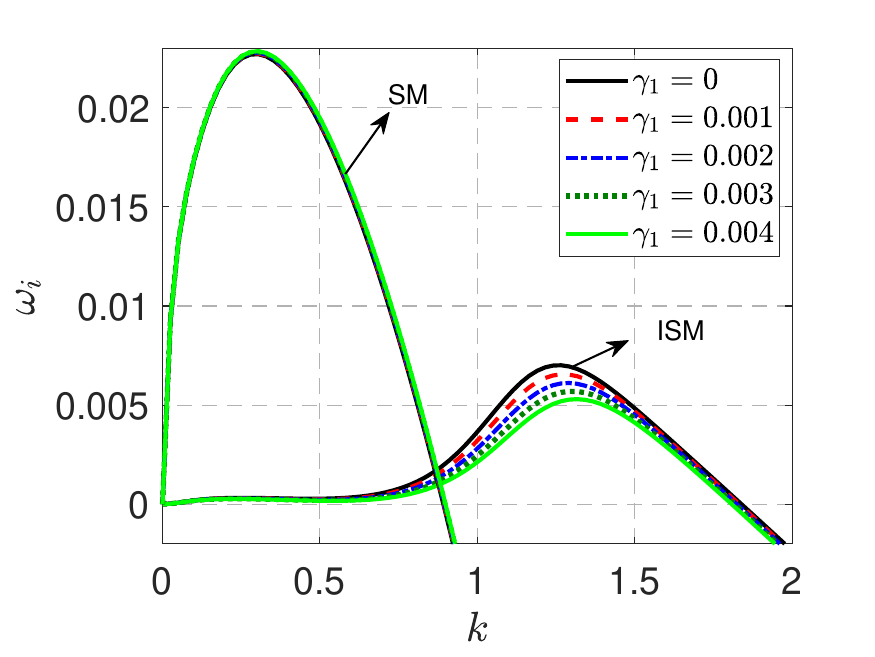}}
\subfigure[]{\includegraphics[width=5.4cm]{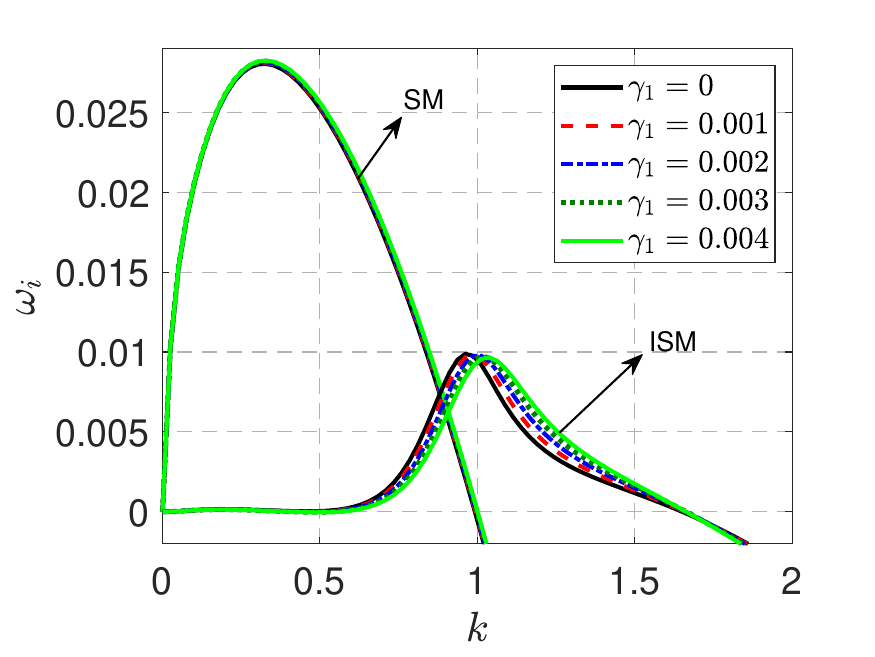}}
\subfigure[]{\includegraphics[width=5.4cm]{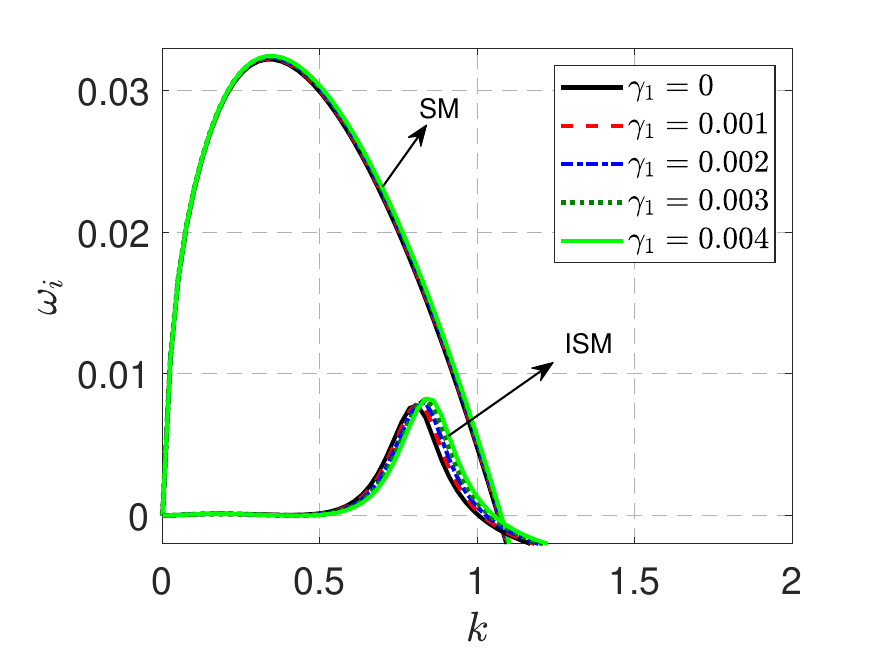}}
\end{center}
\caption{  The growth rate curves related to SM and ISM with varying $\gamma_1$ when (a) $m=1.4$, (b) $m=2.1$, and (c) $m=2.8$.Here $\gamma_2=0.001$ with the remaining parameters are $\delta=1$, $r=1$, $Re_2=50$, $\mathrm{Ca}_2=1$, $\mathrm{Ca}_1=1$, $Ma_1=0.01$, $Ma_2=0.01$, $\theta=0.2~\textrm{rad}$, $Pe_1=\infty$ and $Pe_2=\infty$. }\label{f222}
\end{figure}

Fig.~\ref{f222} shows the change of SM's dominance over the ISM with the change of top layer viscoelasticity $\gamma_1$ when the viscosity ratio $m$ alters. 
When $m=1.5$ (i.e., higher bottom layer viscosity than top layer), the unstable SM dominates the ISM in the longwave region (see Fig.~\ref{f222}(a)). However, there exists a shortwave unstable range for the ISM where SM's growth rate is negative, indicating that the ISM's primary instability occurs in the shortwave zone, while the SM becomes stable in this zone. Now, once we increase the viscosity ratio to $m=2.1$ (i.e., increasing the bottom layer viscosity with $m>1$), the unstable $k-$ domain of ISM shrinks, while it increases for the SM (see Fig.~\ref{f222}(b)). That means with increasing bottom layer viscosity (in $m>1$), the dominance of ISM's instability reduces, while it enhances for the SM. Now, if we increase the viscosity ratio further to $m=2.8$ (i.e., bottom layer viscosity increases further in $m>1$), the unstable $k-$domain of ISM shrinks further, and the SM fully dominates the ISM in the unstable $k-$region (see Fig.~\ref{f222}(c)). The growth rate of the longwave SM instability is much higher than that of ISM, which implies the high intensity of SM instability compared to the ISM in the double-layered viscoelastic fluid. Moreover, the parameter $\gamma_1$ has a dual influence on the ISM in the finite wave number range (i.e., stabilizes up to a certain wave number range and then destabilizes) for $m=2.1$ (Fig.~\ref{f222}(b)) and $m=2.8$ (Fig.~\ref{f222}(b)). That means the effect of parameter $\gamma_1$ on the ISM depends on the choice of viscosity ratio $m$. Another novel finding from Fig.~\ref{f222} is that for all values of $m$, the viscoelastic coefficient $\gamma_1$ has a negligible impact on the SM, but has a significant effect on the ISM.  This implies that the ISM is more sensitive than the SM under the viscoelastic property.

\begin{figure}[ht!]
\begin{center}
\subfigure[]{\includegraphics[width=5.4cm]{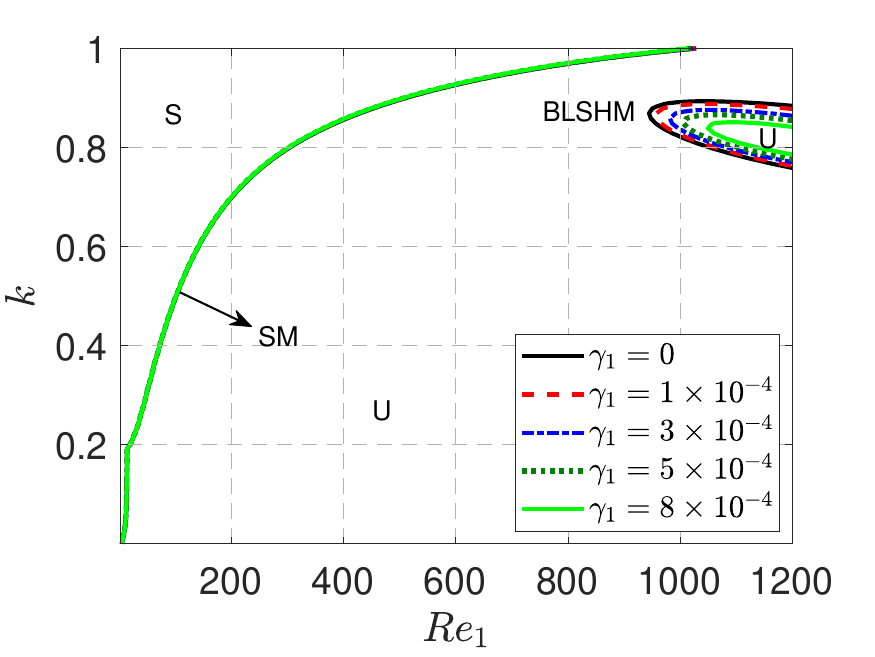}}
\subfigure[]{\includegraphics[width=5.4cm]{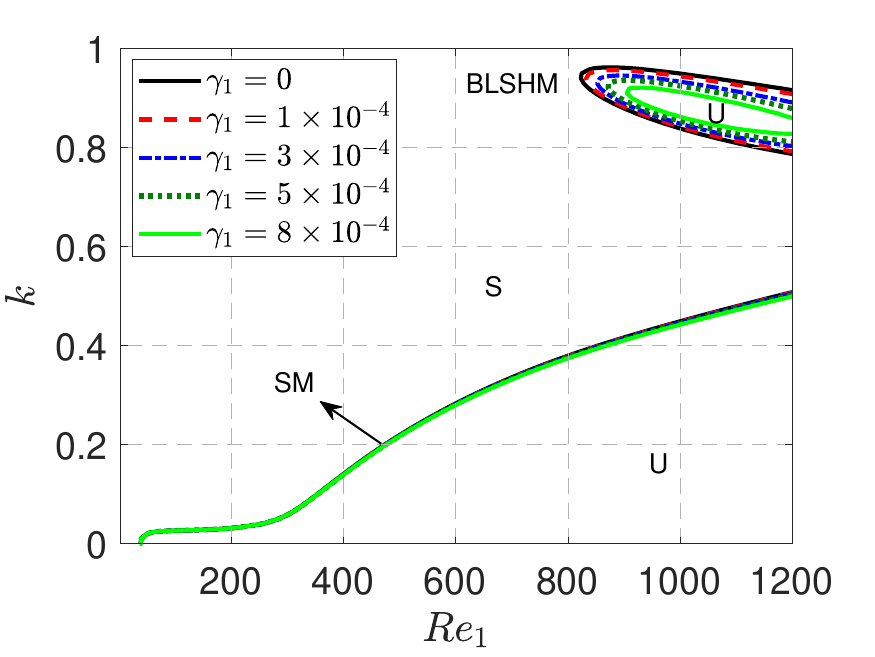}}
\subfigure[]{\includegraphics[width=5.4cm]{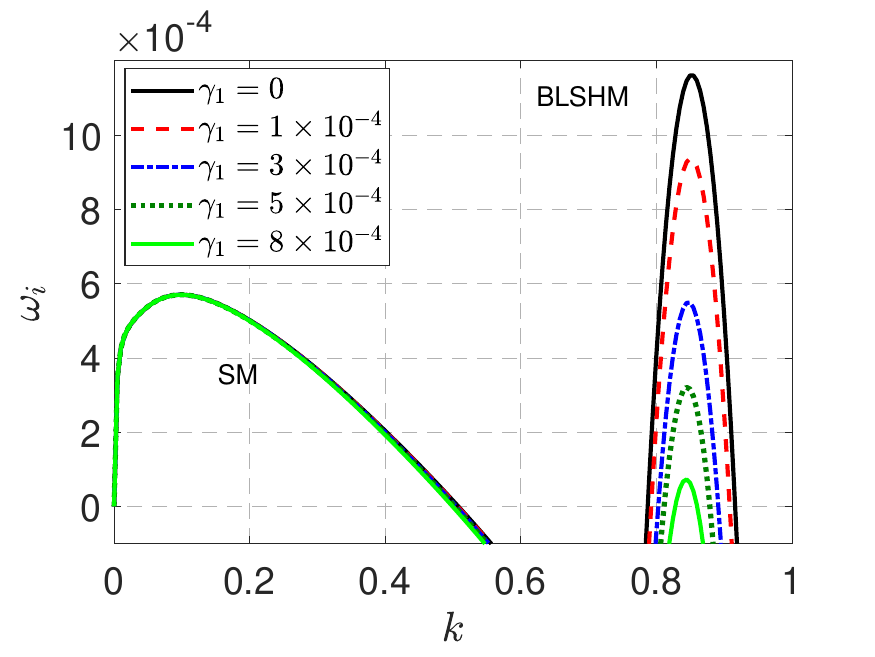}}
\end{center}
\caption{ The stability boundary lines related to SM and BLSHM in $(Re_1,k)$ plane with varying $\gamma_1$ when (a) $\theta=0.08~\textrm{rad}$ and (b) $\theta=0.008~\textrm{rad}$. (c) The growth rate corresponding to (b), when  $Re_1=1200$. Here $\gamma_2=1\times 10^{-5}$ with the remaining parameters are $\delta=1$, $r=6$, $m=1.5$, $\mathrm{Ca}_1=1$, $\mathrm{Ca}_2=1$, $Ma_1=0.1$, $Ma_2=2$, $Pe_1=\infty$ and $Pe_2=\infty$. }\label{f21}
\end{figure}

Furthermore, we have compared the dominance of unstable SM and BLSHM with varying $\gamma_1\sim \mathcal{O}(10^{-4})$, when the inclination angles of the bounding well are $\theta=0.08~\textrm{rad}$ (Fig.~\ref{f21}(a)) and $\theta=0.008~\textrm{rad}$ (Fig.~\ref{f21}(b)). As expected, the top layer viscoelasticity $\gamma_1\sim \mathcal{O}(10^{-4})$ has a negligible impact on the SM, but has a significant stabilizing effect on the BLSHM, which confirms that the BLSHM is highly sensitive compared to the SM. Besides, when $\theta=0.08~\textrm{rad}$, the unstable boundary lines fully occupy the unstable boundary lines of BLSHM in the whole $Re_1-k$ domain. This implies that when the bounding wall is inclined enough, the SM dominates the BLSHM. However, as long as the inclination angle decreases to $\theta=0.008~\textrm{rad}$ (i.e., almost horizontal bottom wall), the BLSHM fully dominates the SM in the higher wavenumber region. This fact of dominance of BLSHM in the higher wavenumber region is further assured by the corresponding temporal growth rate profiles in Fig.~\ref{f21}(c). The parameter $\gamma_1$ has negligible influence on the SM's growth rate, but $\gamma_1$ remarkably mitigates the BLSHM's growth rate (i.e., stabilizing effect). The parameter $\gamma_1$ has a negligible influence on the SM's growth rate, yet it induces a pronounced stabilization of the BLSHM by markedly reducing its growth rate. Moreover, in the shortwave region, BLSHM emerges, where the growth rate of the SM is completely negative, which indicates the dominance of the BLSHM in the shortwave region.

\section{\bf{Conclusions}}\label{CON}
In this work, we focus on the linear stability analysis of a double-layered immiscible viscoelastic (weakly elastic) fluid flowing over an inclined plane, where the insoluble surfactant is present at both the top-layer surface and interface. The weak viscoelasticity of both layers is formulated by Walter's $B^{''}$  model. The corresponding Orr-Sommerfeld equations, along with the boundary conditions, are solved numerically using the Chebyshev spectral collocation technique. The numerical results detect three distinct unstable modes: SM, IM, and ISM. The viscoelasticity of both layers initially strengthens the surface mode (SM) instability near the instability threshold. However, this potent SM instability can be effectively dissipated by applying a surface surfactant, which results in a comparatively slow top-layer flow. Besides, the numerical simulation confirms that the characteristics of the interfacial wave instability depend on the viscosity and density stratification. The nature of IM's primary instability differs in the disjoint regions $m>1$ and $m<1$. When $m>1$, the top-layer viscoelasticity $\gamma_1$ weakens the IM instability, whereas bottom-layer viscoelasticity ($\gamma_2$) enhances it. However, the IM instability intensified by $\gamma_2$ can be significantly suppressed by imposing the interfacial Marangoni force. 
On the other hand, for $m<1$, top-layer viscoelasticity exerts a stabilizing influence on the IM, while the bottom layers' viscoelastic coefficient plays a double role: destabilization in the longwave zone, and stabilization in the shortwave region. An interfacial Marangoni force can dissipate the interfacial instability raised by the viscoelastic property for $m>1$, but can boost the interfacial instability in the case of $m<1$. Moreover, it is found that the viscoelastic property of both layers has a negligible effect on the ISM in the longwave region, but has a substantial influence in the finite wavenumber zone. Specifically, increasing top-layer viscoelasticity ($\gamma_1$) attenuates the ISM in this zone. This stabilizing behavior, however, can be modulated by a suitable choice of viscosity ratio. Furthermore, the ISM exhibits a dual response to the bottom-layer viscoelasticity $\gamma_2$ based on its proximity to the instability threshold: it is destabilized near the onset but stabilized far away from it. 
Additionally, the shear wave instabilities for both viscoelastic layers emerge in the flow configuration with a very strong inertia force and a low inclination angle when the lower layer's viscosity and density are significantly higher than those of the upper layer. The TLSHM can be destabilized by increasing top-layer/bottom-layer viscoelasticity, whereas the BLSHM can be stabilized/destabilized by choosing higher viscoelastic parameter values in the top/bottom-layer. However, the viscoelasticity of the top layer exerts a weak influence on the BLSHM compared to the TLSHM. Furthermore, the interfacial surfactant-induced Marangoni force has a significant destabilizing impact on the TLSHM instability, while it has a very weak impact on the BLSHM. Among all the identified unstable modes, the shear wave instability for both layers demonstrates markedly greater sensitivity to the viscoelastic coefficients of both layers, while the interface and interface surfactant modes are more sensitive than the SM.

%%%%%%%%%%%%%%%%%%%%%%%%%%%%%%%%%%%%%%%%%%%%%%%%%%%%%%%%%%%%%%%%%%%%%%%%%%%%%%%%%%%%%%%%%%%%%%
%%%%%%%%%%%%%%%%%%%%%%%%%%%%%%%%%%%%%%%%%%%%%%%%%%%%%%%%%%%%%%%%%%%%%%%%%%%%%%%%%%%%%%%%%%%%%%
%%%%%%%%%% END OF CONCLUSIONS %%%%%%%%%%%%%%%%%%%%%%%%%%%%%%%%
%--------------------------------------------------------------------------------------------%
%--------------------------------------------------------------------------------------------%
%%%%%%%%%%%%%%%%%%%%%%%%%%%%%%%%%%%%%%%%%%%%%%%%%%%%%%%%%%%%%%%%%%%%%%%%%%%%%%%%%%%%%%%%%%%%%%
%%%%%%%%%%%%%%%%%%%%%%%%%%%%%%%%%%%%%%%%%%%%%%%%%%%%%%%%%%%%%%%%%%%%%%%%%%%%%%%%%%%%%%%%%%%%%%
%%%%%%%%%% ACKNOWLEDGEMENTS%%%%%%%%%%%%%%%%%%%%%%%%%%%%%%%%
\appendix
\section{Expression for the dimensionless stress tensor components } \label{app_1}
\allowdisplaybreaks
\begin{align}
    & \tau_{xx}^{(1)}=-Re_1p^{(1)}+2 u^{(1)}_x-2\gamma_1 Re_1\biggl[u^{(1)}_{xt}+u^{(1)}u^{(1)}_{xx}+v^{(1)}u^{(1)}_{xy} -2 (u^{(1)}_x)^2\nonumber\\
    &\hspace{8cm}-u^{(1)}_y\biggl(u^{(1)}_y+v^{(1)}_x\biggr)\biggr],\tag{A.1}\\
    & \tau_{yy}^{(1)}=-Re_1p^{(1)}+2v^{(1)}_y-2\gamma_1 Re_1\biggl[v^{(1)}_{yt}+v^{(1)}v^{(1)}_{ yy}+u^{(1)}v^{(1)}_{xy} -2( v^{(1)}_y)^2\nonumber\\
    &\hspace{8cm}-v^{(1)}_x\biggl(u^{(1)}_y+v^{(1)}_x\biggr)\biggr],\tag{A.2}\\
    &  \tau_{xy}^{(1)}=\biggl(u^{(1)}_y+ v^{(1)}_x\biggr)-\gamma_1 Re_1\biggl[u^{(1)}_{yt}+ v^{(1)}_{xt}+u^{(1)}\biggl( v^{(1)}_{xx} + u^{(1)}_{xy}\biggr)+ v^{(1)}\biggl(u^{(1)}_{y y} + v^{(1)}_{x  y}\biggr)\nonumber\\
    &\hspace{8cm}-2u^{(1)}_{x}v^{(1)}_{x}-2u^{(1)}_{y}v^{(1)}_{y}\biggr],\tag{A3}\\
    & \tau_{xx}^{(2)}=-mRe_2p^{(2)}+2m u^{(2)}_x-2m\gamma_2 Re_2\biggl[u^{(2)}_{xt}+u^{(2)}u^{(2)}_{xx}+v^{(2)}u^{(2)}_{xy} -2 (u^{(2)}_x)^2\nonumber\\
    &\hspace{8cm}-u^{(2)}_y\biggl(u^{(2)}_y+v^{(2)}_x\biggr)\biggr],\tag{A.4}\\
    & \tau_{yy}^{(2)}=-mRe_2p^{(2)}+2mv^{(2)}_y-2m\gamma_2 Re_2\biggl[v^{(2)}_{yt}+v^{(2)}v^{(2)}_{ yy}+u^{(2)}v^{(2)}_{xy} -2( v^{(2)}_y)^2\nonumber\\
    &\hspace{8cm}-v^{(2)}_x\biggl(u^{(2)}_y+v^{(2)}_x\biggr)\biggr],\tag{A.5}\\
    &  \tau_{xy}^{(2)}=m\biggl(u^{(2)}_y+ v^{(2)}_x\biggr)-m\gamma_2 Re_2\biggl[u^{(2)}_{yt}+ v^{(2)}_{xt}+u^{(2)}\biggl( v^{(2)}_{xx} + u^{(2)}_{xy}\biggr)+ v^{(2)}\biggl(u^{(2)}_{y y} + v^{(2)}_{x  y}\biggr)\nonumber\\
    &\hspace{8cm}-2u^{(2)}_{x}v^{(2)}_{x}-2u^{(2)}_{y}v^{(2)}_{y}\biggr],\tag{A.6} 
   \end{align}
 where $\displaystyle\tau^{(i)}_{lm}=\frac{d^{(1)}}{\mu^{(1)} U_c}\Bar{\tau}^{(i)}_{lm}$ with the dimensional stress components $\Bar{\tau}^{(i)}_{lm}$ and $\displaystyle \gamma_i=\frac{E^{(i)}}{\rho^{(i)} (d^{(i)})^2}$ is the nondimensional viscoelastic parameters ($i=1,~2$) correspond to the top layer for $i=1$ and bottom layer for $i=2$. 
 %%%%
\section{Behavior of IM in the disjoint region \texorpdfstring{$r>1$}{r} and \texorpdfstring{$r<1$}{r} } \label{app_2}
\setcounter{figure}{0}
 \renewcommand{\thefigure}{B.\arabic{figure}}
\begin{figure}[ht!]
\begin{center}
\subfigure[]{\includegraphics[width=5.4cm]{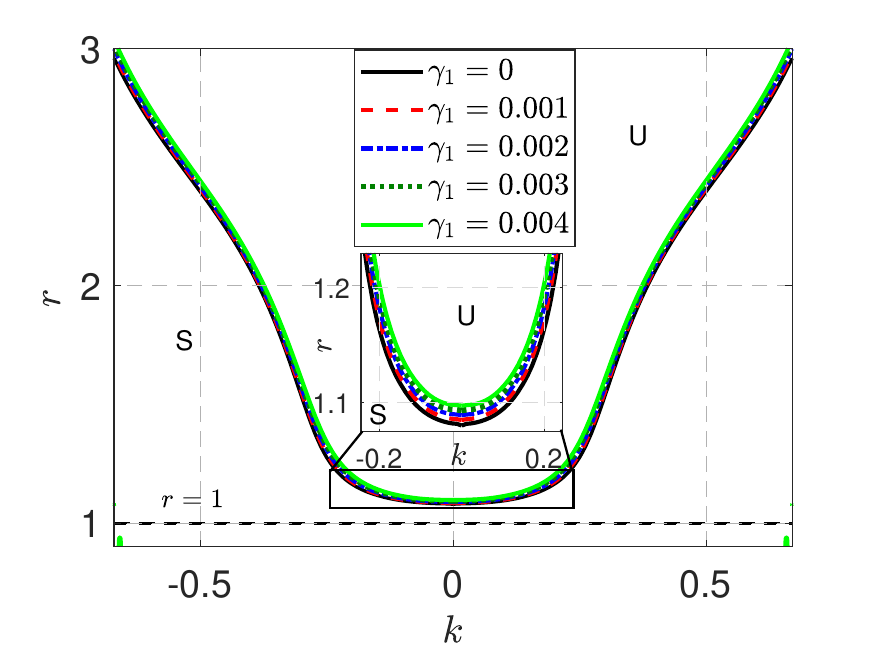}}
\subfigure[]{\includegraphics[width=5.4cm]{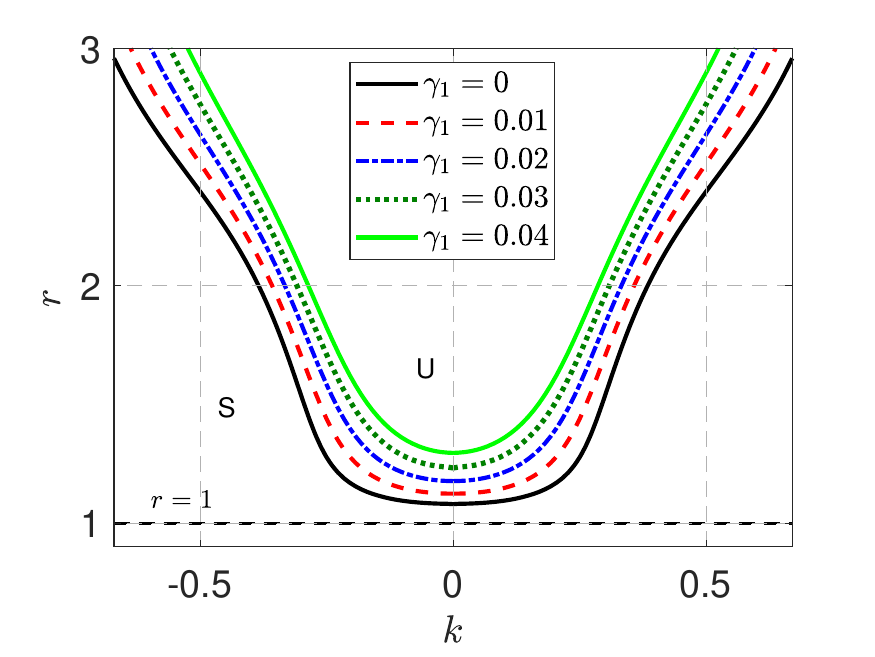}}
\subfigure[]{\includegraphics[width=5.4cm]{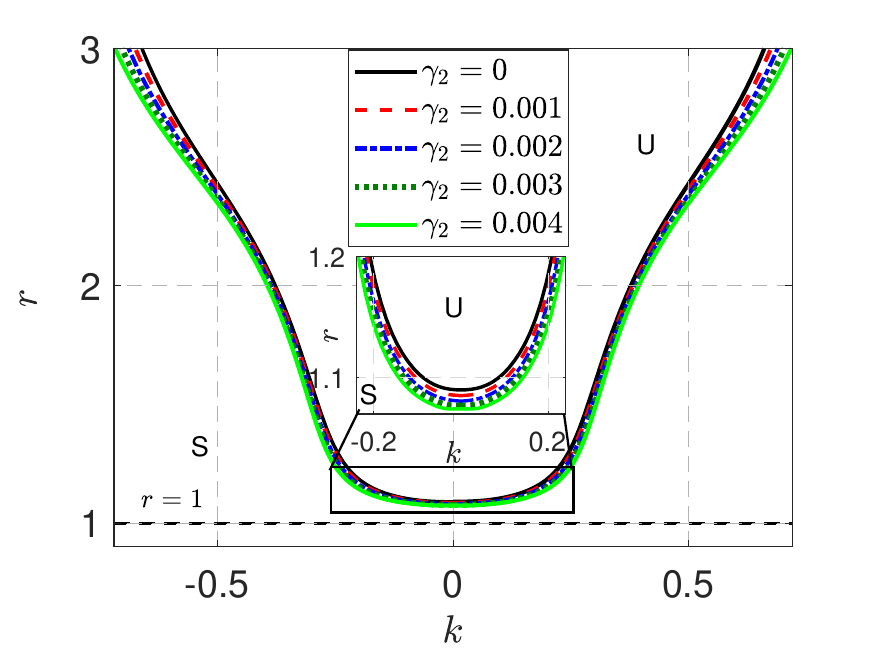}}
\end{center}
\caption{The variation of unstable boundary lines ($\omega_i=0$) of the IM in ($\displaystyle k-r$) plane with the change in top-layer viscoelasticity (a) $\gamma_1\sim O(10^{-3})$ and (b) $\gamma_1\sim \mathcal{O}(10^{-2})$. (c) The variation of unstable boundary lines ($\omega_i=0$) of the IM in ($\displaystyle k-r$) plane with the change in bottom-layer viscoelasticity $\gamma_2\sim \mathcal{O}(10^{-3})$ when $\gamma_1=1\times 10^{-3}$. Here, the results are presented in the case of $r>1$ and $m=1$, with the remaining parameters as shown in Fig.~\ref{f11}. }\label{f13}
\end{figure}
\begin{figure}[ht!]
\begin{center}
\subfigure[]{\includegraphics[width=8.2cm]{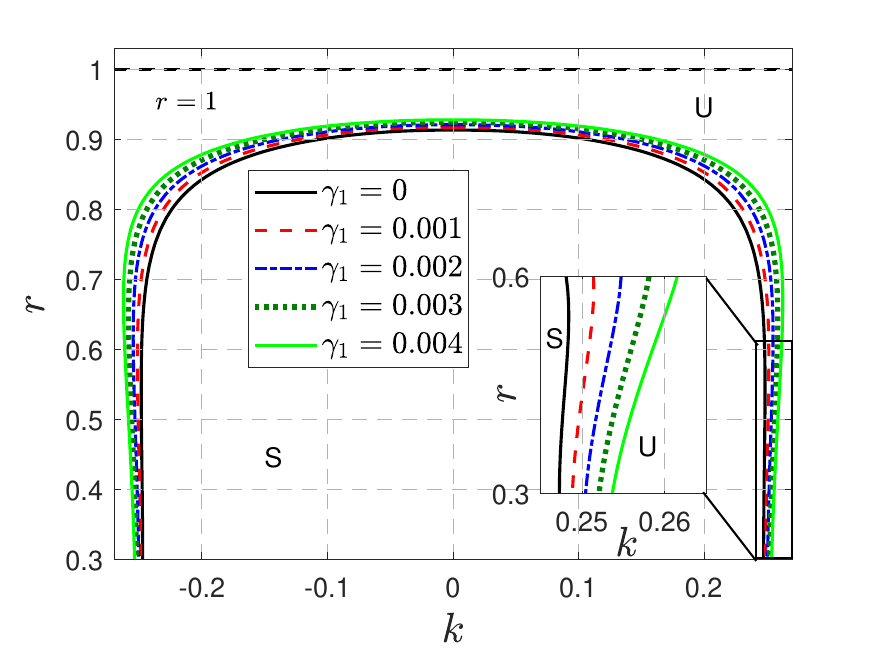}}
\subfigure[]{\includegraphics[width=8.2cm]{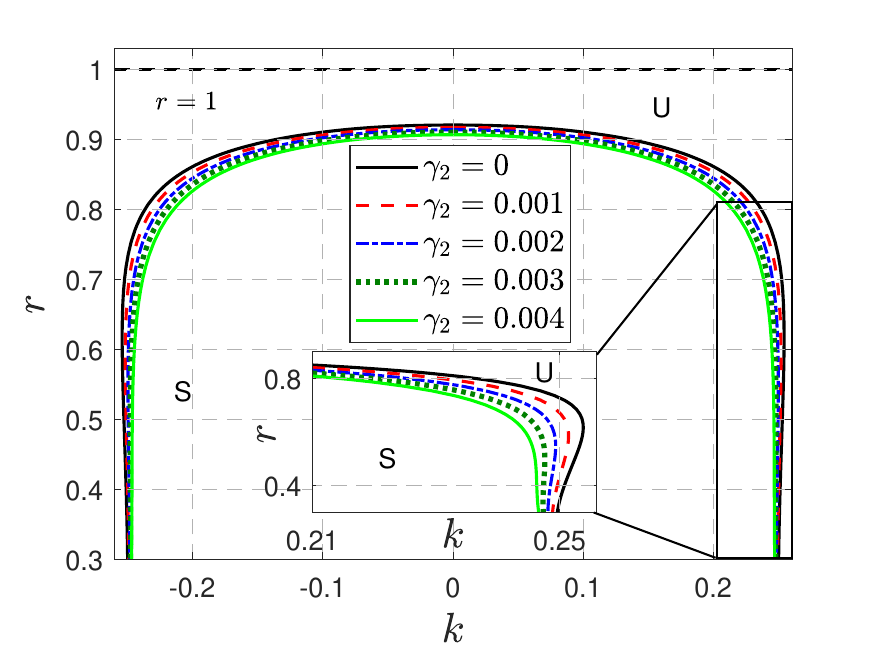}}
\end{center}
\caption{ The variations of the unstable boundary lines ($\omega_i=0$) related to the IM in ($\displaystyle k-r$) plane when (a) $\gamma_1$ alters with $\gamma_2=0.001$ and (b) $\gamma_2$ alters with $\gamma_1=0.001$ in the case of $r<1$. Here $m=1$ with the remaining parameters as in Fig.~\ref{f11}. }\label{f14}
\end{figure}
The instability behavior of unstable IM is analyzed under varying density ratio $r$ when the viscoelasticity of both layers is altered individually. As of the viscosity ratio $m$, the instability characteristic of IM differs significantly between the disjoint regions $r>1$ (i.e., higher bottom layer density than top layer) and $r<1$ (i.e., lower bottom layer density than top layer), as evident in Figs.~\ref{f13} and \ref{f14}, respectively. In the  $r>1$ region, when $\gamma_1\sim \mathcal{O}(10^{-3})$ is increased for a fixed $\gamma_2$, IM instability, as shown in Fig.~\ref{f13}(a), weakly reduces in the longwave regime, accompanied by a progressive contraction. However, one can boost this stabilizing effect by enhancing the changing rate of $\gamma_1$ (see Fig.~\ref{f13}(b), where $\gamma_1\sim \mathcal{O}(10^{-2})$ increasing from $0$ to $0.04$). Thus, in the region $r>1$, the lower-layer viscoelasticity ($\gamma_1$) exhibits a stabilizing nature near the instability onset.  
In contrast, increasing the bottom layer's viscoelasticity ($\gamma_2$) amplifies IM instability (Figs.~\ref{f13}(c). On the other side, for the region $r<1$, IM instability arises exclusively in the longwave regime, unlike the case for 
$r>1$, indicating that interfacial wave instability is observable near the threshold when the bottom layer is less dense than the top layer. Here, higher values of the lower-layer viscoelasticity ($\gamma_1$) suppress longwave IM instability by shrinking the corresponding unstable region (Figs.~\ref{f14}(a)), similar to the behavior observed in the case of $r>1$. However, a reverse trend, i.e., a destabilizing effect, occurs with increasing bottom-layer viscoelasticity ($\gamma_2$), as shown in Figs.~\ref{f14}(b). Thus, irrespective of the regions $r<1$ and $r>1$, the viscoelasticity of the top-layer/bottom-layer has the potential to stabilize/destabilize the unstable IM. 
 \vspace{5in}
\begin{center}
{\bf Declaration of interests}
\end{center}
\vspace{0.1in}
There are no conflicts of interest to report. 
% \section*{Declaration of competing interest}
% The author declares that he has no known competing financial interests or personal relationships that could have appeared to influence the work reported in this paper. % \section*{Acknowledgment}
\vspace{0.25in}
\begin{center}
{\bf Acknowledgements}
\end{center}
\vspace{0.1in}
This work is supported by the Fundamental Research Funds for the Central Universities of China (Grants No. AUGA2160100324 and No. AUGA2160503123). 
% The first author would like to acknowledge the help he received from Dr. Mohamin B.M. Khan, Department of Mathematics, Manarat Riyadh International, Riyadh, Saudi Arabia during the preparation of this manuscript. 
\\	
	 
% \section*{Credit authorship contribution statement}
% \noindent Md. Mouzakkir Hossain: Conceptualization, Methodology, Software, Writing - original draft, Validation, Formal analysis, Investigation.
% \\
% The first author would like to acknowledge the help he received from Dr. George Karapetsas, (Aristotle University of Thessaloniki, Greece) during the preparation of this manuscript.
% \\
% Youchuang Chao: Conceptualization, Review \& Editing.

\section*{Data Availability}
 The data that supports the findings of this study are available
within the article, highlighted in the related figure captions and corresponding discussions.

\bibliographystyle{unsrtnat}
\bibliography{REF}
\end{document}